\documentclass[fleqn,usenatbib]{mnras}

\usepackage{newtxtext,newtxmath}

\usepackage[T1]{fontenc}
\usepackage{rotating}

\DeclareRobustCommand{\VAN}[3]{#2}
\let\VANthebibliography\thebibliography
\def\thebibliography{\DeclareRobustCommand{\VAN}[3]{##3}\VANthebibliography}

\usepackage{graphicx}	
\usepackage{amsmath}	
\usepackage{pdflscape}
\usepackage{subfig}

\title[New CSS low mass ratio binaries]{New low mass ratio contact binaries in the Catalina Sky Survey}

\author[P.E. Christopoulou et al.]{
Panagiota-Eleftheria Christopoulou,$^{1}$\thanks{E-mail: pechris@physics.upatras.gr }
Eleni Lalounta,$^{1}$
Athanasios Papageorgiou,$^{1}$
\newauthor
C.E. Ferreira Lopes,$^{2}$
M\'arcio Catelan,$^{3,4,5}$ 
and Andrew J. Drake$^{6}$
\\
$^{1}$ Department of Physics, University of Patras, 26500, Patra, Greece\\
$^{2}$ Astrophysics Division, National Institute For Space Research (DAS/INPE), Av. dos Astronautas, 1758 - Jardim da Granja, São José dos Campos, SP 12227-010, Brazil\\
$^{3}$ Pontificia Universidad Cat\'olica de Chile, Facultad de F\'{i}sica, Instituto de Astrof\'\i sica, Av. Vicu\~{n}a Mackenna 4860, 7820436 Macul, Santiago, Chile\\
$^{4}$
   Millennium Institute of Astrophysics, Nuncio Monse\~{n}or Sotero Sanz 100, Of. 104, Providencia, Santiago, Chile\\
$^{5}$
Centro de Astroingenier{\'{\i}}a, Pontificia Universidad Cat{\'{o}}lica de Chile, Av. Vicu\~{n}a Mackenna 4860, 7820436 Macul, Santiago, Chile\\
$^{6}$ California Institute of Technology, 1200 East California,  Boulevard, CA 
91225, USA\\
}

\date{Accepted XXX. Received YYY; in original form ZZZ}

\pubyear{2021}

\begin{document}
\label{firstpage}
\pagerange{\pageref{firstpage}--\pageref{lastpage}}
\maketitle

\begin{abstract}
We present the identification and photometric analysis of 30 new low mass ratio (LMR) totally eclipsing contact binaries found in Catalina Sky Survey data. The LMR candidates are identified using Fourier coefficients and visual inspection. We perform a detailed scan in the parameter plane of mass-ratio ($q$) versus inclination ($i$) using Phoebe-0.31 scripter to derive the best $(q,i)$ pair for the initial models. The relative physical parameters are determined from the final model of each system. A Monte-Carlo approach was adopted to derive the parameter errors. The resulting parameters confirm the identification. The approximate absolute physical parameters of the systems are estimated based on the light curve solutions and {\em Gaia} early Data Release 3 distances. Twelve out of 30 new systems have fill-out factors $f>50\%$ and $q\leq0.25$ (deep contact LMR systems), and 8 of them, to within errors, are extreme LMR deep systems with $q\leq0.1$. We discuss the evolutionary status of the 30 LMR systems in comparison with the most updated catalog of LMR systems from the literature. The scenario of the LMR systems as pre-merger candidates forming fast rotating stars is investigated for all systems, new and old, based both on Hut’s stability criteria and critical instability mass ratio ($q_{\rm inst}$) relation. CSS$\_$J075848.2+125656, with $q/q_{\rm inst}=1.23\pm 0.23$, and  CSS$\_$J093010.1-021624, with $q/q_{\rm inst}=1.25\pm 0.23$, can be considered as merger candidates.

\end{abstract}




\section{Introduction}
Eclipsing binary systems of EW light-curve type (EWs) with extreme low mass ratios challenge the current theoretical models since the latter predict coalescence into a single star \citep[][and references therein]{Robertson-1977,Eggleton-2010}. In at least one case, V1309 Sco, such a merger event has been directly observed \citep{2008IAUC.8972....1N}. This defined a distinct new class of luminous red novae that was later attributed, upon analysis of archival photometric data from the Optical Gravitational Lensing Experiment \citep[OGLE;][]{Udalski-2003}, to the merging components of a cool overcontact eclipsing binary system  with a decreasing orbital period \citep{2011A&A...528A.114T}. What triggers the binary to merge is still controversial. The widely accepted scenario is Darwin’s instability model \citep{10.2307/113751}, which implies that the merger happens when the spin angular momentum of the system is more than one third of the orbital angular momentum \citep{1980A&A....92..167H,1995ApJ...444L..41R}. When the binary mass ratio is extremely small and the secondary component has extremely small mass, it can no longer continue to co-rotate synchronously with the primary via the tidal interaction. As a consequence of the secondary’s higher angular velocity, angular momentum is transported rapidly from the orbital motion to the primary’s spin, causing the orbit to shrink and the period to shorten, until the engulfment of the secondary. Models suggest that single stars (FK Com and blue straggler type) could be the result of this coalescence \citep{1995ApJ...444L..41R, 2006AcA....56..199S, 2015A&A...577A.117S}.In addition, both the loss of mass and angular momentum through magnetic winds \citep{2006AcA....56..199S,2009MNRAS.397..857S,2012AcA....62..153S} and the presence of other companion(s) play a crucial role in the merging process. Since most stars are in binaries, and a significant fraction are in triples or higher-order systems \citep{2006A&A...450..681T,2007AJ....134.2353R,2010ApJS..190....1R,2013ApJ...768...33R}, potential stellar mergers may serve as keys to the binary fate. On the other hand, as orbital variations are common in contact binaries, the orbital period decay  at a high rate induced by a third star proved to be crucial in the case of the recently claimed red nova precursor KIC 9832227 \citep{2017ApJ...840....1M} after the revision of  its period variations \citep{2018ApJ...864L..32S,2019A&A...631A.126K}. 

\citet{2015AJ....150...69Y} presented the first compilation of low mass ratio systems (LMRs) with high fill-out factor, and since then many systems have been detected \citep[for reviews, see][]{2020RAA....20..163Q,2021MNRAS.502.2879G}. Ongoing all-sky surveys have significantly increased the number of known systems with extreme low mass ratio (LMR), and many more will be discovered in the Rubin Observatory's Legacy Survey of Space and Time era \citep[LSST;][]{2019ApJ...873..111I}. 

The current work focuses on the Catalina Sky Survey (CSS) and presents 30 new LMRs with total eclipses. In Section~\ref{Sample selection}, we briefly describe the CSS sample and our method for identifying LMRs. In section~\ref{Physical Parameters}, we describe the photometric analysis of the CSS light curves and the physical parameters determination. The methods for estimating the absolute parameters in absence of spectroscopic data are also discussed. Our results are presented  in Section~\ref{Results}, where the possibility that they may be merger candidates is also discussed. Our final conclusions are summarized in Section~\ref{Conclusions}.

\section{Sample selection}\label{Sample selection}
\subsection{The Catalina sample}\label{Catalina sample}
\label{sec:maths} 

Our sample is selected from the Catalina Real-Time Transient Survey Data Release 2 \citep[CRTS DR2;][]{2014yCat..22130009D}, which includes observations spanning 9 years (2004-2013). CRTS searches for optical transients by using the three CSS telescopes that search for rapidly moving Near Earth Objects. CSS covers the sky between declinations $\delta=-75\degr$ to $+65\degr$, avoiding crowded regions near the Galactic plane \citep{2009ApJ...696..870D}. The observations were obtained unfiltered in order to maximize the throughput, and then the magnitudes were transformed to an approximate $V$ magnitude \citep[$V_{\rm CSS}$,][]{2014yCat..22130009D}.

\cite{2014ApJ...790..157D} investigated the properties of 367 ultra-short period binary candidates identified from CSS data. \cite{PA18} reported an updated catalog of 4683 eclipsing Algol-type binaries (EA) from the CSS survey. The physical parameters of the EA systems were explored by \cite{2015MNRAS.454.2946L}, \cite{PA19}, and \cite{2020MNRAS.498.2833C}. Using the same sample of 4683 EA in CSS, \cite{PA21} reported 126 EAs with possible quadratic or cyclic period variations. We focus on the initial sample of \cite{2014yCat..22130009D} of 30,743 eclipsing binaries classified as EWs. This sample was explored by \cite{2017MNRAS.465.4678M}, who characterise 9380 EWs, and more recently by \cite{2020ApJS..247...50S}, who presented the physical parameters of 2335 late-type EWs.

\subsection{Identification of LMRs}\label{Identification}

Our first step was to exclude from the initial sample systems with insufficiently sampled light curves (hereafter, LCs) that have less than 150 observations ($N_{\rm p}$) and periods longer than $\sim 0.8$~d, resulting in a sample of 30,592 systems (sample~1). To remove outliers and poor measurements, we first cleaned the LCs of this sample using $3\sigma$ clipping along the phase-folded LCs, adopting the period from \cite{2014yCat..22130009D}. Then the initial epoch values were determined using an iterative procedure of fitting a 2-degree polynomial to the deeper eclipse. As a result of this procedure, we re-determined the ephemerides of the systems. 

As the components of LMRs are characterized by a large difference between their masses, and hence their radii, it is expected that their LCs should exhibit flat bottoms of long duration, compared to common EWs. Since the main idea was to take advantage of the LC morphology, we compiled the LCs of 42 well-studied LMRs from the literature (sample~2) on the basis of their mass ratios, the totality of the eclipses, and the availability of their $V$-band LCs. These are indicated in Table~\ref{tab:literature} with (*). For 26 systems from sample~2 that are also identified as extreme LMRs by \cite{2016PASA...33...43S}, the LCs from the $\it{Kepler}$ mission \citep{Koch-2010} were used. Although $\it{Kepler}$'s $K_p$ filter is essentially a broad $VR$, its high-accuracy photometry provides valuable information for studies of the LC morphology. 

We performed Fourier decomposition (hereafter, FD) of the phase-folded, normalized flux LCs of both samples, based on the following equation \citep[e.g.,][]{2009A&A...507.1729D}:

\begin{equation}
 \label{eq:1}
m(t)=A_0+\sum\limits_{j=1}^{10}\{a_j  \sin [2 \pi  j  \phi(t)] + b_j \cos [2 \pi  j  \phi(t)]\},   
\end{equation}

\noindent where $m(t)$ is the observed magnitude at time $t$, $A_{0}$ is the mean magnitude, $a_{j}, b_{j}$ are the amplitude components of the $(j- 1)^{\rm th}$ harmonic, and $\phi(t)$ is the phase (in the range [$0-1$]) corresponding to a full cycle, with the zero-point of the phased light curve corresponding to the primary (deeper) minimum. As to the number of terms to be used in the FD, we choose a high order because otherwise the flat part of the total eclipse cannot be properly reproduced.

\begin{figure*}
	\includegraphics[scale=1.8]{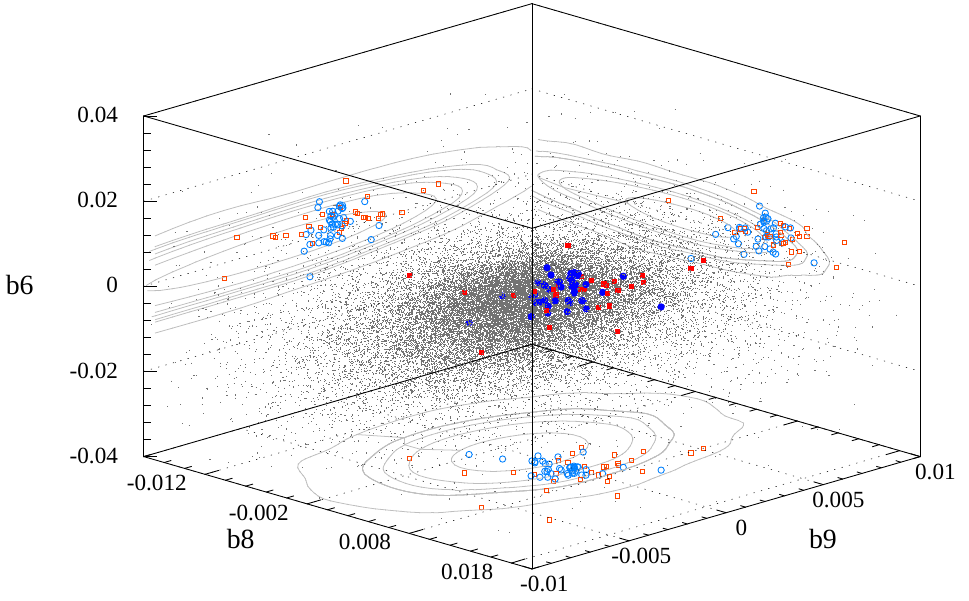}
    \caption{A 3D plot of $b_{6}$, $b_{8}$ and $b_{9}$ FD coefficients of sample~1 EWs represented by grey points and sample 2 (systems from literature, see text for description) represented by solid blue circles. The new CSS candidates (solid red squares) are located on the right top. Contour lines represent the projected distribution of EWs in each plane while the light blue open circles and light red open squares, are the projections of the literature and new LMR systems, respectively. }
	\label{fig:3D}
\end{figure*}

We analysed the full set of $a_j$, $b_j$ coefficients (20 in total), and arrived at the conclusion that the higher-order coefficients, especially $b_{8}$, are more efficient in identifying LMRs than are the lower-order ones. This can be appreciated from Fig.~\ref{fig:3D}, which shows the distribution of the $b_{6}$, $b_{8}$, and $b_{9}$ coefficients, and where the literature and new candidate LMRs (blue solid circles and red solid squares, respectively) occupy a separate position from the other EWs (grey points). After this analysis, sample~1 was reduced to 2101 candidates. The LCs of the stars in this sample were then subjected to visual inspection, focusing on the totality and duration of the eclipses, finally leading to our sample of 30 new LMR candidates. 
Table~\ref{tab:1} lists the coordinates (RA and Dec), reference time of minimum (in Heliocentric Julian Days, $\rm HJD_0$), period ($P$), the CSS magnitude at maximum light ($V_{\rm CSS}({\rm max})$), the mean photometric error ($\sigma_{\rm CSS}$), and the number of available CSS observations ($N_{\rm p}$) of the 30 new CSS LMRs. The overestimated original CSS photometric errors were corrected by using the correction factor provided by \cite{mgea17}, based on the analytical expression derived by \citet[][their footnote~8]{PA18}.

\begin{table*}
\centering
 \caption{The 30 new LMR candidates selected from sample 1 using FD.}
 \label{tab:1}
  \begin{tabular}{ccccccccc}\hline
  ID &  ${\rm RA}_{\rm J2000}$  &  ${\rm Dec}_{\rm J2000}$  &  ${\rm HJD}_0$   & Period & $V_{\rm CSS}({\rm max})$ & $\sigma_{\rm CSS}$ & $N_{p}$    \\ 
 & (h:m:s) & (\degr:\arcmin:\arcsec)  &  (${\rm 2450000+}$) & (days) & (mag) & (mag) & \\
 \hline

CSS$\_$J011848.4+132107	&	01:18:48.49	&	+13:21:07.6	&	6589.79386	&	0.3439788	&	13.21	&	0.01	&	300	\\
CSS$\_$J015301.5+223638	&	01:53:01.55	&	+22:36:38.4	&	3653.82712	&	0.2594140	&	15.92	&	0.03	&	274	\\
CSS$\_$J021552.4+324419	&	02:15:52.41	&	+32:44:19.5	&	5120.91902	&	0.5737980	&	14.68	&	0.02	&	350	\\
CSS$\_$J022044.4+280006	&	02:20:44.42	&	+28:00:06.1	&	4405.81533	&	0.7593753	&	14.42	&	0.02	&	356	\\
CSS$\_$J030702.2+261521	&	03:07:02.24	&	+26:15:21.8	&	5648.62161	&	0.7283894	&	13.98	&	0.02	&	374	\\
CSS$\_$J051156.6+011756	&	05:11:56.63	&	+01:17:56.5	&	3703.75981	&	0.7527221	&	14.82	&	0.02	&	351	\\
CSS$\_$J075839.9+131355	&	07:58:39.98	&	+13:13:55.7	&	6358.70319	&	0.4620180	&	14.88	&	0.02	&	293	\\
CSS$\_$J075848.2+125656	&	07:58:48.24	&	+12:56:56.1	&	4922.69449	&	0.3499800	&	14.08	&	0.02	&	293	\\
CSS$\_$J080724.7+164610	&	08:07:24.78	&	+16:46:10.6	&	5211.84350	&	0.3629629	&	14.39	&	0.02	&	415	\\
CSS$\_$J082140.8+192034	&	08:21:40.85	&	+19:20:34.5	&	5240.76455	&	0.4323400	&	15.15	&	0.02	&	528	\\
CSS$\_$J082850.5+015641	&	08:28:50.52	&	+01:56:41.1	&	5301.61530	&	0.4886486	&	13.37	&	0.01	&	346	\\
CSS$\_$J082916.5+131557	&	08:29:16.54	&	+13:15:57.6	&	5321.64675	&	0.3729305	&	13.98	&	0.02	&	438	\\
CSS$\_$J084222.2+041155	&	08:42:22.24	&	+04:11:55.5	&	4588.65898	&	0.2809483	&	15.67	&	0.03	&	381	\\
CSS$\_$J093010.1-021624	&	09:30:10.15	&	-02:16:24.9	&	5240.80316	&	0.3214746	&	15.56	&	0.03	&	318	\\
CSS$\_$J103653.7-072753	&	10:36:53.79	&	-07:27:53.6	&	5652.74006	&	0.2444613	&	15.75	&	0.03	&	276	\\
CSS$\_$J110526.4+285617	&	11:05:26.44	&	+28:56:17.2	&	4977.70007	&	0.3491059	&	14.93	&	0.02	&	388	\\
CSS$\_$J112643.3-141735	&	11:26:43.32	&	-14:17:35.7	&	4537.84328	&	0.3116524	&	13.60	&	0.01	&	156	\\
CSS$\_$J120945.8-025729	&	12:09:45.87	&	-02:57:29.2	&	5596.86649	&	0.3114363	&	15.29	&	0.02	&	294	\\
CSS$\_$J134010.1+134515	&	13:40:10.17	&	+13:45:15.2	&	5919.99314	&	0.4110526	&	14.29	&	0.02	&	380	\\
CSS$\_$J134512.0+034251	&	13:45:12.07	&	+03:42:51.1	&	3762.98614	&	0.4240236	&	15.22	&	0.02	&	349	\\
CSS$\_$J145437.2+060239	&	14:54:37.21	&	+06:02:39.8	&	4917.95998	&	0.5436659	&	15.28	&	0.02	&	351	\\
CSS$\_$J155637.0+060949	&	15:56:37.09	&	+06:09:49.3	&	4477.02450	&	0.3605211	&	15.81	&	0.03	&	357	\\
CSS$\_$J161753.6+205014	&	16:17:53.67	&	+20:50:14.1	&	4954.82770	&	0.5098331	&	14.60	&	0.02	&	379	\\
CSS$\_$J163819.6+034852	&	16:38:19.65	&	+03:48:52.0	&	6158.79806	&	0.2053320	&	14.26	&	0.02	&	339	\\
CSS$\_$J210300.1+050345	&	21:03:00.14	&	+05:03:45.6	&	4235.88388	&	0.5921606	&	13.90	&	0.01	&	387	\\
CSS$\_$J211420.2-142710	&	21:14:20.28	&	-14:27:10.8	&	5530.59888	&	0.5981238	&	15.32	&	0.02	&	297	\\
CSS$\_$J233821.8+200518	&	23:38:21.81	&	+20:05:18.2	&	6167.79999	&	0.3545902	&	13.23	&	0.01	&	324	\\
CSS$\_$J234145.7+233158	&	23:41:45.73	&	+23:31:58.6	&	4391.81504	&	0.5986466	&	13.91	&	0.01	&	309	\\
CSS$\_$J234324.8+211100	&	23:43:24.84	&	+21:11:00.4	&	5585.57623	&	0.6310680	&	14.50	&	0.02	&	332	\\
CSS$\_$J234807.2+193717	&	23:48:07.21	&	+19:37:17.8	&	5590.60823	&	0.3800130	&	15.46	&	0.03	&	331	\\

 \hline
\end{tabular}
\end{table*}


\section{Light curve modelling}\label{Physical Parameters}
\begin{figure}
	\includegraphics[width=\columnwidth]{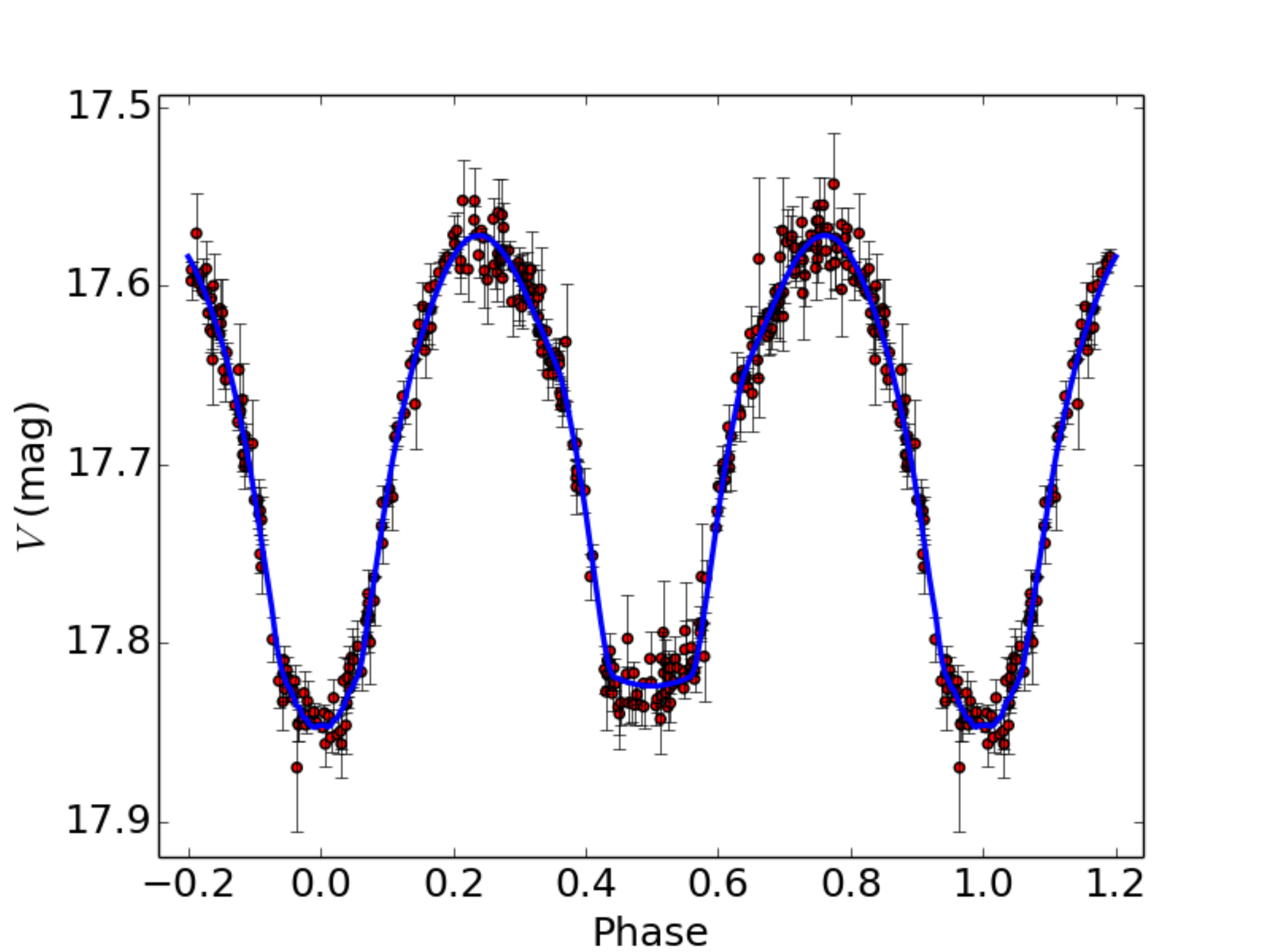}
	\includegraphics[width=\columnwidth]{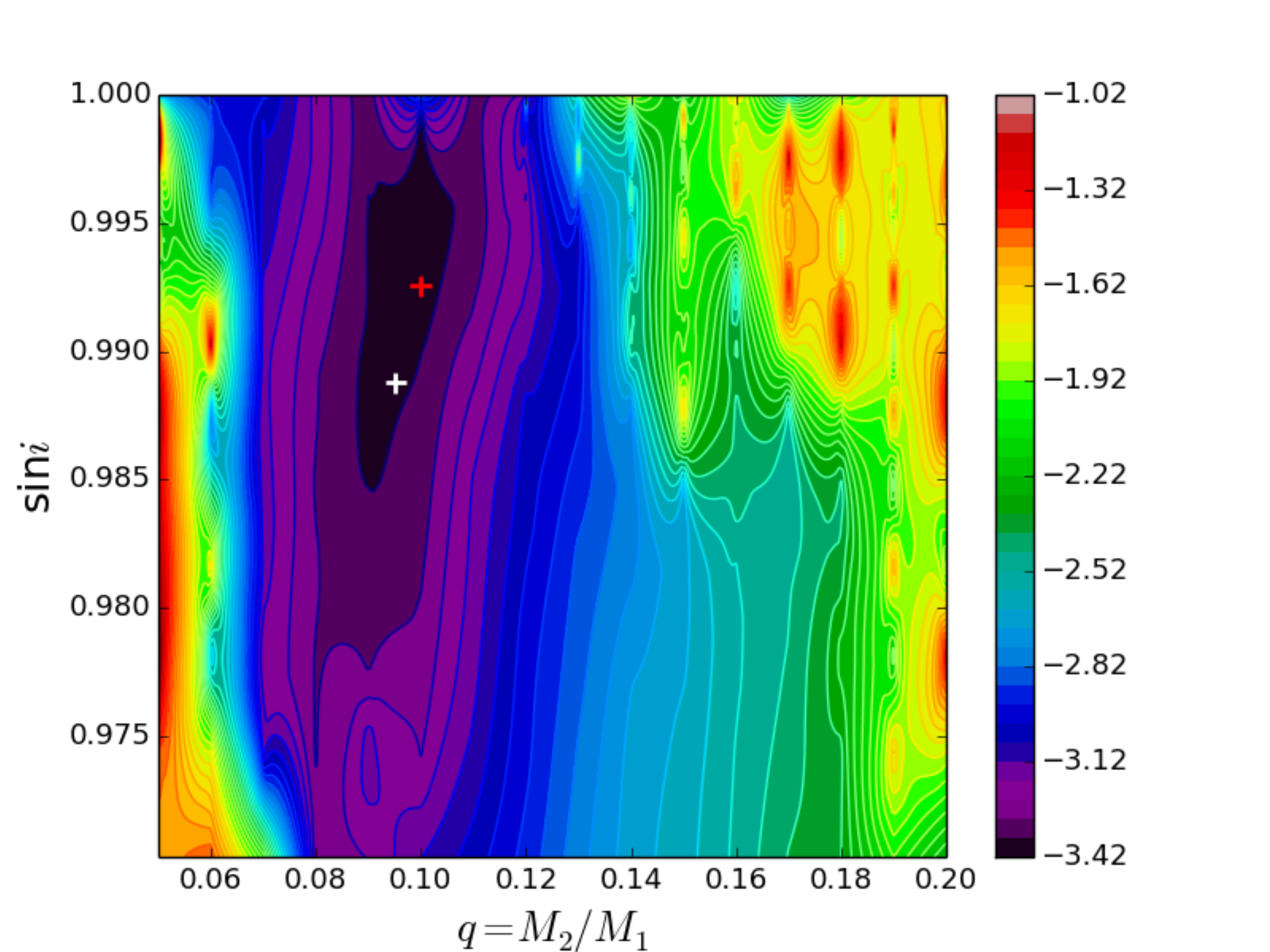}
    \caption{Top panel: Synthetic LC with photometric errors generated using PHOEBE-scripter (red points). The blue line shows the solution given by our method. Bottom panel: Contour plot of $\log\chi^{2}$ (color-coded according to the scale on the right) in the $(q,\sin i)$ plane of the synthetic system. The white cross represents the real solution of the synthetic system, while the red cross the resulting solution.}
	\label{synthetics}
\end{figure}

The photometric data of the 30  CSS LMR candidates were analyzed using PHOEBE-scripter \citep{2005ApJ...628..426P} to determine the systems' physical parameters. In the absence of spectroscopic data (as it is often the case for relatively dim, $V>13$~mag, EWs), the totality provides the most reliable photometric value for the mass ratio \citep{2005Ap&SS.296..221T,2013CoSka..43...27H,2016PASA...33...43S}. This is usually performed by initializing a 2D grid search on the mass ratio ($q$) - inclination ($i$) plane, setting ``Overcontact not in thermal contact'' mode where $q=M_2/M_1$ is the mass ratio of the system \footnote{Throughout this work we use number 1 (2) for the more (less) massive component, which is considered to be the primary (secondary).}. As we wanted  to explore the systems parameter space of the solutions and investigate how well the area of global minimum is mapped in the case of LMRs given the photometry cadence and precision of CSS survey, a synthetic LC of a LMR system ($q=0.096$, $i=81.4\degr$) was generated using PHOEBE-scripter. The LC mimics the observational data of CSS having 350 data points and photometric errors around 0.01~mag. 
The synthetic LC (red points) with the photometric errors and the solution (blue line) are shown in Fig.~\ref{synthetics} (top panel), whereas the bottom panel of Fig.~\ref{synthetics} represents the 2D distribution of $\log\chi^{2}$ values in the $(q, \sin i)$ plane. 
The solution ($q_{\rm min},i_{\rm min}$), as it emerges from the $q-i$ scan, is ($0.1, 83\degr$), which is in very good agreement with the real solution of the synthetic system, thus confirming the applicability of the method on this sample. 

\begin{figure*}

\minipage{0.5\textwidth}
\centering
\center \textbf{CSS$\_$J084222.2+041155}
\includegraphics[width=\linewidth]{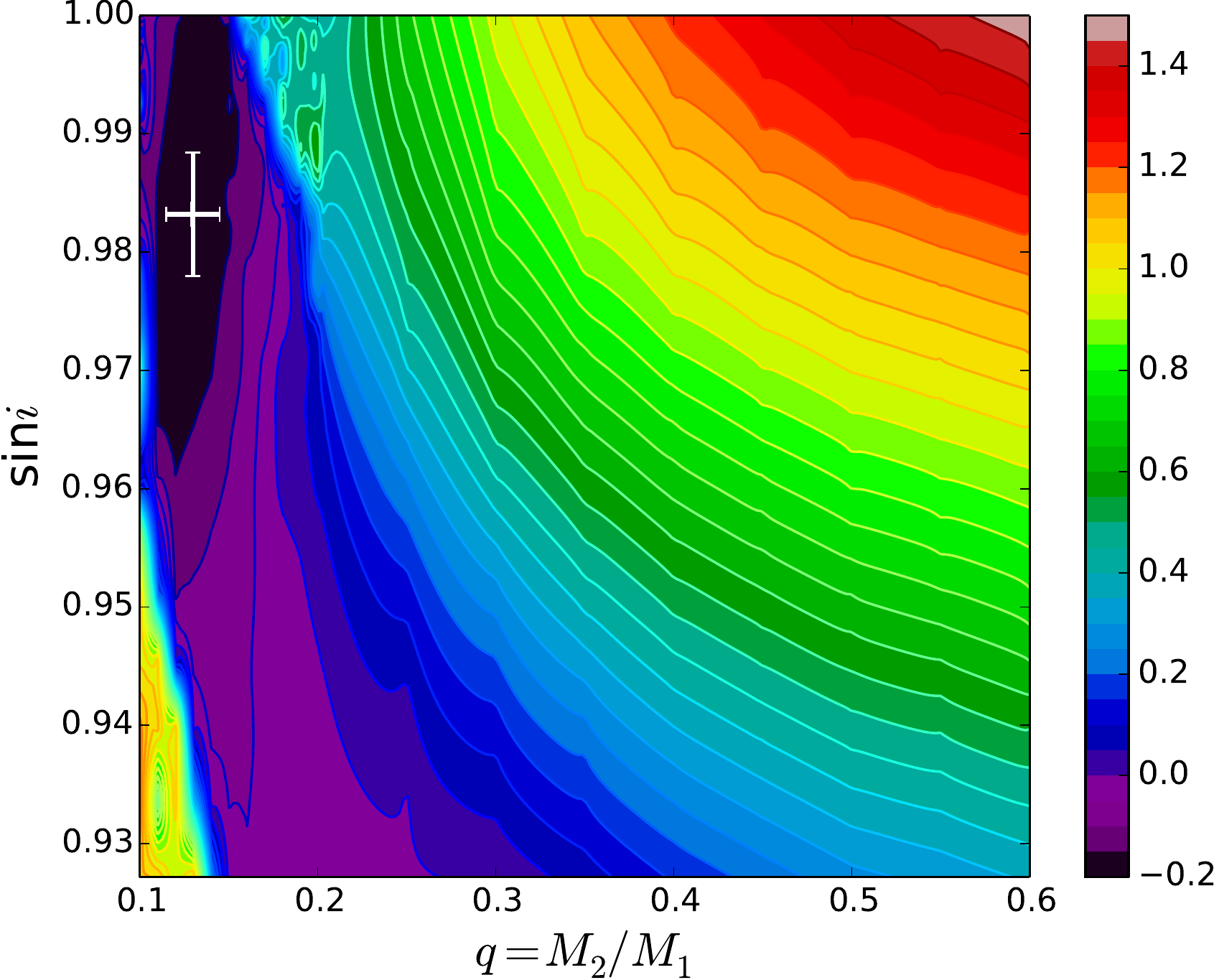} %

\endminipage\hfill
\minipage{0.5\textwidth}
\center \textbf{CSS$\_$J134010.1+134515}
\includegraphics[width=\linewidth]{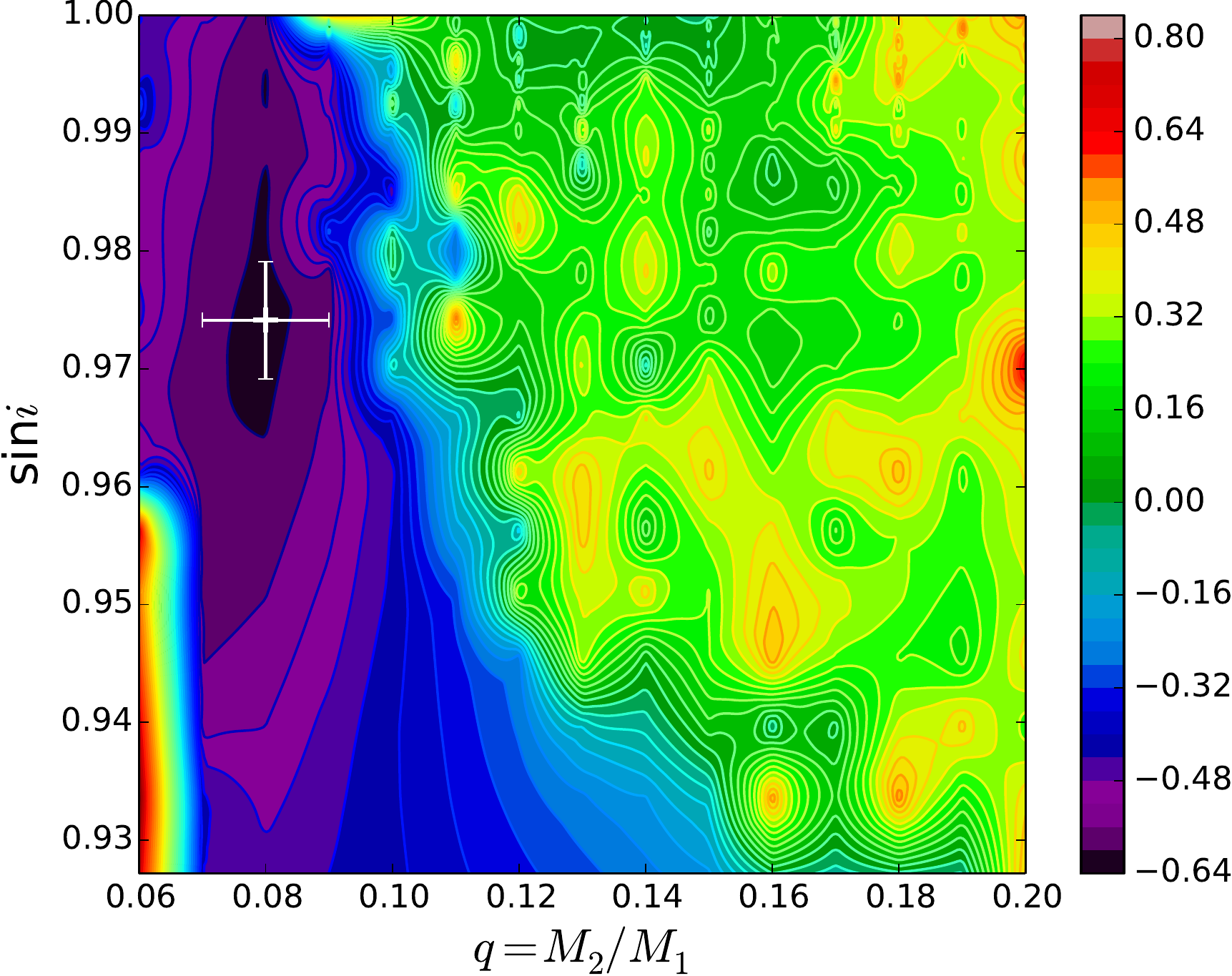} %
\endminipage
\vspace{+0.3cm}

\minipage{0.5\textwidth}
\centering
\center \textbf{CSS$\_$J075848.2+125656}
\includegraphics[width=\linewidth]{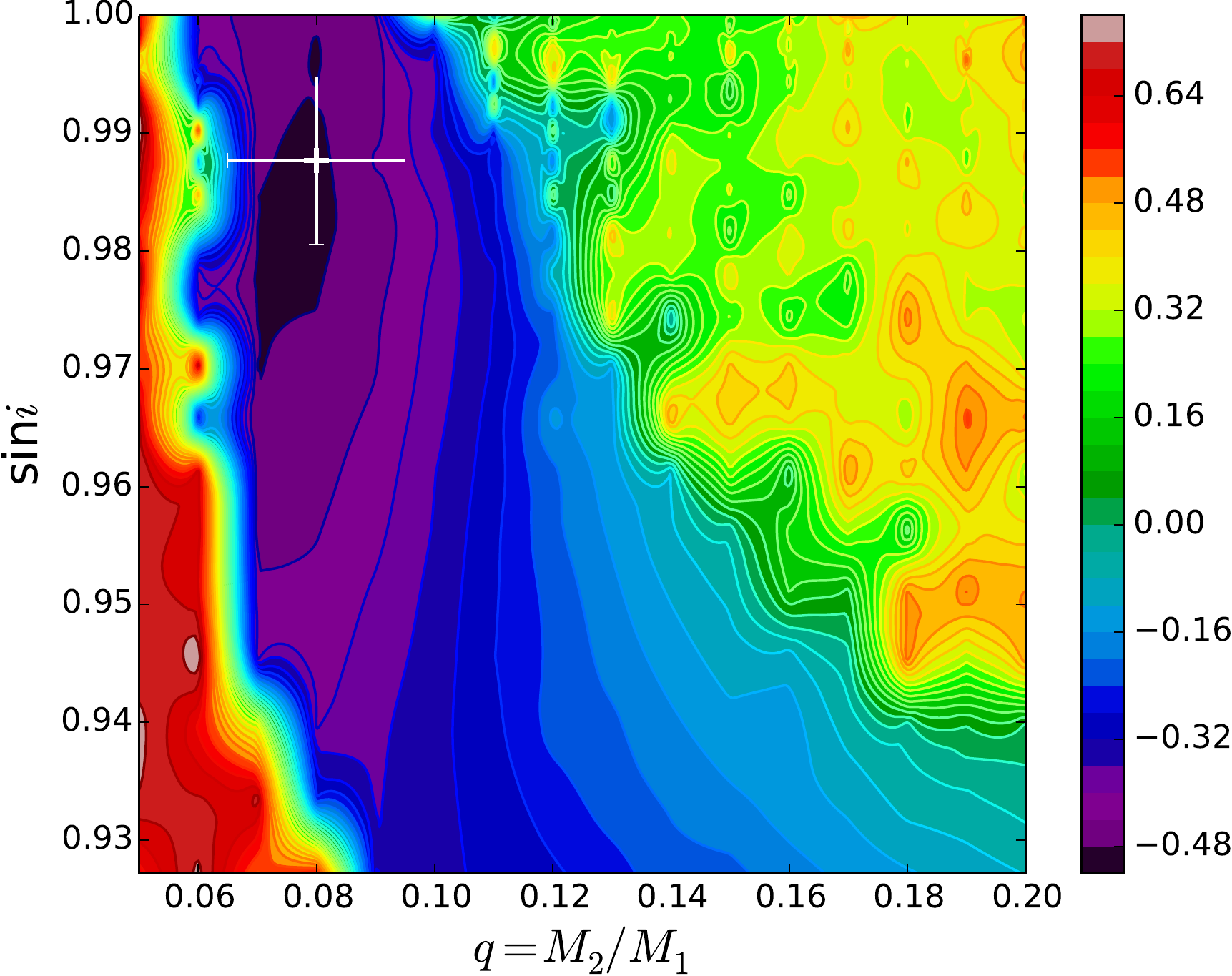} %
\endminipage\hfill
\minipage{0.5\textwidth}
\center \textbf{CSS$\_$J134512.0+034251}
\includegraphics[width=\linewidth]{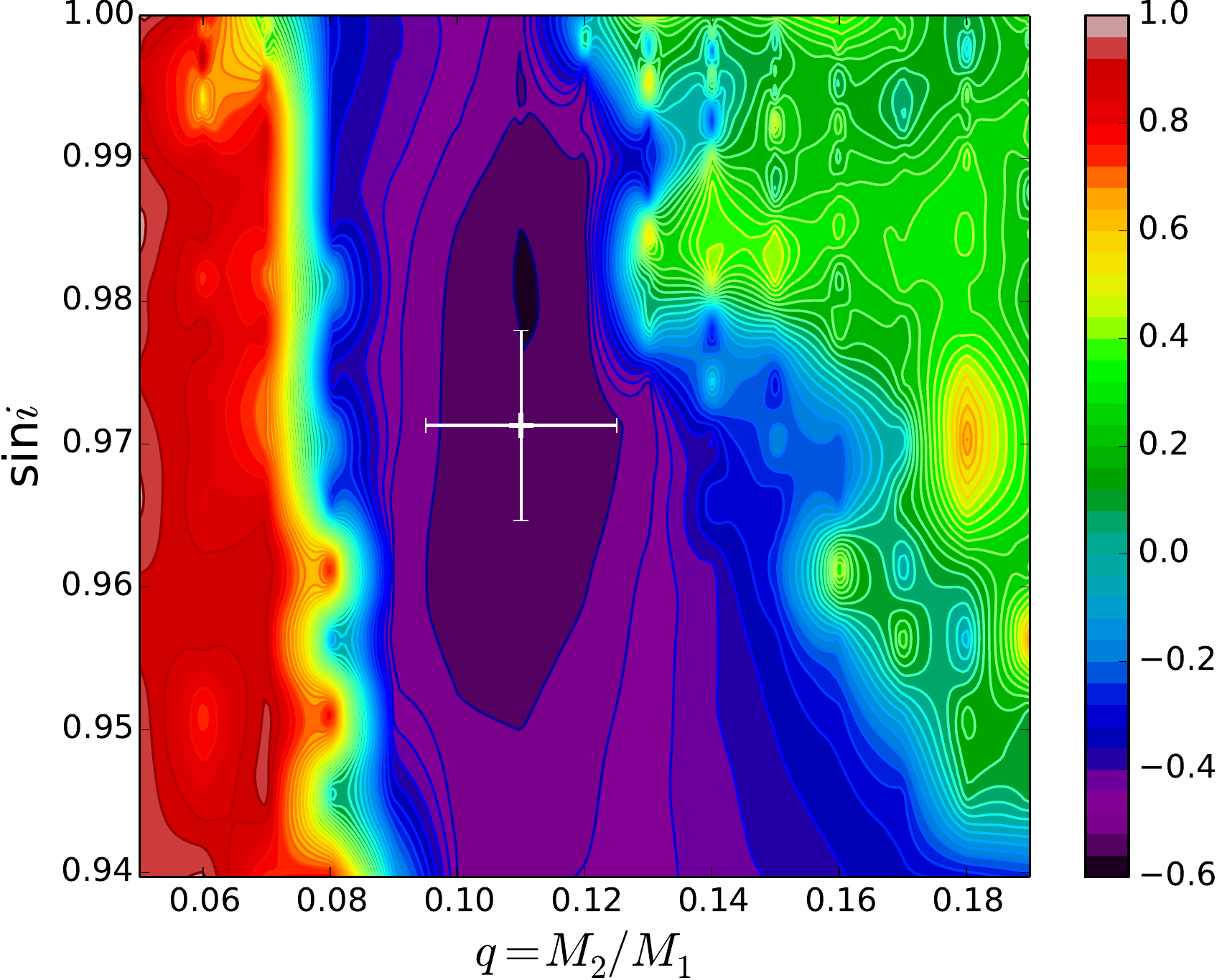} %
\endminipage
\caption{Contour plots of $\log{\chi^2}$ (color-coded according to the scale on the right) in the ($q$, $\sin\it{i}$) plane of four CSS LMRs. The white crosses represent the solutions of the systems with the error bars derived from the MC procedure.}\label{fig:CSS_qi_plots}
\end{figure*}

Following the results of the above test, for a rough estimation of the solution ($q_{\rm min},i_{\rm min}$), which was obtained by $\chi^{2}$ minimization, the range of explored $q$ values was set to $[0.1-0.6]$, and that of $i$ values to $[68\degr-90\degr]$, with step sizes of 0.05 and $1\degr$, respectively. Then, depending on the results, the range was set to $q_{\rm min}\pm 0.05$ with a step 0.01 for two runs, one without phase shift and another with phase shift 0.5.  During this $q-i$ scan, the effective temperature of the primary star $T_{1}$, the mass ratio $q$, and the orbital inclination $i$ were set as fixed parameters, while the effective temperature $T_{2}$ of the secondary, the modified potential $\Omega=\Omega_{\rm 1,2}$, and the passband luminosity of the primary $L_{1}$ were set as adjustable parameters. Fig.~\ref{fig:CSS_qi_plots} exhibits representative examples of the $q-i$ scan for four new CSS LMRs.

\subsection{Initial models for PHOEBE-scripter}\label{InitModels}
To initialize the models, an estimation of the system’s temperature is needed since the LC morphology can only constrain the ratio of the components’ temperatures ($T_2/T_1$). The effective temperature of the primary was fixed at the system's value ($T_{\rm sys}$) given by \cite{2019AJ....158..138S}. This was derived from either spectroscopy or from empirical relations between $T_{\rm eff}$ and the {\em Gaia} $G_{\rm BP} - G_{\rm RP}$ color \citep[DR2;][]{GAIADR2}, based on a set of 19962 stars having
spectroscopicaly determined $T_{\rm eff}$ and being within 100 pc so as to avoid reddening. Taking into account that $T_{\rm sys}$ is dominated by the temperature of the primary (hotter) component, we set $T_{\rm sys}=T_{1}$. Limb darkening coefficients were interpolated from \cite{1993AJ....106.2096V} tables for the given temperatures, while gravity darkening coefficients $g$ and surface albedos $A$ were accordingly assumed for convective ($g=0.32, A=0.5$) or radiative envelopes ($g=1, A=1$). The synchronization parameters were set as $F=1$, assuming synchronous rotation of the components. Having set the above parameters and before the $q-i$ scan is performed, the ephemeris of each system was refined by adjusting the reference time of minimum ($\rm HJD_0$) in PHOEBE. The resulting values of $q$ and $i$, along with corresponding $T_{2}$, $\Omega_{\rm 1,2}$, and $L_{1}$ acquired from the fine scan of the parameter space, were used as input parameters to check the solution for each system. The final parameters $q$, $T_2/T_1$, $\Omega_{\rm 1,2}$,  $r_{1}$, $r_{2}$, $i$,  $L_{1}/L_{tot}$ and $f$\footnote{$f=\frac{\Omega-\Omega_{\rm in}}{\Omega_{\rm out}-\Omega_{\rm in}}$, where $f$ is the fill-out factor, and $\Omega_{\rm in}$ and $\Omega_{\rm out}$ are the modified potential of the inner and the outer Lagrangian points, respectively.} are presented in Table~\ref{tab:models} where $r_{1}$, $r_{2}$ are the mean volume radii of the components and $L_{1}/L_{tot}$ is the ratio of the bandpass luminosity of the primary component to the total bandpass luminosity. The LCs of the improved final models are shown in Fig.~\ref{fig:LCsWsynt}. Four systems (CSS$\_$J080724.7+164610, CSS$\_$J082850.5+015641, CSS$\_$J110526.4+285617 and CSS$\_$J155637.0+060949) show significant period variation of the order of $10^{-6} \, {\rm d} \, {\rm yr}^{-1}$ during the 9 years of CSS data. The third light contribution was also investigated and found to be negligible ($\leq 1\%$) in the context of our photometric accuracy in one filter. However in the past, several systems showing low-amplitude light curves on the {\em Hipparcos} mission \citep{2004A&A...416.1097S} were later spectroscopically found to be triples and the true mass
ratio and/or inclination angle was higher.

\begin{table*}
\caption{The physical parameters of the 30 new CSS LMRs derived by LC modeling.}
\label{tab:models}
\scriptsize
\centering
\begin{tabular}{ccccccccc}
\hline
ID &  $q$ &  $\frac{T_{2}}{T_{1}}$ &  $\Omega_{\rm 12}$ &  $r_{1}$  & $r_{2}$  & $i (\degr)$ &   $\frac{L_{1}}{L_{tot}}$   &  $f$  \\ \hline 
CSS$\_$J011848.4+132107 	&	0.110	$\pm$	0.015	&	0.975	$\pm$	0.003	&	1.940	$\pm$	0.002	&	0.594	$\pm$	0.001	&	0.237	$\pm$	0.001	&	89.6	$\pm$	2.0	&	0.883	&	0.68	$\pm$	0.03	\\
CSS$\_$J015301.5+223638 	&	0.200	$\pm$	0.015	&	1.026	$\pm$	0.006	&	2.186	$\pm$	0.008	&	0.539	$\pm$	0.003	&	0.269	$\pm$	0.003	&	80.0	$\pm$	1.0	&	0.782	&	0.36	$\pm$	0.06	\\
CSS$\_$J021552.4+324419 	&	0.110	$\pm$	0.010	&	0.920	$\pm$	0.006	&	1.971	$\pm$	0.005	&	0.580	$\pm$	0.002	&	0.221	$\pm$	0.003	&	79.3	$\pm$	2.0	&	0.909	&	0.25	$\pm$	0.07	\\
CSS$\_$J022044.4+280006 	&	0.150	$\pm$	0.010	&	0.944	$\pm$	0.005	&	2.063	$\pm$	0.007	&	0.563	$\pm$	0.003	&	0.249	$\pm$	0.003	&	84.7	$\pm$	1.0	&	0.870	&	0.41	$\pm$	0.07	\\
CSS$\_$J030702.2+261521 	&	0.090	$\pm$	0.015	&	0.952	$\pm$	0.005	&	1.903	$\pm$	0.002	&	0.599	$\pm$	0.001	&	0.213	$\pm$	0.001	&	81.2	$\pm$	0.7	&	0.910	&	0.42	$\pm$	0.04	\\
CSS$\_$J051156.6+011756 	&	0.150	$\pm$	0.010	&	0.925	$\pm$	0.006	&	2.072	$\pm$	0.009	&	0.560	$\pm$	0.004	&	0.245	$\pm$	0.004	&	83.1	$\pm$	2.4	&	0.882	&	0.32	$\pm$	0.10	\\
CSS$\_$J075839.9+131355 	&	0.070	$\pm$	0.010	&	0.869	$\pm$	0.012	&	1.834	$\pm$	0.009	&	0.621	$\pm$	0.004	&	0.203	$\pm$	0.006	&	70.0	$\pm$	2.1	&	0.947	&	0.61	$\pm$	0.19	\\
CSS$\_$J075848.2+125656 	&	0.080	$\pm$	0.015	&	0.974	$\pm$	0.006	&	1.870	$\pm$	0.003	&	0.609	$\pm$	0.002	&	0.207	$\pm$	0.002	&	81.0	$\pm$	2.6	&	0.909	&	0.48	$\pm$	0.07	\\
CSS$\_$J080724.7+164610 	&	0.150	$\pm$	0.015	&	1.006	$\pm$	0.003	&	2.058	$\pm$	0.002	&	0.565	$\pm$	0.001	&	0.251	$\pm$	0.001	&	88.5	$\pm$	1.2	&	0.836	&	0.46	$\pm$	0.03	\\
CSS$\_$J082140.8+192034 	&	0.100	$\pm$	0.010	&	0.966	$\pm$	0.006	&	1.921	$\pm$	0.002	&	0.597	$\pm$	0.001	&	0.226	$\pm$	0.001	&	75.1	$\pm$	1.0	&	0.891	&	0.59	$\pm$	0.03	\\
CSS$\_$J082850.5+015641 	&	0.180	$\pm$	0.015	&	1.001	$\pm$	0.002	&	2.132	$\pm$	0.003	&	0.550	$\pm$	0.001	&	0.265	$\pm$	0.001	&	85.7	$\pm$	0.9	&	0.816	&	0.44	$\pm$	0.02	\\
CSS$\_$J082916.5+131557 	&	0.190	$\pm$	0.010	&	1.037	$\pm$	0.002	&	2.158	$\pm$	0.003	&	0.545	$\pm$	0.001	&	0.267	$\pm$	0.001	&	83.3	$\pm$	1.1	&	0.786	&	0.41	$\pm$	0.02	\\
CSS$\_$J084222.2+041155 	&	0.130	$\pm$	0.015	&	1.016	$\pm$	0.006	&	2.000	$\pm$	0.007	&	0.579	$\pm$	0.003	&	0.245	$\pm$	0.003	&	79.5	$\pm$	1.6	&	0.845	&	0.56	$\pm$	0.08	\\
CSS$\_$J093010.1-021624 	&	0.110	$\pm$	0.020	&	1.031	$\pm$	0.006	&	1.921	$\pm$	0.003	&	0.603	$\pm$	0.001	&	0.249	$\pm$	0.002	&	78.0	$\pm$	2.3	&	0.842	&	0.96	$\pm$	0.03	\\
CSS$\_$J103653.7-072753 	&	0.180	$\pm$	0.010	&	1.070	$\pm$	0.005	&	2.175	$\pm$	0.011	&	0.535	$\pm$	0.004	&	0.246	$\pm$	0.004	&	82.9	$\pm$	2.2	&	0.768	&	0.06	$\pm$	0.04	\\
CSS$\_$J110526.4+285617 	&	0.110	$\pm$	0.015	&	1.031	$\pm$	0.005	&	1.923	$\pm$	0.003	&	0.602	$\pm$	0.001	&	0.248	$\pm$	0.002	&	85.8	$\pm$	2.4	&	0.839	&	0.93	$\pm$	0.04	\\
CSS$\_$J112643.3-141735 	&	0.120	$\pm$	0.020	&	1.033	$\pm$	0.004	&	1.950	$\pm$	0.003	&	0.595	$\pm$	0.001	&	0.253	$\pm$	0.002	&	79.3	$\pm$	1.5	&	0.834	&	0.88	$\pm$	0.04	\\
CSS$\_$J120945.8-025729 	&	0.140	$\pm$	0.015	&	1.073	$\pm$	0.005	&	1.989	$\pm$	0.003	&	0.588	$\pm$	0.001	&	0.270	$\pm$	0.002	&	81.2	$\pm$	2.1	&	0.787	&	0.95	$\pm$	0.03	\\
CSS$\_$J134010.1+134515 	&	0.080	$\pm$	0.010	&	0.964	$\pm$	0.006	&	1.876	$\pm$	0.003	&	0.607	$\pm$	0.001	&	0.204	$\pm$	0.001	&	76.9	$\pm$	1.3	&	0.913	&	0.37	$\pm$	0.05	\\
CSS$\_$J134512.0+034251 	&	0.110	$\pm$	0.015	&	0.966	$\pm$	0.007	&	1.970	$\pm$	0.006	&	0.581	$\pm$	0.003	&	0.221	$\pm$	0.003	&	79.2	$\pm$	1.6	&	0.889	&	0.27	$\pm$	0.08	\\
CSS$\_$J145437.2+060239 	&	0.110	$\pm$	0.010	&	0.914	$\pm$	0.009	&	1.959	$\pm$	0.008	&	0.586	$\pm$	0.003	&	0.227	$\pm$	0.004	&	79.3	$\pm$	1.6	&	0.909	&	0.43	$\pm$	0.11	\\
CSS$\_$J155637.0+060949 	&	0.120	$\pm$	0.025	&	1.033	$\pm$	0.007	&	1.944	$\pm$	0.009	&	0.598	$\pm$	0.002	&	0.257	$\pm$	0.003	&	75.9	$\pm$	1.6	&	0.833	&	0.96	$\pm$	0.04	\\
CSS$\_$J161753.6+205014 	&	0.080	$\pm$	0.010	&	0.980	$\pm$	0.008	&	1.894	$\pm$	0.003	&	0.598	$\pm$	0.001	&	0.193	$\pm$	0.002	&	78.0	$\pm$	1.3	&	0.912	&	0.04	$\pm$	0.02	\\
CSS$\_$J163819.6+034852 	&	0.150	$\pm$	0.015	&	0.999	$\pm$	0.004	&	2.034	$\pm$	0.003	&	0.575	$\pm$	0.001	&	0.263	$\pm$	0.002	&	87.1	$\pm$	1.8	&	0.833	&	0.71	$\pm$	0.03	\\
CSS$\_$J210300.1+050345 	&	0.100	$\pm$	0.015	&	0.912	$\pm$	0.005	&	1.937	$\pm$	0.003	&	0.590	$\pm$	0.001	&	0.217	$\pm$	0.001	&	82.8	$\pm$	1.2	&	0.917	&	0.34	$\pm$	0.04	\\
CSS$\_$J211420.2-142710 	&	0.110	$\pm$	0.020	&	1.126	$\pm$	0.006	&	1.976	$\pm$	0.010	&	0.578	$\pm$	0.004	&	0.218	$\pm$	0.004	&	74.3	$\pm$	1.2	&	0.814	&	0.18	$\pm$	0.10	\\
CSS$\_$J233821.8+200518 	&	0.220	$\pm$	0.015	&	1.003	$\pm$	0.002	&	2.236	$\pm$	0.009	&	0.531	$\pm$	0.003	&	0.275	$\pm$	0.003	&	86.3	$\pm$	2.6	&	0.790	&	0.33	$\pm$	0.06	\\
CSS$\_$J234145.7+233158 	&	0.090	$\pm$	0.015	&	1.000	$\pm$	0.006	&	1.882	$\pm$	0.002	&	0.610	$\pm$	0.001	&	0.226	$\pm$	0.001	&	77.7	$\pm$	0.9	&	0.884	&	0.78	$\pm$	0.04	\\
CSS$\_$J234324.8+211100 	&	0.110	$\pm$	0.010	&	0.994	$\pm$	0.005	&	1.941	$\pm$	0.002	&	0.594	$\pm$	0.001	&	0.237	$\pm$	0.001	&	80.6	$\pm$	1.0	&	0.870	&	0.68	$\pm$	0.03	\\
CSS$\_$J234807.2+193717 	&	0.180	$\pm$	0.020	&	1.044	$\pm$	0.005	&	2.137	$\pm$	0.006	&	0.551	$\pm$	0.002	&	0.265	$\pm$	0.003	&	87.9	$\pm$	1.8	&	0.786	&	0.40	$\pm$	0.05	\\

\hline
\end{tabular}
\end{table*}


\subsection{Error estimation of the physical parameters}\label{Error estimation}
 In our analysis we consider that the uncertainty of the mass ratio $(\delta q)$ is determined by the step of the grid search, thus we set $\delta q= 0.010$. In some cases, the global minimum of the $\chi^2$ curve is broader and flatter, and so the uncertainty in $q$ is underestimated. For these systems we consider the values of q around the global minimum that resulted to $\chi^2$ increased up to $5\%$ of the minimum value ($\chi^2_{min}$) and adopt the standard deviation of this area as $\delta q$.
 Thus, for 11 out of 30 systems $\delta q=0.010$, while for the remainder it falls in the range $0.010 < \delta q\leq 0.025$. The temperature uncertainty of the systems is taken from \cite{2019AJ....158..138S}.
The uncertainties in the derived physical parameters are estimated by performing Monte-Carlo simulations \citep{2015AJ....149..168P,2019ApJS..242....6P}, since the formal errors from the fitting of the LC are heavily underestimated. During this procedure, each photometric point of the observed LC was randomly displaced 1000 times according to its photometric error $\sigma_{\rm CSS}$, drawn from a normal distribution with zero mean and $\sigma_{\rm CSS}$ as standard deviation. The 1000 synthetic LCs, generated in this way, for each system were then fitted by only adjusting $T_{2}$, $\Omega_{\rm 1,2}$, $L$, and $i$. We finally extracted the lower and the upper error boundaries from each parameter distribution ($q$, $T_2/T_1$, $\Omega_{\rm 1,2}$, $r_{1}$, $r_{2}$, $i$, $L_{1}/L_{tot}$ and $f$). The mean of these is the final error provided for each of these quantities in the final catalog (Table~\ref{tab:models}). 

Additionally one of the 30 candidates in our catalog, namely CSS$\_$J011848.4+132107 (also known as EM Psc), has already been studied by \cite{2005Ap&SS.300..337Y}, who found, using a $V$-band LC, ($q_{\rm min},i_{\rm min}) = (0.100, 83.3\degr$). \cite{2008AJ....136.1940Q}, using $R,I$ band LCs, later found ($0.1487,88.6\degr$) for the same system.
Our solution, ($q_{\rm min},i_{\rm min})= (0.110\pm0.015,89.6\degr\substack{+1.8\degr\\-2.2\degr}$) (see Table~\ref{tab:models}),  
is  in accordance with \cite{2005Ap&SS.300..337Y}, showing that $V_{\rm CSS}$ LCs can be used to estimate the systems' physical parameters.

\subsection{Absolute parameters}\label{sec:ReviewAbsPars}
In the absence of spectroscopic data, given the photometric mass ratio of EWs with total eclipses, the usual and simplest way to estimate the primary mass (and consequently $M_{2}=q M_{1}$) and semimajor axis ($\alpha$) is using a spectral type-temperature and spectral type-mass calibration, under the assumption that the primary is on the main sequence \citep{2000asqu.book.....C, 2013ApJS..208....9P}. Other ways include an indirect path of statistically derived relations such as period-mass, period-semimajor axis, or equivalently the total mass of the system \citep{2016PASA...33...43S,2015MNRAS.448.2890D, 2015AJ....150...69Y}. Another indirect method comes from the  published 2D and 3D correlations involving combinations of $M_{1}, M_{2}, R_{1}, R_{2}, L_{1}, L_{2}, P, q$, as given in \cite{2009CoAst.159..129G} based on the data in \cite{2008MNRAS.390.1577G}.

Alternatively, the absolute parameters of a binary can be estimated out of the absolute magnitude of the primary $\mathcal{M_{\rm V1}}$. This can be done using the passband luminosity ratio of the photometric solution and the maximum magnitude of the system $\mathcal{M_{\rm Vsys}}$. The latter can be estimated directly from the data using the $\it{Gaia}$ distance corrected for the interstellar extinction $A_{V}$. In the case of differential photometry, $\mathcal{M_{\rm Vsys}}$ can be derived from a period–color–magnitude relation, such as the one derived by \cite{2017AJ....154..125M} based on {\em Gaia} Data Release 1 (DR1) data (318 EWs) or \cite{2019gaia.confE..60C} based on {\em Gaia} DR2 measurements. 
Next, $\mathcal{M_{\rm V1}}$ is converted to $\mathcal{M_{\rm bol}}$ based on the BC correction corresponding to the temperature derived from the photometric solution. Finally, the primary mass $M_{1}$ is estimated if a mass-luminosity (M-L) relation $L\sim M^{\it{a}}$ is used, under the assumption, well supported by the empirical data, that primary components of contact systems lie in general between the zero-age main sequence (ZAMS) and the terminal-age main sequence \citep[TAMS;][]{2005ApJ...629.1055Y}. \cite{2020ApJS..247...50S} used $\alpha=4.216$ \citep{2013MNRAS.430.2029Y} to infer the mass of the primary. \cite{2020RAA....20..196L} suggested a method that combined the PARSEC code \citep[PAdova and TRieste Stellar Evolution Code;][]{PARSEC} with the Roche geometric model to determine the masses of EWs. They applied their method on a sample of 140 spectroscopic studied binaries, and found it to be efficient in high-mass ratio and short-period systems with low effective temperatures. 

In this study, we tested all the above methods and empirical relations, as we wanted to select those that provide a better description of the data. For this, we compiled a sample of 161 EWs from \citet{2013MNRAS.430.2029Y} and \citet{2021ApJS..254...10L} with spectroscopically derived mass ratios, since spectroscopic studies are known to produce results with the highest accuracy. This provides us with an extensive and highly heterogeneous sample. We performed a linear fitting in the $\log M-\log L$ plane of the massive component of this spectroscopic sample. By randomly selecting the mass and luminosity errors from the available literature errors, this fitting was repeated $10^{5}$ times. Then, the best pair ($b$, $a$) in the relation $\log L=\log b+a\log M$ is represented by the mean value of the region where the above fittings have the lowest absolute percentage error (up to $10\%$) between the calculated and spectroscopic masses. The corresponding uncertainties were derived from the standard deviation of the $b$ and $a$ values in the same region. Thus, we derived $b= 0.63 \pm 0.04$ and $a= 4.8 \pm 0.2$. With these values, the above  relation predicts the observed masses within $\Delta M_{1}\leq 10\%$ for $65\%$ of the systems and within $\Delta M_{1}\leq 20\%$ for $90\%$  of the systems, where $\Delta M_{1}=M_{\rm 1,obs}-M_{\rm 1,rel}$ is the average fractional mass difference, $M_{\rm 1,obs}$ represents the mass from the analysis of spectroscopic data, and $M_{\rm 1,rel}$ is the mass obtained from the relation.
Nevertheless, we caution that there are several factors that, if not properly accounted for in the analysis, may affect the inferred luminosities of the components, and thereby their masses. This includes, among the others, spots due to magnetic activity, presence of circumstellar material and third light contributions. In the case of a spotted system, for instance, the inferred luminosities (hence masses) may very well depend on the phase of the magnetic cycle during which the observations are carried out.

We decided to follow the second path and estimate the absolute magnitude of the primary. We cross-matched the LMR sample with {\em Gaia} EDR3 \citep[EDR3;][]{2021yCat.1352....0B} using a matching radius of 2 arcsec to find the corresponding distances. The distances of $60\%$ of the sources are $\leq 2$~kpc. For each source, we also calculate the total line of sight Galactic extinction from the 3D dust reddening map of \cite{2019ApJ...887...93G}, derived from {\em Gaia} parallaxes and stellar photometry from Pan-STARRS 1 and 2MASS data. The magnitude of the primary is estimated from the magnitude at quadratures of the system using the results of Table~\ref{tab:models} $(L_1/L_{tot})$. The magnitudes were transformed to Johnson $V$ based on the relation of \cite{2013ApJ...763...32D}. The temperatures of the individual components were derived from disentangling the value of the binary temperature (see Section~\ref{InitModels}) using equation (3) of \cite{2003A&A...404..333Z}, the ratio of relative radii $r_2/r_1$ and the temperature ratio $T_2/T_1$ from the light-curve analysis (see Table~\ref{tab:models}). Then the absolute bolometric magnitude is derived based on the bolometric correction \citep[from][]{1998A&A...333..231B} corresponding to the effective temperature of the primary, and the primary luminosity is obtained. Finally, the mass of the primary components was estimated from our derived mass-luminosity relation, and consequently the mass of the secondaries from the mass ratio, the semimajor axis from Kepler's law, and the mean radii from the corresponding relative radii.

To calculate the uncertainties of the component’s effective temperature ($\sigma_{T_{1}}$, $\sigma_{T_{2}}$), we considered the uncertainty of the systems’ effective temperatures from TIC-8 ($\it{TESS}$ Input Catalog, v-8.0) as the dominant source of error, since the temperature ratio $T_2/T_1$ uncertainty contribution is of the order $\sim 10$~K while the ratio of relative radii $r_2/r_1$ uncertainty contribution is negligible. The remaining uncertainties were calculated considering the photometric error from the light curves, the uncertainty in the $B-V$ colors (calculated from $\sigma_B$, $\sigma_V$, as given by TIC-8), and the uncertainty of the distance estimation combined with the estimated errors of the physical parameters (Section~\ref{Error estimation}). The results are given in Table~\ref{tab:absparams}.

\begin{table*}
 \caption{The absolute parameters of the 30 new CSS LMRs.}\label{tab:absparams}
 \tiny
 \centering

 \begin{tabular}{cccccccccccc}
  \hline
  ID  &  $T_{\rm sys}$ &   $T_{1}$  &   $T_{2}$ &  $M_{1}$   & $M_{2}$ & $R_{1}$ &  $R_{2}$  & $L_{1}$  &  $L_{2}$ & $\frac{J_{s}}{J_{o}}$ & $(\frac{J_{s}}{J_{o}})_k$ \\
 &  (K) & (K) & (K) &  ($M{\sun}$) &  ($M{\sun}$)  & ($R{\sun}$) & ($R{\sun}$) & ($L{\sun}$) & ($L{\sun}$) &  &  \\ 
 \hline 
  CSS$\_$J011848.4+132107 	&	5723	$\pm$	100					&	5742	$\pm$	101	&	5600	$\pm$	99	&	1.25	$\pm$	0.05	&	0.14	$\pm$	0.02	&	1.37	$\pm$	0.02	&	0.55	$\pm$	0.01	&	1.85	$\pm$	0.34	&	0.26	$\pm$	0.02	&	0.217	&	0.195	\\
  CSS$\_$J015301.5+223638 	&	4811	$\pm$	175					&	4786	$\pm$	174	&	4909	$\pm$	180	&	0.97	$\pm$	0.02	&	0.19	$\pm$	0.02	&	0.97	$\pm$	0.01	&	0.48	$\pm$	0.01	&	0.55	$\pm$	0.06	&	0.12	$\pm$	0.02	&	0.110	&	0.171	\\
  CSS$\_$J021552.4+324419 	&	6801	$\pm$	176					&	6863	$\pm$	178	&	6315	$\pm$	169	&	1.70	$\pm$	0.07	&	0.19	$\pm$	0.02	&	2.08	$\pm$	0.03	&	0.79	$\pm$	0.01	&	8.03	$\pm$	1.29	&	0.89	$\pm$	0.10	&	0.207	&	0.117	\\
  CSS$\_$J022044.4+280006 	&	6702	$\pm$	229					&	6760	$\pm$	231	&	6382	$\pm$	220	&	1.89	$\pm$	0.09	&	0.28	$\pm$	0.02	&	2.56	$\pm$	0.04	&	1.13	$\pm$	0.02	&	13.64	$\pm$	2.37	&	1.90	$\pm$	0.27	&	0.150	&	0.092	\\
  CSS$\_$J030702.2+261521 	&	6646	$\pm$	63					&	6680	$\pm$	63	&	6362	$\pm$	70	&	1.84	$\pm$	0.07	&	0.17	$\pm$	0.03	&	2.57	$\pm$	0.03	&	0.91	$\pm$	0.01	&	11.72	$\pm$	1.62	&	1.23	$\pm$	0.06	&	0.264	&	0.148	\\
  CSS$\_$J051156.6+011756 	&	6344	$\pm$	156					&	6414	$\pm$	158	&	5936	$\pm$	152	&	1.73	$\pm$	0.08	&	0.26	$\pm$	0.02	&	2.45	$\pm$	0.04	&	1.07	$\pm$	0.02	&	8.76	$\pm$	1.58	&	1.28	$\pm$	0.14	&	0.148	&	0.089	\\
  CSS$\_$J075839.9+131355 	&	6643	$\pm$	138					&	6714	$\pm$	140	&	5833	$\pm$	145	&	1.51	$\pm$	0.06	&	0.11	$\pm$	0.02	&	1.83	$\pm$	0.03	&	0.60	$\pm$	0.02	&	4.57	$\pm$	0.70	&	0.37	$\pm$	0.04	&	0.356	&	0.187	\\
  CSS$\_$J075848.2+125656 	&	6095	$\pm$	25					&	6111	$\pm$	25	&	5951	$\pm$	41	&	1.32	$\pm$	0.04	&	0.11	$\pm$	0.02	&	1.43	$\pm$	0.01	&	0.49	$\pm$	0.01	&	2.39	$\pm$	0.24	&	0.27	$\pm$	0.01	&	0.303	&	0.229	\\
  CSS$\_$J080724.7+164610 	&	5991	$\pm$	22					&	5984	$\pm$	22	&	6022	$\pm$	31	&	1.25	$\pm$	0.03	&	0.19	$\pm$	0.02	&	1.37	$\pm$	0.01	&	0.61	$\pm$	0.01	&	1.86	$\pm$	0.20	&	0.44	$\pm$	0.01	&	0.151	&	0.140	\\
  CSS$\_$J082140.8+192034 	&	7097	$\pm$	158					&	7126	$\pm$	159	&	6882	$\pm$	159	&	1.56	$\pm$	0.07	&	0.16	$\pm$	0.02	&	1.72	$\pm$	0.02	&	0.65	$\pm$	0.01	&	5.38	$\pm$	0.94	&	0.86	$\pm$	0.08	&	0.239	&	0.131	\\
  CSS$\_$J082850.5+015641 	&	6428	$\pm$	35					&	6426	$\pm$	35	&	6433	$\pm$	38	&	1.47	$\pm$	0.05	&	0.26	$\pm$	0.02	&	1.73	$\pm$	0.02	&	0.83	$\pm$	0.01	&	4.04	$\pm$	0.45	&	1.06	$\pm$	0.03	&	0.124	&	0.076	\\
  CSS$\_$J082916.5+131557 	&	6051	$\pm$	24					&	6006	$\pm$	24	&	6226	$\pm$	28	&	1.27	$\pm$	0.04	&	0.24	$\pm$	0.01	&	1.36	$\pm$	0.01	&	0.67	$\pm$	0.01	&	1.98	$\pm$	0.26	&	0.60	$\pm$	0.02	&	0.117	&	0.109	\\
  CSS$\_$J084222.2+041155 	&	6325	$\pm$	168					&	6310	$\pm$	168	&	6408	$\pm$	174	&	1.19	$\pm$	0.05	&	0.16	$\pm$	0.02	&	1.15	$\pm$	0.02	&	0.49	$\pm$	0.01	&	1.49	$\pm$	0.27	&	0.36	$\pm$	0.04	&	0.179	&	0.181	\\
  CSS$\_$J093010.1-021624 	&	5746	$\pm$	224					&	5720	$\pm$	223	&	5894	$\pm$	232	&	1.17	$\pm$	0.04	&	0.13	$\pm$	0.02	&	1.30	$\pm$	0.02	&	0.54	$\pm$	0.01	&	1.35	$\pm$	0.22	&	0.31	$\pm$	0.05	&	0.224	&	0.236	\\
  CSS$\_$J103653.7-072753 	&	5446	$\pm$	227					&	5487	$\pm$	229	&	5287	$\pm$	222	&	1.05	$\pm$	0.04	&	0.19	$\pm$	0.01	&	0.94	$\pm$	0.01	&	0.43	$\pm$	0.01	&	0.74	$\pm$	0.14	&	0.13	$\pm$	0.02	&	0.117	&	0.159	\\
  CSS$\_$J110526.4+285617 	&	5297	$\pm$	210					&	5273	$\pm$	209	&	5435	$\pm$	216	&	1.17	$\pm$	0.06	&	0.13	$\pm$	0.02	&	1.37	$\pm$	0.02	&	0.57	$\pm$	0.01	&	1.37	$\pm$	0.32	&	0.25	$\pm$	0.04	&	0.224	&	0.234	\\
  CSS$\_$J112643.3-141735 	&	5557	$\pm$	148					&	5528	$\pm$	147	&	5709	$\pm$	153	&	1.14	$\pm$	0.03	&	0.14	$\pm$	0.02	&	1.25	$\pm$	0.01	&	0.53	$\pm$	0.01	&	1.17	$\pm$	0.11	&	0.27	$\pm$	0.03	&	0.203	&	0.229	\\
  CSS$\_$J120945.8-025729 	&	5640	$\pm$	159					&	5563	$\pm$	157	&	5967	$\pm$	171	&	1.12	$\pm$	0.05	&	0.16	$\pm$	0.01	&	1.23	$\pm$	0.02	&	0.57	$\pm$	0.01	&	1.09	$\pm$	0.21	&	0.36	$\pm$	0.04	&	0.174	&	0.206	\\
  CSS$\_$J134010.1+134515 	&	6924	$\pm$	35					&	6949	$\pm$	36	&	6695	$\pm$	53	&	1.52	$\pm$	0.06	&	0.12	$\pm$	0.02	&	1.66	$\pm$	0.02	&	0.56	$\pm$	0.01	&	4.71	$\pm$	0.70	&	0.56	$\pm$	0.02	&	0.301	&	0.159	\\
  CSS$\_$J134512.0+034251 	&	7006	$\pm$	188					&	7035	$\pm$	189	&	6797	$\pm$	190	&	1.62	$\pm$	0.07	&	0.18	$\pm$	0.03	&	1.68	$\pm$	0.02	&	0.64	$\pm$	0.01	&	6.47	$\pm$	1.06	&	0.78	$\pm$	0.09	&	0.208	&	0.116	\\
  CSS$\_$J145437.2+060239 	&	6861	$\pm$	192					&	6930	$\pm$	194	&	6336	$\pm$	189	&	1.68	$\pm$	0.10	&	0.18	$\pm$	0.02	&	2.02	$\pm$	0.04	&	0.78	$\pm$	0.02	&	7.65	$\pm$	1.98	&	0.89	$\pm$	0.12	&	0.211	&	0.119	\\
  CSS$\_$J155637.0+060949 	&	6343	$\pm$	168					&	6309	$\pm$	167	&	6516	$\pm$	178	&	1.26	$\pm$	0.06	&	0.15	$\pm$	0.03	&	1.43	$\pm$	0.02	&	0.61	$\pm$	0.01	&	1.93	$\pm$	0.37	&	0.61	$\pm$	0.07	&	0.205	&	0.182	\\
  CSS$\_$J161753.6+205014 	&	6993	$\pm$	22					&	7006	$\pm$	22	&	6864	$\pm$	66	&	1.75	$\pm$	0.06	&	0.14	$\pm$	0.02	&	1.99	$\pm$	0.02	&	0.64	$\pm$	0.01	&	9.22	$\pm$	1.04	&	0.82	$\pm$	0.04	&	0.292	&	0.160	\\
  CSS$\_$J163819.6+034852 	&	6662	$\pm$	162					&	6665	$\pm$	162	&	6649	$\pm$	164	&	1.13	$\pm$	0.03	&	0.17	$\pm$	0.02	&	0.92	$\pm$	0.01	&	0.42	$\pm$	0.00	&	1.13	$\pm$	0.10	&	0.31	$\pm$	0.03	&	0.157	&	0.183	\\
  CSS$\_$J210300.1+050345 	&	6723	$\pm$	143					&	6787	$\pm$	144	&	6188	$\pm$	136	&	1.67	$\pm$	0.05	&	0.17	$\pm$	0.03	&	2.15	$\pm$	0.02	&	0.79	$\pm$	0.01	&	7.46	$\pm$	0.58	&	0.82	$\pm$	0.07	&	0.233	&	0.129	\\
  CSS$\_$J211420.2-142710 	&	6778	$\pm$	164					&	6876	$\pm$	166	&	6171	$\pm$	157	&	1.51	$\pm$	0.09	&	0.17	$\pm$	0.03	&	2.05	$\pm$	0.04	&	0.77	$\pm$	0.02	&	4.59	$\pm$	1.31	&	0.78	$\pm$	0.09	&	0.206	&	0.112	\\
  CSS$\_$J233821.8+200518 	&	5991	$\pm$	175					&	5987	$\pm$	175	&	6006	$\pm$	176	&	1.30	$\pm$	0.04	&	0.29	$\pm$	0.02	&	1.30	$\pm$	0.01	&	0.68	$\pm$	0.01	&	2.23	$\pm$	0.28	&	0.53	$\pm$	0.06	&	0.099	&	0.090	\\
  CSS$\_$J234145.7+233158 	&	6853	$\pm$	151					&	6853	$\pm$	151	&	6852	$\pm$	155	&	1.71	$\pm$	0.05	&	0.15	$\pm$	0.03	&	2.24	$\pm$	0.02	&	0.83	$\pm$	0.01	&	8.39	$\pm$	0.69	&	1.37	$\pm$	0.13	&	0.273	&	0.152	\\
  CSS$\_$J234324.8+211100 	&	6635	$\pm$	152					&	6640	$\pm$	152	&	6603	$\pm$	155	&	1.73	$\pm$	0.06	&	0.19	$\pm$	0.02	&	2.29	$\pm$	0.03	&	0.91	$\pm$	0.01	&	8.86	$\pm$	1.10	&	1.42	$\pm$	0.14	&	0.217	&	0.124	\\
  CSS$\_$J234807.2+193717 	&	5553	$\pm$	164					&	5504	$\pm$	163	&	5749	$\pm$	172	&	1.19	$\pm$	0.04	&	0.21	$\pm$	0.03	&	1.36	$\pm$	0.02	&	0.66	$\pm$	0.01	&	1.45	$\pm$	0.24	&	0.42	$\pm$	0.05	&	0.125	&	0.133	\\

  \hline
 \end{tabular}
\end{table*}

\section{Results and discussion}\label{Results}
It can be seen from Table~\ref{tab:absparams} that the new LMRs have mean primary mass $1.42\pm 0.26\, M_{\sun}$, and mean secondary mass $0.177\pm 0.048 \, M_{\sun}$. Half of the systems have temperatures in the range 5000-6500~K, while the hottest reaches 7280~K, and there is no preference between A or W subtypes. There are 5 systems with temperature difference between the components $(\Delta T)$ above 500~K, with the mean being about 260~K. We also find no trend in which $\Delta T$ increases with increasing temperature or with decreasing fill-out factor.

\begin{figure*}
	\includegraphics[scale=1.0 ]{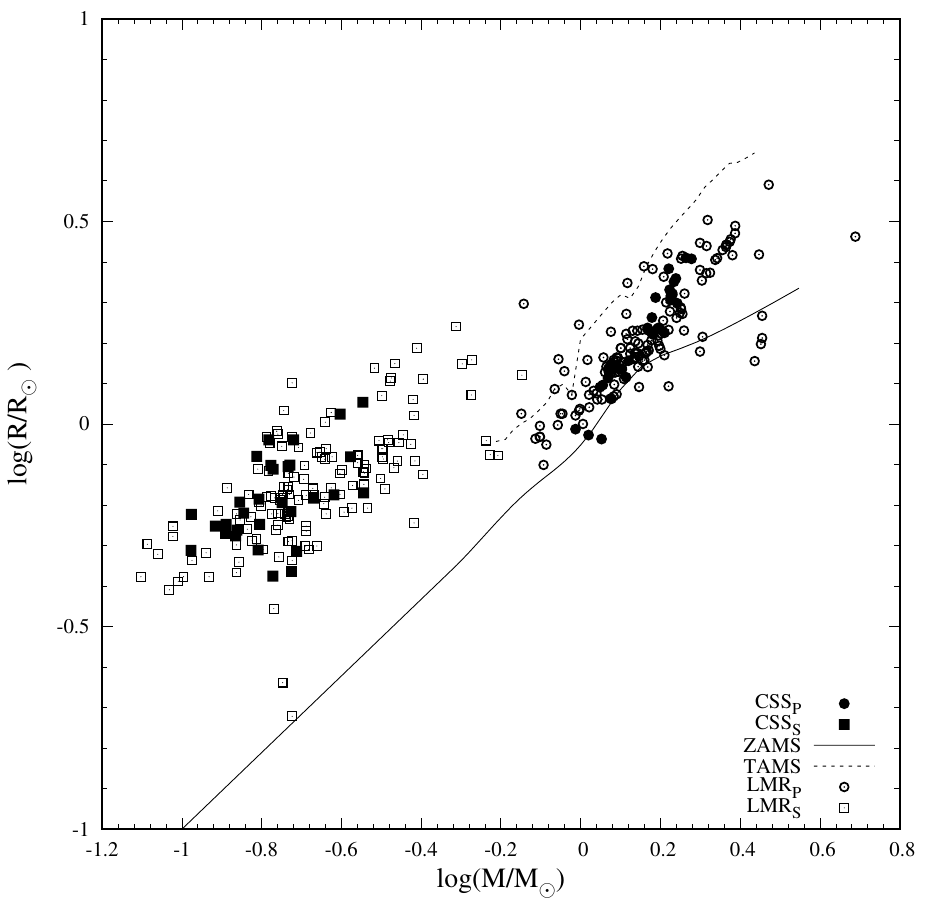}
    \caption{The primary and secondary components of CSS LMRs (solid circles and solid squares, respectively), combined with primary and secondary components from Previous LMRs from the literature (open circles and open squares, respectively), plotted on the $\log \it{M}-\log \it{R}$ diagram. ZAMS (solid) and TAMS (dotted) lines at solar metallicity, as obtained using the BSE code \citep{Hurley2002}, are overplotted.}
	\label{Evolution}
\end{figure*}

The evolutionary status of these new CSS LMRs can be compared in the mass-radius plane with 173 LMRs compiled from the literature having $q\leq\ 0.25$ derived spectroscopically or/and having total eclipse (listed in Table~\ref{tab:literature}). The main source of Table~\ref{tab:literature} comes from \cite{2021ApJS..254...10L}, but we have excluded systems that have unreliable or peculiarly large or small masses, corrected or updated the absolute parameters of others, and also included new systems. We also did not include LMRs from the  automated modeling of \cite{2020ApJS..247...50S} and \cite{2020PASJ...72...66L}, as we wanted to focus on dedicated studies of LMRs. Fig.~\ref{Evolution} shows their location in the $\log M- \log R$ diagram together with ZAMS and TAMS loci calculated for solar metallicity using the Binary Star Evolution code \citep[BSE,][]{Hurley2002}. The primary and secondary components are gathered in two different areas, above ZAMS and TAMS respectively indicating their different status but also their difference with the main sequence stars of the same mass. 

\begin{figure*}
	\includegraphics[scale=0.5]{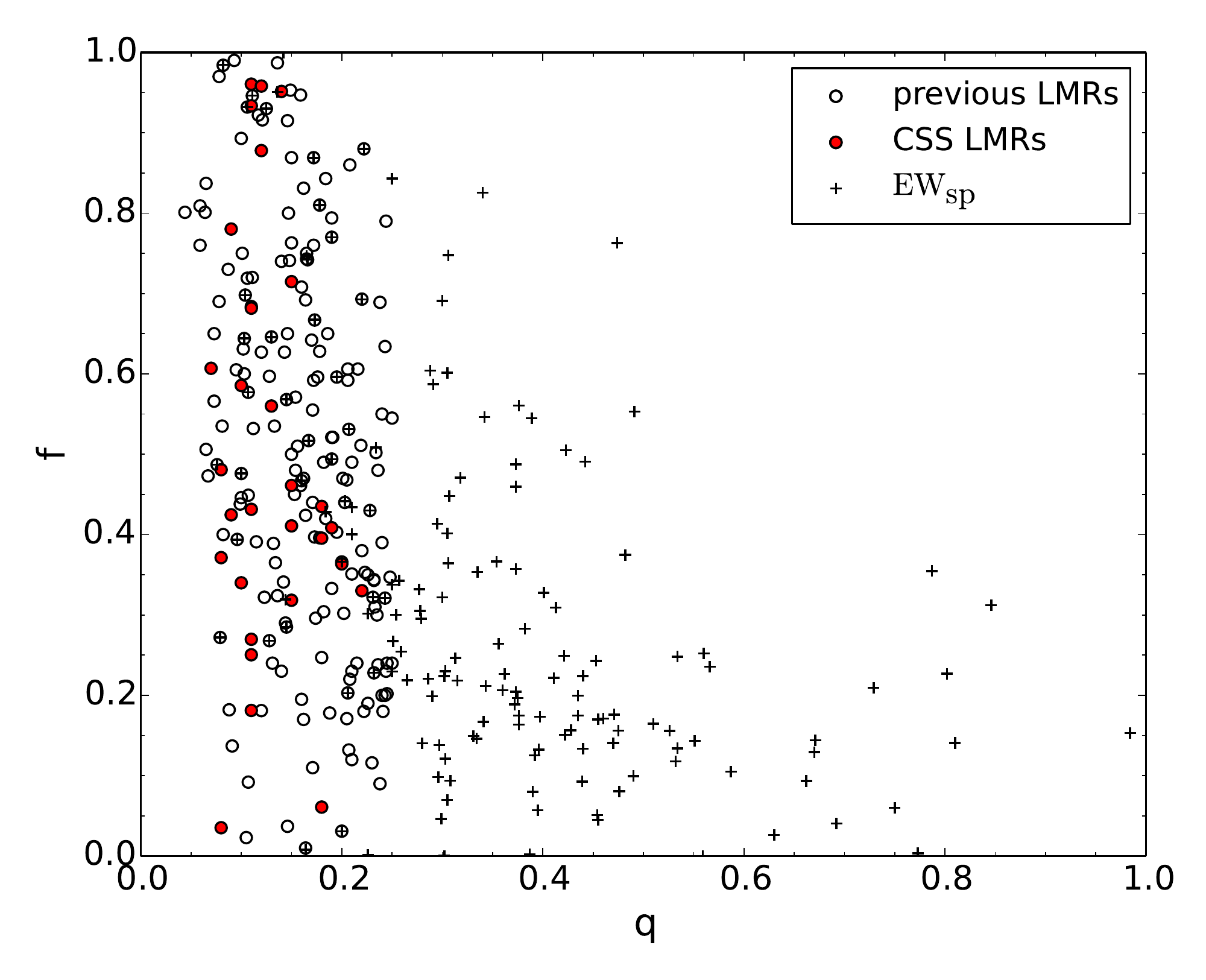}
    \caption{The relation $\it{q-f}$ for all previous LMRs (open circles) from Table~\ref{tab:literature}, new CSS LMRs (solid red circles), and EWs with spectroscopically derived mass ratios (EW$_{sp}$, crosses). In previous LMRs having both total eclipses and spectroscopic mass ratio, the symbols coincide.}
	\label{fig:qF}
\end{figure*}

It can be seen from Table~\ref{tab:models} that $\sim 40\%$ of the 30 new LMRs are deep contact $(f\geq 50 \%)$ with $q\leq0.25$, and that 8 of them, within the errors, belong to the class of extreme LMRs with $q\leq0.1$. Thus, with our 17 new discoveries, there are 43 extreme LMRs, only 6 of which have spectroscopic mass ratios (listed in Table~\ref{tab:literature}). The latter include the peculiar AW UMa, which shows the largest difference between photometric and spectroscopic mass ratio, $q=0.076$ \citep{2016MNRAS.457..836E} and $q = 0.099$ \citep{2015AJ....149...49R,2020AJ....160..104R}, respectively. The remaining 5 systems are SX CrV, V870 Ara, KR Com, FP Boo, and XX Sex. The system with the highest fill-out factor ($f\sim99\%$) is KR Com, which also has a large third light contribution at the level of $60\%$, while  SX CrV is the system with the lowest fill-out factor ($f\sim27\%$). As can be seen in Fig.~\ref{fig:qF}, where all previously known LMRs are plotted together with EWs with spectroscopic mass ratio (EW$_{sp}$) and new CSS LMRs, the smallest the $q$ value, the larger the fill-out factor distribution range. It is obvious that in previous LMRs which have both total eclipses and spectroscopic mass ratio, the symbols coincide.

\subsection{Premerger candidates} \label{Premerger cand}
According to \cite{1980A&A....92..167H}, a binary system becomes unstable if the ratio of the sum of the spin angular momenta of the stars ($J_{s}$)
to the orbital angular momentum ($J_{o}$) exceeds $\sim 1/3$ (Darwin’s instability), and is heading towards merging because the companion star can no longer keep the primary star synchronously rotating via the tidal interaction. 

\begin{figure*}
	\includegraphics[scale=0.5]{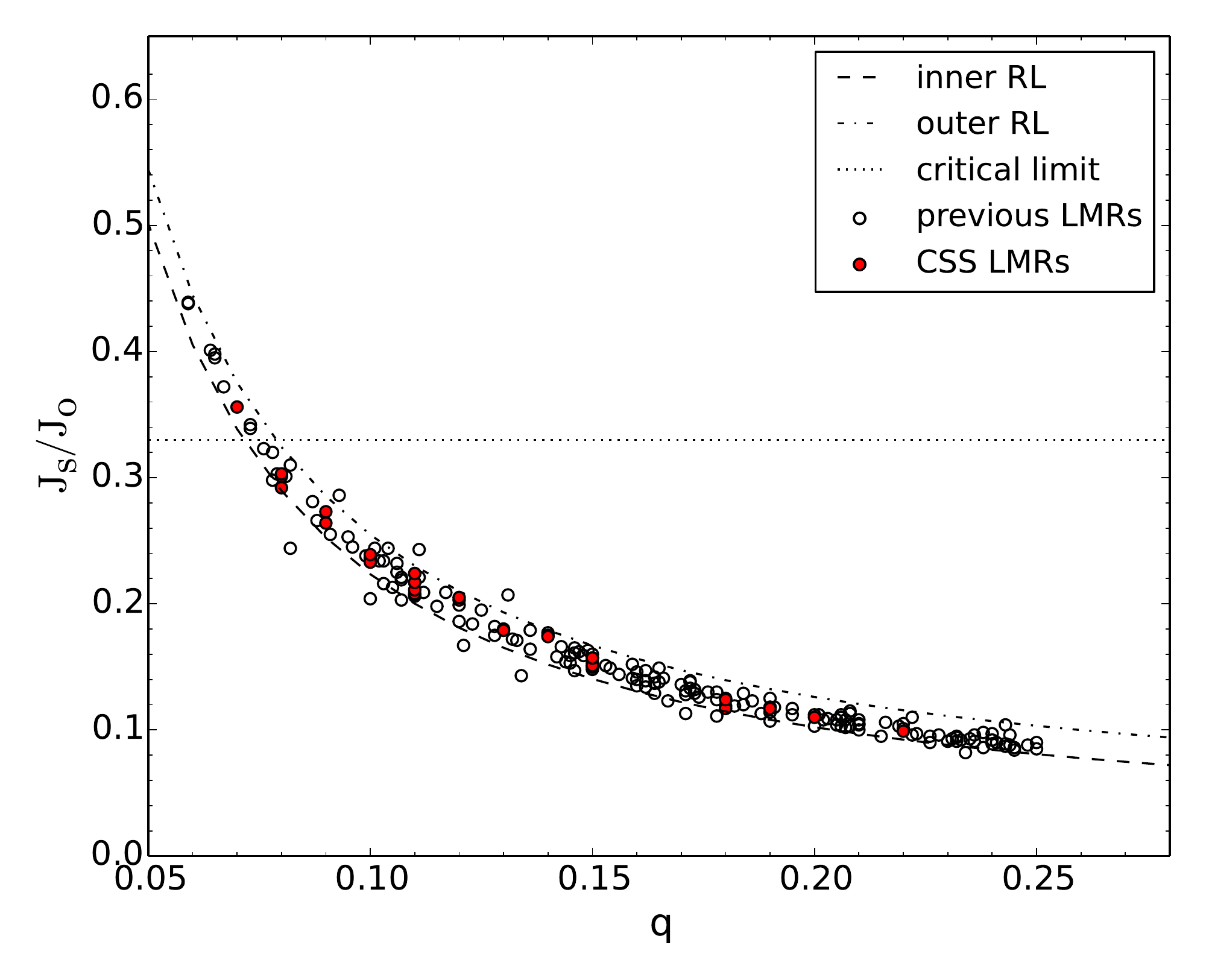}
    \caption{The angular momentum ratio $\frac {\it{J_s}}{\it{J_o}}$ against the mass ratio $\it{q}$ assuming $\it{k^2}$=0.06 for both components of all previous LMRs (open circles) from Table~\ref{tab:literature} and new CSS LMRs (solid red circles). The dashed and dashed-dotted lines correspond to theoretical systems filling the inner and outer Roche lobes, respectively, and the dotted line to the instability limit $\frac {\it{J_s}}{\it{J_o}}= 0.33$.}
	\label{Ratio0.06}
\end{figure*}

To investigate the stability of the 30 new LMRs listed in Tables~\ref{tab:1}-\ref{tab:absparams}, we determine this ratio using the formula derived from eq. (1) and (2) of \cite{2006MNRAS.369.2001L}
\begin{equation}
\frac{J_{s}}{J_{o}}= \frac{(1+q)}{q} (k_{1} r_{1})^{2} \left[1+q\left(\frac{k_{2}}{k_{1}}\right)^{2} \left(\frac{r_{2}}{r_{1}}\right)^2  \right],
\label{eq:2}
\end{equation}
\noindent where $q$ is the mass ratio of the system and $k_{1}, k_{2}$ are the dimensionless gyration radii of both components. The estimation of the gyration radii is crucial for the above estimation even when the mass ratio of the system is known spectroscopically, since it depends on the internal structure of the star. Main sequence stars tend to have gyration radii from 0.075 (fully radiative) to 0.205 (fully convective). Assuming $k_{1}^2= k_{2}^2=k^2=0.06$ (Sun like) as in \cite{1995ApJ...444L..41R} and \citet{2006MNRAS.369.2001L}, the results for the 30 new LMRs are plotted in Fig.~\ref{Ratio0.06}.

\begin{figure*}
	\includegraphics[scale=0.5 ]{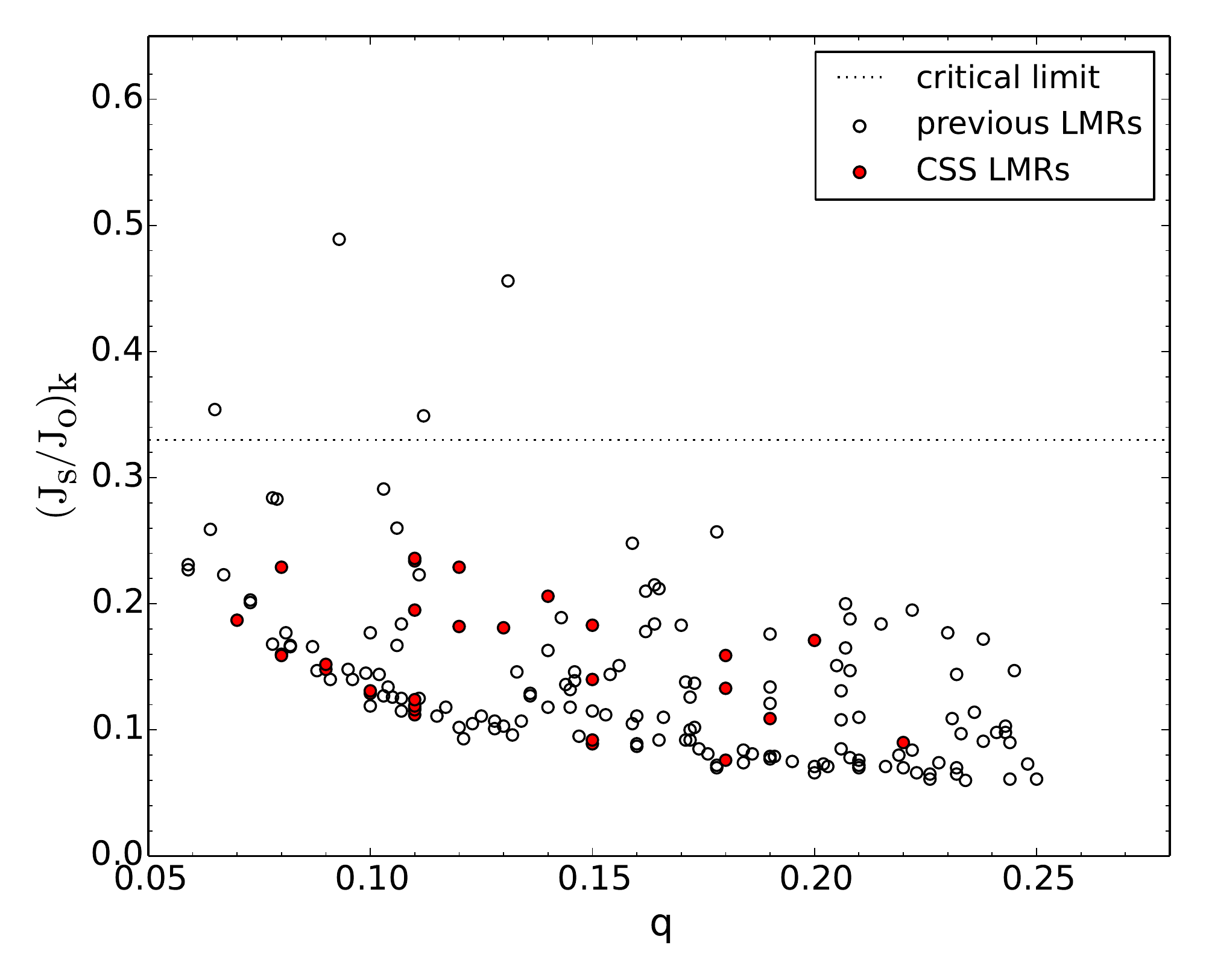}
    \caption{The angular momentum ratio $(\frac {\it{J_s}}{\it{J_o}})_{k}$ against the mass ratio $\it{q}$ assuming different values of $\it{k_{1}}$$^{2}$ for the massive component and $\it{k_{2}}$$^{2}$$\sim$ 0.205 for the less massive component. The symbols are as in Fig.~\ref{Ratio0.06}.}
	\label{fig:Ratiok}
\end{figure*}

It is seen in Fig.~\ref{Ratio0.06} and in Tables~\ref{tab:absparams} and \ref{tab:literature} that the majority of the systems lies in the region limited between the inner and outer Roche lobes. Nevertheless, CSS$\_$J075839.9+131355, V1187 Her, V857 Her, ASAS 083241+2332, V53 (in the globular cluster M4~= NGC~6121), and six contact binaries from Kepler Input catalogue (KICs) (KIC 4244929, KIC 9151972, KIC 11097678, KIC 8539720, KIC 3127873, KIC 12352712) are in the instability area, i.e., have surpassed the $J_{s}/J_{o}=1/3$ limit, and accordingly should have already merged. Alternative solutions that take into account phase smearing, as proposed by \citet{2017MNRAS.466.2488Z}, may be possible for the KICs. Three of them with periods $P>0.67$ days may have different $q,i$ model  when third light is included  in the $\it{Kepler}$ light curve with binned data. V53 is a member of the globular cluster M4, and a blue straggler with a decreasing period \citep{Kaluzny1997, 2017PASJ...69...79L}.
We therefore re-estimate the ratio $\frac{J_s}{J_o}$ assuming $k_{1}^2\neq k_{2}^2$ using the tabulated results of \cite{2009A&A...494..209L}, that take into account the combined effects of tidal and rotational distortions on a star in a binary system. For different masses of the primary components in the ZAMS, we calculate $k_{1}^2$ from the linear relationships we derived as $k_{1}=-0.250 \, M+0.539$ for stars with $M=0.5-1.4 \, M_{\sun}$ and $k_{1}=0.014 \, M+0.152$ for stars with $M>1.4M_{\sun}$, although in this range it can be considered approximately constant with a value of $k_{1} \approx 0.18$. \cite{2010MNRAS.405.2485J} also found that $k^2$ decreases with increasing mass and age of the star if the star's mass is less than about $1.3 \, M_{\sun}$, and that above this mass $k^2$ is roughly constant. For the secondary we assume $k_{2}^2\sim 0.205$, in accordance with a fully convective star \citep{2007MNRAS.377.1635A}, which is a good assumption as it is a very low-mass star ($M_{2}<0.3\, M_{\sun}$), although we know that it is oversized compared to a main sequence star of the same mass due to energy transfer. The results of the angular momentum ratio against the mass ratio $q$ assuming different values of gyration radius for the two components are presented in Fig.~\ref{fig:Ratiok}.

As seen from Fig.~\ref{fig:Ratiok} and Tables~\ref{tab:absparams} and \ref{tab:literature}, the scatter is larger than Fig.~\ref{Ratio0.06} but the majority of the LMRs has not surpassed the dynamical stability limit, with the exceptions of KR Com (0.489), ASAS J102556+2049.3 (0.456),  ASAS J083241+2332.4 (0.354) and V1222 Tau (0.349), that should have already merged. Although these systems have different fill-out factors ($99\%$, $24\%$, $51-70\%$ and $53\%$ respectively) and primary masses, all of them show indications of the presence of a third body. ASAS J102556+2049.3 has a large period decrease over 11 years ($dP/dt =-3.4 \times 10^{-6} \, {\rm d} \, {\rm yr}^{-1}$) \citep{2019AJ....158..186K}, ASAS J083241+2332.4 shows a sinusoidal period variation \citep{2016AJ....151...69S}, the spectroscopically studied KR Com has a third light contribution at the level of $60\%$ \citep{2021MNRAS.501.2897G} and V1222 Tau has one of the largest period variations $dP/dt = +8.19 \times 10^{-6} \, {\rm d} \, {\rm yr}^{-1}$ and an extreme O'Connell effect. Like in many LMRs, the long term period variations are associated with the presence of an additional component \citep{2006AJ....131.2986P} that plays a particular role in angular momentum evolution \citep{2006Ap&SS.304...75E}. Some information on the possible triplicity could be obtained from the so-called astrometric over-noise parameter RUWE (Renormalized Unit Weight Error) in the {\em Gaia} EDR3 \citep{2021ApJ...907L..33S}. This is an indicator for the multiplicity of a source when $\geq 1.4$, although according to \cite{2021ApJ...907L..33S} RUWE values even slightly greater than 1.0 may signify unresolved binaries in {\em Gaia} EDR3. Only two systems, CSS$\_$J011848.4+132107 (EM Psc) and CSS$\_$J233821.8+200518 have RUWE 3.6 and 3.7 respectively whereas 17 CSS LMRs  have slightly greater than 1.0 as it is the case of CSS$\_$J110526.4+285617 (1.112) and CSS$\_$J155637.0+060949 (1.024) that have also large orbital period variation. However, {\em Gaia} is insensitive to short-period triples and systems where photocentre does not move significantly \citep[see Fig.4][]{2020MNRAS.496.1922B}. This is the case of systems detected by TESS with RUWE $\leq 1.4$ like TIC 388459317 (0.952), TIC 52041148 (1.052) \citep{2022MNRAS.510.1352B} or even quadruples like TIC 278956474 (1.06) \citep{2020AJ....160...76R} where the low RUWE value suggests the two binaries are likely to be tightly bound.

Nevertheless, the uncertainty in the determination of the mass of the primary component, and consequently $k_{1}$, determines the uncertainty of the $\frac{J_s}{J_o}$ ratio. Under the assumption that the primary is a typical main sequence star with an effective temperature of, e.g, 5880~K–5440~K, corresponding to spectral types G0–G8, a mass uncertainty of $16\%$ results to an uncertainty of $\pm 24\%$ of the $\frac{J_s}{J_o}$ ratio. Thus, we consider additionally as potential premergers, systems with $(J_s/J_o)_k \sim 0.3$, like ZZ PsA, SX CrV, and ASAS J165139+2255.7. For ZZ PsA, our result is consistent with the recent study of \cite{2021MNRAS.501..229W}, who characterized ZZ PsA as a bright nova progenitor by developing a new relationship among the mass of the primary, the instability mass ratio $q_{\rm inst}$, and the degree of contact. Following their methodology, we estimated for V1222 Tau, SX CrV, and ASAS J165139+2255.7 the ratio of the reported photometric mass ratio to the instability mass ratio $q/q_{\rm inst}$ to be 0.89, 1.08, and 0.96, respectively, at a separation and period ($1.95 \, R_{\sun}$, 0.3143~d) for V1222 Tau, ($0.204 \, R_{\sun}$, 0.292~d) for SX CrV, and ($2.24 \, R_{\sun}$, 0.363251~d) for ASAS J165139+2255.7, suggesting that these systems will enter an unstable phase. 
 ASAS J083241+2332.4 and KR Com have $q/q_{\rm inst}=(0.85,0.65)$, as well as theoretical instability separation and period ($2.29 \, R_{\sun}$, 0.3515~d) and ($2.763 \, R_{\sun}$, 0.543028~d),  respectively. Among the new CSS LMRs, CSS$\_$J075848.2+125656, with $q/q_{\rm inst}=1.23\pm 0.23$ and  CSS$\_$J093010.1-021624, with $q/q_{\rm inst}=1.25\pm 0.23$, can be considered as merging system candidates.
 
\section{Conclusions}\label{Conclusions}

In this paper, we mine and model 30 new totally-eclipsing CSS LMRs. We present their physical parameters ($T_2/T_1$, $R_2/R_1$, $\Omega_{\rm 1,2}$, $i$, $f$) as well as the systems' global parameters, as determined from the light curve solutions and {\em Gaia} early DR3 distances. 

Normally, LMRs should have a deep contact configuration, but the existence also of systems far away from this expected state may indicate different evolutionary states. LMRs presenting a shallow or medium degree of contact may be the result of mass transfer from the less massive to the more massive component, which causes the orbit to enlarge without breaking the contact. In the case of a high contact degree, the surface of the binary is very close to its outer critical Roche lobe, and  a long-term period decrease due to mass transfer or angular momentum loss via magnetic braking, can cause the orbit to tighten, the inner and outer critical Roche lobes to shrink,  and the degree of contact to increase. In the case of a high contact degree with continuous mass transfer from the more massive to the less massive (increasing period), and mass loss from the second Lagrange point ($L_2$), the orbit of the system is contracting, with the mass ratio decreasing, and the evolution ends with a rapidly rotating single star. Tidal friction in the presence of a circumstellar disc or a third (or more) companion(s) can constitute another possible mechanism of angular momentum or mass loss \citep{Tutukov2004, Martin2011}.

Why do systems with a lower mass ratio than theoretically predicted for stability exist? If the observations do not support the Darwin instability criterion in such systems, maybe the mechanism of merger is a different one, or we need to reconsider the parameters involved. Primarily since the primary's dimensionless gyration radius $k_{1}$ depends on the primary's mass $M_{1}$, in the case of a photometric solution with total eclipse, $M_{1}$ estimation is based on various assumptions that need to be reexamined. In addition, more realistic calculations should include the decrease of $k_{1}$ with evolution \citep{2010MNRAS.405.2485J}, or the decrease of $\frac{k_2}{k_1}$ due to the differential rotation of the components \citep{Hilditch2001}. The latter has been proposed as a mechanism for the energy transfer between the components \citep{2005ApJ...629.1055Y,2010ASPC..435..151E}. 

Looking at the prototype of mergers, V1309 Sco, that was dominated in its final stage by $L_2$ mass loss, another possible explanation is that the non-conservative mass and angular momentum loss drives the Darwin criterion  and the instability mass ratio to lower values, as proposed by \citet{2017ApJ...850...59P}. Nevertheless, since the progenitor of V1309 Sco is proposed to be an eclipsing contact binary with period of about 1.4 d \citep{2011A&A...528A.114T} maybe larger period systems need to be observed in contact phase. These are few and represent the very end of contact binaries with periods $1.3-1.5$~d in the period distribution as initially noticed by \cite{1998Rucinski} and is still confirmed in the plethora of new discoveries of surveys \citep{2019Li,2020Jayasinghe}. However, in the model proposed by \cite{Stepien2011} the primary of the preburst contact binary V1309 Sco is a giant that recently filled its Roche lobe and whose contact phase is very short. Thus, the proposed progenitors of mergers may differ from EWs in both evolutionary state and duration of the contact phase. Considering the rapid period decay of V1309 Sco preceding the merging, we suggest to search for systems with large orbital period change rates ($dP/dt$) in photometric surveys,  in combination with detailed study of their evolutionary state through high-accuracy spectroscopic observations \citep{2020AJ....160..104R}, reconsidering the energy transfer models between the components and including systematic monitoring of times of minima of the already proposed merger candidates.

\section*{Acknowledgements}
AP gratefully acknowledges the support provided by the grant co-financed by Greece and the European Union (European Social Fund- ESF) through the Operational Programme «Human Resources Development, Education and Lifelong Learning» in the context of the project “Reinforcement of Postdoctoral Researchers - 2nd Cycle” (MIS-5033021), implemented by the State Scholarships Foundation (IKY). 
CEFL acknowledges a post-doctoral fellowship from the CNPq , MCTIC/FINEP (CT-INFRA grant 0112052700), and the Embrace Space Weather Program for the computing facilities at INPE. 
This work made use of data products from the CSS survey. 
Support for MC is provided by ANID's Millennium Science Initiative through grant ICN12\textunderscore 12009, awarded to the Millennium Institute of Astrophysics (MAS), and by ANID's Basal project FB210003.  
The CSS survey is funded by the National Aeronautics and Space
Administration under Grant No. NNG05GF22G issued through the Science Mission Directorate Near-Earth Objects Observations
Program. The CRTS survey is supported by the US National
Science Foundation under grants AST-0909182, AST-1313422,
AST-1413600, and AST-1518308.

We would like to thank the referee for constructive comments and recommendations that have improved the paper.

\section*{Data availability}

The data underlying this article are available in the article and in its online supplementary material




\bibliographystyle{mnras}
\bibliography{LMRatCSS}



\appendix
\section{Previous LMRs from the literature}

In this Appendix, we provide a compilation of LMRs, along with their parameters, from the literature and derived in this work (Table~\ref{tab:literature}). The columns are organized as follows (from left to right): 
Name, where (n) represents LMRs not included in \citet{2021ApJS..254...10L} and (*) LMRs with LCs  used for sample 2; 
period (in days); 
mass ratio ($q$); 
method used to determine the mass ratio: photometric with total eclipses (T) or spectroscopic  (SP);
primary mass ({$M_1$}); 
primary radius ({$R_1$}); 
secondary radius ({$R_2$}); 
fill-out factor ($f$);
the derived ratio of total spin angular to orbital momentum $\frac{J_{s}}{J_{o}}$ for  gyration radii $k=0.06$ {} (this work); the derived ratio of total spin angular to orbital momentum  $(\frac{J_{s}}{J_{o}})_{k}$ for different values  of $k_{1}^2$ for the massive component and $k_{2}^2\sim 0.205$ for the less massive component (this work); and references.

\begin{table*}
\centering
 \caption{Parameters of previously known LMRs.}  \label{tab:literature}
 \scriptsize
 \begin{tabular}{l*{11}{c}r}
 \hline
 \text{Name} & \text{Period} & \text{$q$} & \text{type} & \text{$M_1$}  & \text{$R_1$} & \text{$R_2$} &\text{$f$}&\text {$\frac{J_{s}}{J_{o}}$} & $\left(\frac{J_{s}}{J_{o}}\right)_k$ &  \text{Reference} \\ 
 &(days) & & &       ($M_{\sun}$) & ($R_{\sun})$ & ($R_{\sun})$ & &  \\
 \hline
V1187 Her	&	0.31076	&	0.044	&	T	&		&		&		&	0.801	&	0.612	&		&	\cite{2019PASP..131e4203C}	\\
KIC 4244929*	&	0.34140	&	0.059	&	T	&	1.481	&	1.521	&	0.477	&	0.809	&	0.439	&	0.227	&	\cite{2016PASA...33...43S}	\\
KIC 9151972*	&	0.38680	&	0.059	&	T	&	1.606	&	1.696	&	0.528	&	0.760	&	0.438	&	0.231	&	\cite{2016PASA...33...43S}	\\
KIC 11097678*	&	0.99972	&	0.064 (0.097)	&	T	&	2.960	&	3.897	&	1.264	&	0.801 (0.87)	&	0.401	&	0.259	&	\cite{2016PASA...33...43S};\cite{2017MNRAS.466.2488Z}	\\
ASAS J083241+2332.4*	&	0.31132	&	0.065	&	T	&	1.220	&	1.34	&	0.42	&	0.506	&	0.398	&	0.354	&	\cite{2016AJ....151...69S}	\\
V857 Her*	&	0.38223	&	0.065	&	T	&		&		&		&	0.837	&	0.395	&		&	\cite{2005AJ....130.1206Q}	\\
KIC 8539720*	&	0.74450	&	0.067 (0.158)	&	T	&	2.438	&	2.955	&	0.929	&	0.473 (0.860)	&	0.372	&	0.223	&	\cite{2016PASA...33...43S}; \cite{2017MNRAS.466.2488Z}	\\
KIC 3127873*	&	0.67153	&	0.073 (0.109)	&	T	&	2.268	&	2.69	&	0.899	&	0.65 (0.88)	&	0.342	&	0.201	&	\cite{2016PASA...33...43S}; \cite{2017MNRAS.466.2488Z}	\\
KIC 12352712*	&	0.72207	&	0.073	&	T	&	2.377	&	2.859	&	0.944	&	0.566	&	0.339	&	0.203	&	\cite{2016PASA...33...43S}	\\
AW UMa	&	0.43872	&	0.076 (0.099)	&	SP	&		&		&		&	0.487	&	0.323	&		&	\cite{2016MNRAS.457..836E};\cite{2015AJ....149...49R}	\\
ZZ PsA	&	0.37388	&	0.078	&	T	&	1.213	&	1.422	&	0.559	&	0.970	&	0.298	&	0.284	&	\cite{2021MNRAS.501..229W}	\\
 M4 V53	&	0.30845	&	0.078	&	T	&	1.472	&	1.383	&	0.481	&	0.690	&	0.320	&	0.168	&	\cite{2017PASJ...69...79L}	\\
SX Crv	&	0.31660	&	0.079	&	SP	&	1.246	&	1.347	&	0.409	&	0.272	&	0.303	&	0.283	&	\cite{2004AcA....54..299Z}	\\
KIC 10007533*	&	0.64806	&	0.081 (0.101)	&	T	&	2.199	&	2.566	&	0.881	&	0.535 (0.178)	&	0.301	&	0.177	&	\cite{2016PASA...33...43S}; \cite{2017MNRAS.466.2488Z}	\\
V870 Ara*	&	0.39972	&	0.082	&	SP	&	1.503	&	1.61	&	0.61	&	0.984	&	0.310	&	0.166	&	\cite{2007Szalai}	\\
KIC 8145477*	&	0.56578	&	0.082 (0.102)	&	T	&	2.012	&	2.26	&	0.767	&	0.40 (0.650)	&	0.244	&	0.167	&	\cite{2016PASA...33...43S}; \cite{2017MNRAS.466.2488Z}	\\
KIC 11144556*	&	0.64298	&	0.087 (0.161)	&	T	&	2.174	&	2.542	&	0.927	&	0.730 (0.97)	&	0.281	&	0.166	&	\cite{2016PASA...33...43S}; \cite{2017MNRAS.466.2488Z}	\\
KIC 10596883*	&	0.46891	&	0.088	&	T	&	1.772	&	1.882	&	0.641	&	0.182	&	0.266	&	0.147	&	\cite{2016PASA...33...43S}	\\
KIC 8804824*	&	0.45740	&	0.091 (0.111)	&	T	&	1.738	&	1.829	&	0.628	&	0.137 (0.67)	&	0.255	&	0.140	&	\cite{2016PASA...33...43S}; \cite{2017MNRAS.466.2488Z}	\\
KR Com (n)	&	0.40797	&	0.093	&	SP	&	0.880	&	1.445	&	0.505	&	0.990	&	0.286	&	0.489	&	\cite{2021MNRAS.501.2897G}	\\
KIC 7698650*	&	0.59916	&	0.095 (0.123)	&	T	&	2.064	&	2.357	&	0.876	&	0.605 (0.700)	&	0.253	&	0.148	&	\cite{2016PASA...33...43S}; \cite{2017MNRAS.466.2488Z}	\\
FP Boo	&	0.64048	&	0.096	&	SP	&	1.614	&	2.31	&	0.774	&	0.394	&	0.245	&	0.140	&	\cite{2006AcA....56..127G}	\\
KIC 9453192*	&	0.71884	&	0.099 (0.155)	&	T	&	2.314	&	2.734	&	1.01	&	0.438 (0.620)	&	0.238	&	0.145	&	\cite{2016PASA...33...43S}; \cite{2017MNRAS.466.2488Z}	\\
XX Sex	&	0.54011	&	0.1	&	SP	&	1.301	&	1.87	&	0.696	&	0.476	&	0.236	&	0.177	&	\cite{2011MNRAS.412.1787D}	\\
UCAC4 479-113711	&	0.35292	&	0.1	&	T	&	1.400	&	1.47	&	0.58	&	0.893	&	0.204	&	0.119	&	\cite{2019NewA...69...21E}	\\
NW Aps	&	1.06556	&	0.1	&	T	&		&		&		&	0.446	&	0.237	&		&	\cite{2005Wadhwa}	\\
AW CrB*	&	0.36094	&	0.101	&	T	&		&		&		&	0.750	&	0.244	&		&	\cite{2013MNRAS.430.3070B}	\\
KIC 9350889*	&	0.72595	&	0.102 (0.106)	&	T	&	2.322	&	2.775	&	1.068	&	0.631 (0.870)	&	0.234	&	0.144	&	\cite{2016PASA...33...43S}; \cite{2017MNRAS.466.2488Z}	\\
ASAS J165139+2255.7*	&	0.35321	&	0.103	&	T	&	1.030	&	1.27	&	0.46	&	0.600	&	0.216	&	0.291	&	\cite{2018JAVSO..46....3A}	\\
DN Boo	&	0.44757	&	0.103	&	SP	&	1.428	&	1.71	&	0.67	&	0.644	&	0.234	&	0.127	&	\cite{2008NewA...13..468S}	\\
FG Hya	&	0.32783	&	0.104	&	SP	&	1.445	&	1.438	&	0.515	&	0.698	&	0.244	&	0.134	&	\cite{2010MNRAS.408..464Z}	\\
KIC 10229723*	&	0.62872	&	0.105 (0.145)	&	T	&	2.110	&	2.363	&	0.852	&	0.023 (0.35)	&	0.213	&	0.126	&	\cite{2016PASA...33...43S}; \cite{2017MNRAS.466.2488Z}	\\
ASAS J082243+1927.0	&	0.28000	&	0.106	&	T	&	1.100	&	1.15	&	0.42	&	0.719	&	0.225	&	0.266	&	\cite{2015MNRAS.446..510K}	\\
GR Vir	&	0.34697	&	0.106	&	SP	&	1.376	&	1.49	&	0.55	&	0.932	&	0.232	&	0.167	&	\cite{2005AcA....55..123G}	\\
V1191 Cyg*	&	0.31339	&	0.107	&	SP	&	1.283	&	1.292	&	0.503	&	0.577	&	0.221	&	0.184	&	\cite{2014NewA...31...14O}	\\
ASAS J025115-2525.4	&	0.55927	&	0.107	&	T	&	1.680	&	2.07	&	0.7	&	0.092	&	0.203	&	0.115	&	\cite{2016NewA...46...94S}	\\
KIC 7601767*	&	0.48673	&	0.107	&	T	&	1.784	&	1.925	&	0.74	&	0.449	&	0.219	&	0.125	&	\cite{2016PASA...33...43S}	\\
CK Boo	&	0.35515	&	0.111	&	SP	&	1.584	&	1.533	&	0.65	&	0.946	&	0.221	&	0.125	&	\cite{2011MNRAS.412.1787D}	\\
TYC 6995-813-1 (n)	&	0.38318	&	0.111	&	T	&	1.230	&	1.46	&	0.6	&	0.720	&	0.243	&	0.223	&	\cite{2021arXiv210512927W}	\\
V1222 Tau*	&	0.29173	&	0.112	&	T	&	0.900	&	1.06	&	0.42	&	0.532	&	0.209	&	0.349	&	\cite{2015PASJ...67...74L}	\\
NSVS 3198272	&	0.35228	&	0.115	&	T	&	1.621	&	1.479	&	0.583	&	0.391	&	0.198	&	0.111	&	\cite{2019AJ....158..186K}	\\
KIC 6118779*	&	0.36425	&	0.117	&	T	&	1.465	&	1.512	&	0.657	&	0.922	&	0.209	&	0.118	&	\cite{2016PASA...33...43S}	\\
V2787 Ori	&	0.81098	&	0.12	&	T	&	1.440	&	2.45	&	0.96	&	0.181	&	0.186	&	0.102	&	\cite{2019PASP..131h4203T}	\\
AL Lep	&	0.44864	&	0.12	&	T	&		&		&		&	0.627	&	0.199	&		&	\cite{2005Wadhwa}	\\
KIC 10395609*	&	0.36425	&	0.121	&	T	&	1.460	&	1.504	&	0.657	&	0.916	&	0.167	&	0.093	&	\cite{2016PASA...33...43S}	\\
Cl* NGC 6121 SAW V66	&	0.26997	&	0.123	&	T	&	1.660	&	1.24	&	0.5	&	0.322	&	0.184	&	0.105	&	\cite{2011MNRAS.415.1509L}	\\
KIC 10618253*	&	0.43740	&	0.125	&	SP	&	1.476	&	1.703	&	0.756	&	0.930	&	0.195	&	0.111	&	\cite{2016PASA...33...43S}	\\
eps CrA	&	0.59143	&	0.128	&	SP	&	1.700	&	2.1	&	0.85	&	0.268	&	0.175	&	0.101	&	\cite{2005PASJ...57..983Y}	\\
KIC 12055014	&	0.49991	&	0.128 (0.16)	&	T	&	1.781	&	1.939	&	0.819	&	0.597 (0.670)	&	0.182	&	0.107	&	\cite{2016PASA...33...43S}; \cite{2017MNRAS.466.2488Z}	\\
V776 Cas	&	0.44042	&	0.13	&	SP	&	1.550	&	1.71	&	0.73	&	0.646	&	0.180	&	0.103	&	\cite{2016ApJ...817..133Z}	\\
ASAS J102556+2049.3	&	0.28498	&	0.131	&	T	&	0.710	&	1.06	&	0.39	&	0.240	&	0.207	&	0.456	&	\cite{2019AJ....158..186K}	\\
V902 Sgr	&	0.29394	&	0.132	&	T	&	1.400	&	1.234	&	0.513	&	0.389	&	0.172	&	0.096	&	\cite{2002Obs...122...22S}	\\
ASAS J050334-2521.9	&	0.41406	&	0.133	&	T	&	1.260	&	1.54	&	0.6	&	0.535	&	0.171	&	0.146	&	\cite{2016NewA...44...40G}	\\
KIC 8265951*	&	0.77996	&	0.134 (0.154)	&	T	&	2.364	&	2.816	&	1.175	&	0.365 (0.380)	&	0.143	&	0.107	&	\cite{2016PASA...33...43S}; \cite{2017MNRAS.466.2488Z}	\\
ASAS J040633-4729.4	&	0.40637	&	0.136	&	T	&	1.330	&	1.54	&	0.6	&	0.324	&	0.164	&	0.127	&	\cite{2016NewA...46...94S}	\\
DZ Psc*	&	0.36613	&	0.136	&	SP	&	1.370	&	1.46	&	0.67	&	0.987	&	0.179	&	0.129	&	\cite{2013AJ....146...35Y}	\\
HV Aqr	&	0.37450	&	0.14	&	SP	&	1.240	&	1.456	&	0.601	&	0.740	&	0.175	&	0.163	&	\cite{2021MNRAS.501.2897G}	\\
HV Aqr	&	0.37446	&	0.145	&	SP	&	1.355	&	1.448	&	0.648	&	0.568	&	0.159	&	0.118	&	\cite{2013NewA...21...46L}	\\
ASAS J142124+1813.1 (n)	&	0.24300	&	0.14	&	T	&	2.730	&	1.43	&	0.57	&	0.230	&	0.177	&	0.118	&	\cite{2019AJ....158..186K}	\\
V677 Cen	&	0.32500	&	0.142	&	T	&		&		&		&	0.341	&	0.158	&		&	\cite{1993ApJ...407..237B}	\\
V710 Mon	&	0.40520	&	0.143	&	T	&	1.140	&	1.46	&	0.66	&	0.627	&	0.166	&	0.189	&	\cite{2014NewA...31...60L}	\\
V410 Aur*	&	0.36636	&	0.144	&	SP	&	1.270	&	1.37	&	0.59	&	0.290	&	0.154	&	0.136	&	 \cite{2017AJ....154...99L}	\\
HN UMa	&	0.38260	&	0.145	&	SP	&	1.279	&	1.42	&	0.61	&	0.285	&	0.153	&	0.132	&	\cite{2007ASPC..362...82O}	\\
KIC 9776718 (n)	&	0.5444	&	0.146	&	T	&	4.880	&	2.9	&	1.32	&	0.915	&	0.165	&	0.139	&	\cite{2020PASJ...72...66L}	\\
GSC 1042-2191	&	0.42380	&	0.146	&	T	&	1.260	&	1.54	&	0.69	&	0.650	&	0.161	&	0.146	&	\cite{2016NewA...44...35B}	\\
NSVS 1917038	&	0.31807	&	0.146	&	T	&		&		&		&	0.037	&	0.147	&		&	\cite{2020MNRAS.497.3381G}	\\
KIC 2159783*	&	0.37388	&	0.147	&	T	&	1.451	&	1.496	&	0.694	&	0.800	&	0.162	&	0.095	&	\cite{2016PASA...33...43S}	\\
XY LMi*	&	0.43689	&	0.148	&	T	&		&		&		&	0.741	&	0.159	&		&	\cite{2011AJ....141..151Q}	\\
EM Psc	&	0.34396	&	0.149	&	T	&		&		&		&	0.953	&	0.163	&		&	\cite{2008AJ....136.1940Q}	\\
ASAS J113031-0101.9 (n)	&	0.27100	&	0.15	&	T	&		&		&		&	0.500	&	0.153	&		&	\cite{2009IBVS.5886....1P}	\\
TYC 4157-0683-1	&	0.39607	&	0.15	&	T	&	1.367	&	1.499	&	0.667	&	0.763	&	0.157	&	0.115	&	\cite{2014NewA...31....1A}	\\
Mis V1395	&	0.73930	&	0.15	&	T	&		&		&		&	0.869	&	0.160	&		&	\cite{2020MNRAS.497.3381G}	\\
V1179 Her (n)	&	0.38551	&	0.153	&	T	&	1.300	&		&		&	0.450	&	0.151	&	0.112	&	\cite{2021MNRAS.501.4935B}	\\
V1511 Her (n)	&	0.35008	&	0.154	&	T	&		&		&		&	0.480	&		&		&	\cite{2021MNRAS.501.4935B}	\\
ASAS J063546+1928.6	&	0.47553	&	0.154	&	T	&	1.230	&	1.184	&	0.19	&	0.571	&	0.149	&	0.144	&	\cite{2018AJ....155..172S}	\\

 \hline
 
\end{tabular}
\end{table*}

\begin{table*}
\centering
\contcaption{Parameters of previous LMRs.}\label{tab:continued}
 \scriptsize

\begin{tabular}{l*{15}{c}r}
 
 \hline
 \text{Name} & \text{Period} & \text{$q$} & \text{type} & \text{$M_1$}  & \text{$R_1$} & \text{$R_2$} &\text{$f$}&\text {$\frac{J_{s}}{J_{o}}$} & $\left(\frac{J_{s}}{J_{o}}\right)_k$ &  \text{Reference} \\ 
 &(days) & & & ($M_{\sun}$) & ($R_{\sun})$ & ($R_{\sun})$ & &  \\
 \hline

AH Cnc	&	0.36046	&	0.156	&	T	&	1.188	&	1.332	&	0.592	&	0.510	&	0.144	&	0.151	&	\cite{2016RAA....16..157P}	\\
NGC 6397 V8	&	0.27124	&	0.159	&	T	&	0.876	&	0.995	&	0.456	&	0.461	&	0.141	&	0.248	&	\cite{2013NewA...25...12L}	\\
KIC 3104113*	&	0.84679	&	0.159 (0.167)	&	T	&	2.440	&	3.085	&	1.538	&	0.947 (0.910)	&	0.152	&	0.105	&	\cite{2016PASA...33...43S}; \cite{2017MNRAS.466.2488Z}	\\
Cl* NGC 6715 SAW V144	&	0.72159	&	0.16	&	T	&	1.310	&	2.23	&	0.95	&	0.195	&	0.135	&	0.111	&	\cite{2013NewA...22...57L}	\\
NSVS 1926064	&	0.40747	&	0.16	&	T	&	1.558	&	1.605	&	0.755	&	0.708	&	0.146	&	0.089	&	\cite{2020NewA...7701352K}	\\
EF Dra	&	0.42403	&	0.16	&	SP	&	1.815	&	1.702	&	0.777	&	0.467	&	0.140	&	0.087	&	\cite{2012RAA....12..419Y}	\\
V902 Cep	&	0.32870	&	0.162	&	T	&	1.077	&	1.208	&	0.563	&	0.470	&	0.139	&	0.178	&	\cite{2019AJ....158..186K}	\\
NSVS 2256852	&	0.34888	&	0.162	&	T	&	0.950	&	1.18	&	0.52	&	0.170	&	0.134	&	0.210	&	\cite{2019AJ....158..186K}	\\
V1695 Aql	&	0.41283	&	0.162	&	T	&		&		&		&	0.831	&	0.147	&		&	\cite{2017JAVSO..45..140S} 	\\
V972 Her	&	0.44309	&	0.164	&	SP	&	0.910	&	1.35	&	0.59	&	0.01	&	0.129	&	0.215	&	\cite{2018Selam}	\\
V1115 Cas	&	0.32329	&	0.164	&	T	&	1.049	&	1.181	&	0.548	&	0.424	&	0.137	&	0.184	&	\cite{2019AJ....158..186K}	\\
GSC 03517-00663	&	0.29502	&	0.164	&	T	&		&		&		&	0.692	&	0.142	&		&	\cite{2015RAA....15..889G}	\\
Cl* NGC 104 WSB V95	&	0.27890	&	0.165	&	T	&	0.970	&	1.05	&	0.49	&	0.743	&	0.138	&	0.212	&	\cite{2014NewA...26..116L}	\\
AH Aur	&	0.49411	&	0.165	&	SP	&	1.674	&	1.897	&	0.837	&	0.750	&	0.149	&	0.092	&	\cite{2005AcA....55..123G}	\\
TV Mus	&	0.44568	&	0.166	&	SP	&	1.350	&	1.7	&	0.83	&	0.742	&	0.141	&	0.110	&	\cite{2005AJ....130..224Q}	\\
V445 Cep	&	0.44878	&	0.167	&	SP	&		&		&		&	0.517	&	0.123	&		&	\cite{2010JASS...27...69O}	\\
EPIC 211957146*	&	0.35502	&	0.17	&	T	&	1.050	&	1.1	&	0.23	&	0.642	&	0.136	&	0.183	&	\cite{2017AJ....153..231S}	\\
V345 Gem	&	0.27477	&	0.171	&	SP	&1.371		&1.134		&0.486		&	0.11	&0.128		&	00.092	&	\cite{2021MNRAS.501.2897G}	\\
NSVS 13602901 (n)	&	0.52389	&	0.171	&	T	&	1.190	&	1.69	&	0.79	&	0.440	&	0.131	&	0.138	&	\cite{2021arXiv210512927W}	\\
V1542 Aql	&	0.41754	&	0.171	&	T	&		&		&		&	0.555	&	0.113	&		&	\cite{2004JAVSO..32...95W}	\\
TYC 1337-1137-1 (n)*	&	0.47550	&	0.172	&	T	&	1.386	&	1.7	&	0.83	&	0.760	&	0.138	&	0.100	&	\cite{2017PASP..129l4204L}	\\
AS CrB*	&	0.38066	&	0.172	&	T	&	1.250	&	1.4	&	0.67	&	0.592	&	0.133	&	0.126	&	\cite{2017NewA...51....1L}	\\
II UMa	&	0.82522	&	0.172	&	SP	&	1.990	&	2.8	&	2.41	&	0.869	&	0.139	&	0.092	&	\cite{2016AJ....151...67Z}	\\
OU Ser	&	0.29677	&	0.173	&	SP	&	1.187	&	1.155	&	0.544	&	0.667	&	0.129	&	0.137	&	\cite{2011MNRAS.412.1787D} 	\\
ASAS J002821-1453.3	&	0.40266	&	0.173	&	T	&	1.330	&	1.49	&	0.6	&	0.397	&	0.132	&	0.102	&	\cite{2016NewA...44...40G}	\\
KIC 5439790*	&	0.79609	&	0.174 (0.192)	&	T	&	2.314	&	2.764	&	1.29	&	0.296 (0.360)	&	0.126	&	0.085	&	\cite{2016PASA...33...43S}; \cite{2017MNRAS.466.2488Z}	\\
KIC 8496820*	&	0.43697	&	0.176	&	T	&	1.566	&	1.655	&	0.803	&	0.596	&	0.130	&	0.081	&	\cite{2016PASA...33...43S}	\\
PZ UMa	&	0.26267	&	0.178	&	T	&	0.770	&	0.92	&	0.43	&	0.396	&	0.124	&	0.257	&	\cite{2019PASJ...71...39Z}	\\
V728 Her	&	0.47129	&	0.178	&	SP	&	1.800	&	1.87	&	0.82	&	0.810	&	0.111	&	0.072	&	\cite{2016NewA...46...73E}	\\
CSS J075258.0+382035 	&	0.42991	&	0.178	&	T	&	1.040	&	1.44	&	0.7	&	0.628	&	0.130	&	0.070	&	\cite{2019AJ....158..186K}	\\
GSC 3599-2569	&	0.40291	&	0.18	&	T	&		&		&		&	0.247	&	0.119	&		&	\cite{2015AstBu..70..109G}	\\
TYC 3700-1384 (n)	&	0.40747	&	0.182	&	T	&	1.450	&		&		&	0.490	&		&		&	\cite{2021MNRAS.501.4935B}	\\
GSC 4778-152	&	0.51746	&	0.182	&	T	&		&		&		&	0.304	&	0.119	&		&	\cite{2008BaltA..17...79T}	\\
TY Pup (n)	&	0.81920	&	0.184	&	T	&	1.650	&	2.636	&	1.373	&	0.843	&	0.129	&	0.084	&	\cite{2018AJ....156..199S}	\\
CN Hyi	&	0.45611	&	0.184	&	SP	&	1.370	&	1.6	&	0.77	&	0.420	&	0.120	&	0.074	&	\cite{2009NewA...14..461O}	\\
V2388 Oph (n)	&	0.80230	&	0.186	&	SP	&	1.800	&	2.6	&	1.3	&	0.650	&	0.123	&	0.081	&	\cite{2000Yakut}	\\
GV Leo	&	0.26673	&	0.188	&	T	&		&		&		&	0.178	&	0.113	&		&	\cite{2013RAA....13.1330K}	\\
TYC 3836-0854-1 (n)	&	0.41557	&	0.19	&	T	&	1.200	&	1.46	&	0.75	&	0.794	&	0.125	&	0.134	&	\cite{2017PASP..129l4204L}	\\
BO Ari*	&	0.31819	&	0.19	&	SP	&	0.995	&	1.09	&	0.515	&	0.494	&	0.118	&	0.176	&	\cite{2015NewA...39....9G}	\\
	&	0.31819	&	0.207	&	SP	&	1.095	&	1.19	&	0.636	&	0.757	&	0.113	&	0.147	&	\cite{2021NewA...8601571P}	\\
V619 Peg	&	0.38872	&	0.19	&	T	&	2.020	&	1.642	&	0.811	&	0.521	&	0.118	&	0.079	&	\cite{2019AJ....158..186K}	\\
HV UMa	&	0.35539	&	0.19	&	SP	&	2.800	&	2.62	&	1.18	&	0.770	&	0.107	&	0.077	&	\cite{2000Csak}	\\
V1853 Ori	&	0.38300	&	0.19	&	T	&	1.200	&	1.36	&	0.66	&	0.333	&	0.114	&	0.121	&	 \cite{2019RAA....19...56H} 	\\
IK Per	&	0.67603	&	0.191	&	T	&	1.990	&	2.4	&	1.15	&	0.521	&	0.118	&	0.079	&	\cite{2005AJ....129.2806Z}	\\
Y Sex	&	0.48774	&	0.195	&	SP	&	1.471	&	1.568	&	0.795	&	0.596	&	0.117	&	0.075	&	\cite{2011MNRAS.412.1787D}	\\
VW Vul	&	0.38451	&	0.195	&	T	&		&		&		&	0.403	&	0.112	&		&	\cite{2008IBVS.5824....1C}	\\
EX Leo	&	0.40860	&	0.2	&	SP	&	1.573	&	1.56	&	0.734	&	0.366	&	0.112	&	0.071	&	\cite{2010MNRAS.408..464Z}	\\
V402 Aur	&	0.60350	&	0.2	&	SP	&	1.638	&	1.997	&	0.915	&	0.031	&	0.103	&	0.066	&	\cite{2004AcA....54..299Z}	\\
KIC 5809868	&	0.43939	&	0.201	&	T	&		&		&		&	0.470	&	0.112	&		&	\cite{2017MNRAS.466.2488Z}	\\
NSVS 9045055	&	0.35459	&	0.202	&	T	&	1.990	&	1.51	&	0.751	&	0.302	&	0.108	&	0.073	&	\cite{2019AJ....158..186K}	\\
EL Aqr	&	0.48141	&	0.203	&	SP	&	1.570	&	1.73	&	0.87	&	0.440	&	0.109	&	0.071	&	\cite{2004Wadhwa}	\\
NSVS 7328383	&	0.27208	&	0.205	&	T	&	1.013	&	1	&	0.49	&	0.171	&	0.104	&	0.151	&	\cite{2019NewA...68...20K}	\\
DN Aur	&	0.61689	&	0.205	&	T	&		&		&		&	0.468	&	0.108	&		&	\cite{1996Goderya}	\\
UY UMa	&	0.37602	&	0.206	&	T	&	1.151	&	1.342	&	0.696	&	0.606	&	0.112	&	0.131	&	\cite{2019JASS...36..265K}	\\
HI Pup	&	0.43257	&	0.206	&	SP	&	1.210	&	1.44	&	0.67	&	0.203	&	0.103	&	0.108	&	\cite{2014NewA...31...56U}	\\
TYC 3836-0854-1*	&	0.41556	&	0.206	&	T	&	1.383	&	1.5	&	0.763	&	0.592	&	0.110	&	0.085	&	\cite{2014NewA...31....1A}	\\
ASAS J212236+0657.3	&	0.29395	&	0.207	&	T	&	0.807	&	0.793	&	0.665	&	0.132	&	0.102	&	0.200	&	\cite{2020BlgAJ..32...71K}	\\
TZ Boo*	&	0.29716	&	0.207	&	SP	&	0.990	&	1.08	&	0.56	&	0.531	&	0.108	&	0.165	&	\cite{2011AJ....142...99C}	\\
TYC 2402-0643-1 (n)	&	0.39943	&	0.208	&	T	&	0.860	&	1.22	&	0.67	&	0.220	&	0.103	&	0.188	&	\cite{2020JAVSO..48...62S}	\\
NSVS 6859986	&	0.38357	&	0.208	&	T	&	1.870	&	1.63	&	0.84	&	0.860	&	0.115	&	0.078	&	\cite{2019AJ....158..186K}	\\
GSC-03950-00707 (n)	&	0.41200	&	0.21	&	T	&	2.850	&	1.85	&	0.91	&	0.120	&	0.100	&	0.076	&	\cite{2019AJ....158..186K}	\\
GM Dra	&	0.33875	&	0.21	&	SP	&	1.213	&	1.252	&	0.606	&	0.230	&	0.105	&	0.110	&	\cite{2005AcA....55..123G}	\\
CV Cyg	&	0.98343	&	0.21	&	T	&	1.600	&		&		&	0.490	&	0.108	&	0.072	&	\cite{1996MNRAS.280..489V}	\\
RR Cen	&	0.60569	&	0.21	&	SP	&	1.820	&	2.1	&	1.05	&	0.351	&	0.104	&	0.070	&	\cite{2005PASJ...57..983Y}	\\
MM Com (n)	&	0.30199	&	0.215	&	T	&	0.790	&	0.99	&	0.35	&	0.240	&	0.095	&	0.184	&	\cite{2019AJ....158..186K}	\\
V409 Hya	&	0.47227	&	0.216	&	T	&	1.500	&	1.69	&	0.9	&	0.606	&	0.106	&	0.071	&	\cite{2014NewA...30..105N}	\\
V816 Cep	&	0.31141	&	0.219	&	T	&	2.830	&	1.576	&	0.836	&	0.511	&	0.103	&	0.080	&	\cite{2019AJ....158..186K}	\\
MW Pav*	&	0.79499	&	0.22	&	SP	&	1.514	&	2.412	&	1.277	&	0.693	&	0.105	&	0.070	&	\cite{2015PASP..127..742A}	\\
V604 Car	&	0.47229	&	0.22	&	T	&		&		&		&	0.380	&	0.101	&		&	\cite{2006WadhwaSurjit}	\\
PY Boo	&	0.27805	&	0.222	&	T	&	0.790	&	0.93	&	0.47	&	0.180	&	0.096	&	0.195	&	\cite{2019AJ....158..186K}	\\
FN Cam	&	0.67713	&	0.222	&	SP	&	2.400	&	2.61	&	1.44	&	0.880	&	0.110	&	0.084	&	\cite{2002IBVS.5258....1P}	\\
NSVS 7051868	&	0.51760	&	0.223	&	T	&	1.610	&	1.8	&	0.94	&	0.353	&	0.097	&	0.066	&	\cite{2017NewA...50...73B}	\\
AQ Psc	&	0.47561	&	0.226	&	SP	&	1.661	&	1.708	&	0.891	&	0.350	&	0.095	&	0.065	&	\cite{2011MNRAS.412.1787D}	\\
V921 Her	&	0.87738	&	0.226	&	SP	&	1.784	&	2.56	&	1.29	&	0.190	&	0.090	&	0.061	&	\cite{2016AdAst2016E...7Z}	\\
V921 Her	&	0.87738	&	0.244	&	SP	&	2.068	&	2.752	&	1.407	&	0.230	&	0.088	&	0.061	&	\cite{2006AcA....56..127G}	\\
KIC 9832227	&	0.45796	&	0.228	&	SP	&	1.395	&	1.581	&	0.83	&	0.430	&	0.096	&	0.074	&	\cite{2017ApJ...840....1M}	\\
1SWASP J000437.82+033301.2 (n)	&	0.26150	&	0.23	&	T	&	0.820	&	0.89	&	0.46	&	0.116	&	0.091	&	0.177	&	\cite{2021PASJ...73..132L}	\\
V2357 Oph	&	0.41557	&	0.231	&	SP	&	1.160	&	1.386	&	0.705	&	0.322	&	0.093	&	0.109	&	\cite{2011MNRAS.412.1787D}	\\

\hline
\end{tabular}
\end{table*}

\begin{table*}
\centering
\contcaption{Parameters of previous LMRs.}\label{tab:continued}
 \scriptsize

\begin{tabular}{l*{15}{c}r}
 \hline
 \text{Name} & \text{Period} & \text{$q$} & \text{type} & \text{$M_1$}  & \text{$R_1$} & \text{$R_2$} &\text{$f$}&\text {$\frac{J_{s}}{J_{o}}$} & $\left(\frac{J_{s}}{J_{o}}\right)_k$ &  \text{Reference} \\ 
 &(days) & & &       ($M_{\sun}$) & ($R_{\sun})$ & ($R_{\sun})$ & &  \\
 \hline

ASAS J035020-8017.4	&	0.62240	&	0.232	&	T	&	0.990	&	1.76	&	0.87	&	0.344	&	0.094	&	0.144	&	\cite{2016NewA...46...94S}	\\
YY CrB	&	0.37656	&	0.232	&	SP	&	1.393	&	1.385	&	0.692	&	0.228	&	0.091	&	0.070	&	\cite{2005AcA....55..123G} 	\\
YY CrB	&	0.37656	&	0.243	&	SP	&	1.430	&	1.43	&	0.81	&	0.634	&	0.104	&	0.098	&	\cite{2004BaltA..13..151V}	\\
TYC 5532-1333-1	&	0.47449	&	0.232	&	T	&	1.510	&	1.67	&	0.9	&	0.343	&	0.095	&	0.065	&	\cite{2020MNRAS.493.1565D}	\\
GSC 0763-0572	&	0.42640	&	0.233	&	T	&	1.230	&	1.434	&	0.757	&	0.310	&	0.092	&	0.097	&	\cite{2012PASJ...64...83W}	\\
DU Boo	&	1.05589	&	0.234	&	SP	&	2.080	&	3.19	&	1.74	&	0.502	&	0.082	&	0.060	&	\cite{2013AJ....145...80D}	\\
V1094 Cas	&	0.51400	&	0.235	&	T	&		&		&		&	0.300	&	0.093	&		&	\cite{2019JAVSO..47...40W}	\\
V789 Her	&	0.32004	&	0.236	&	T	&	1.130	&	1.15	&	0.62	&	0.238	&	0.091	&	0.114	&	\cite{2018PASP..130g4201L}	\\
V2741 Cyg	&	0.31600	&	0.236	&	T	&		&		&		&	0.480	&	0.096	&		&	\cite{2019JAVSO..47...40W}	\\
NSVS 4803568 (n)	&	0.28657	&	0.238	&	T	&	0.790	&	0.93	&	0.43	&	0.090	&	0.086	&	0.172	&	\cite{2019AJ....158..186K}	\\
FO Hya	&	0.46956	&	0.238	&	T	&	1.310	&	1.62	&	0.91	&	0.689	&	0.098	&	0.091	&	\cite{2013NewA...20...52P}	\\
GSC 3208-1986	&	0.40466	&	0.24	&	T	&		&		&		&	0.390	&	0.092	&		&	\cite{2015AJ....149...90S}	\\
KIC 10267044	&	0.43004	&	0.24	&	T	&		&		&		&	0.550	&	0.097	&		&	\cite{2017MNRAS.466.2488Z}	\\
V507 Lyr	&	0.36691	&	0.24	&	T	&		&		&		&	0.200	&	0.089	&		&	\cite{2004IBVS.5547....1W}	\\
HR Boo*	&	0.31597	&	0.241	&	T	&	1.210	&	1.17	&	0.62	&	0.180	&	0.090	&	0.098	&	\cite{2019AJ....158..186K}	\\
V404 Peg	&	0.41918	&	0.243	&	SP	&	1.175	&	1.346	&	0.71	&	0.321	&	0.089	&	0.103	&	\cite{2011AN....332..690G}	\\
GSC 0804-0118	&	0.32369	&	0.243	&	T	&		&		&		&	0.200	&	0.087	&		&	\cite{2005YangQian}	\\
KIC 3221207	&	0.47383	&	0.244	&	T	&	1.300	&	1.67	&	0.87	&	0.790	&	0.096	&	0.090	&	\cite{doi:10.1063/1.4976436}	\\
T-Dra0-00959 (n)	&	0.32933	&	0.245	&	T	&	0.890	&	1.06	&	0.5	&	0.240	&	0.084	&	0.147	&	\cite{2019AJ....158..186K}	\\
UW CVn	&	0.29247	&	0.245	&	T	&		&		&		&	0.202	&	0.086	&		&	\cite{1995AcA....45..753K}	\\
TYC 01963-0488-1	&	0.42704	&	0.248	&	T	&	1.370	&	1.46	&	0.78	&	0.347	&	0.088	&	0.073	&	\cite{2016JAVSO..44...87A}	\\
HI Dra	&	0.59742	&	0.25	&	SP	&	0.720	&	1.98	&	1.08	&	0.240	&	0.085	&	0.061	&	\cite{2015AJ....149..168P}	\\
KN Per	&	0.86650	&	0.25	&	T	&		&		&		&	0.545	&	0.090	&		&	\cite{2015AJ....150...69Y}	\\
\hline
\end{tabular}
\end{table*}

\begin{figure*}
\caption{The observed LCs (black points) and the best fitted models (red line) for the 30 new CSS LMR systems.}
\label{fig:LCsWsynt}
\minipage{0.32\textwidth}
\includegraphics[width=\linewidth]{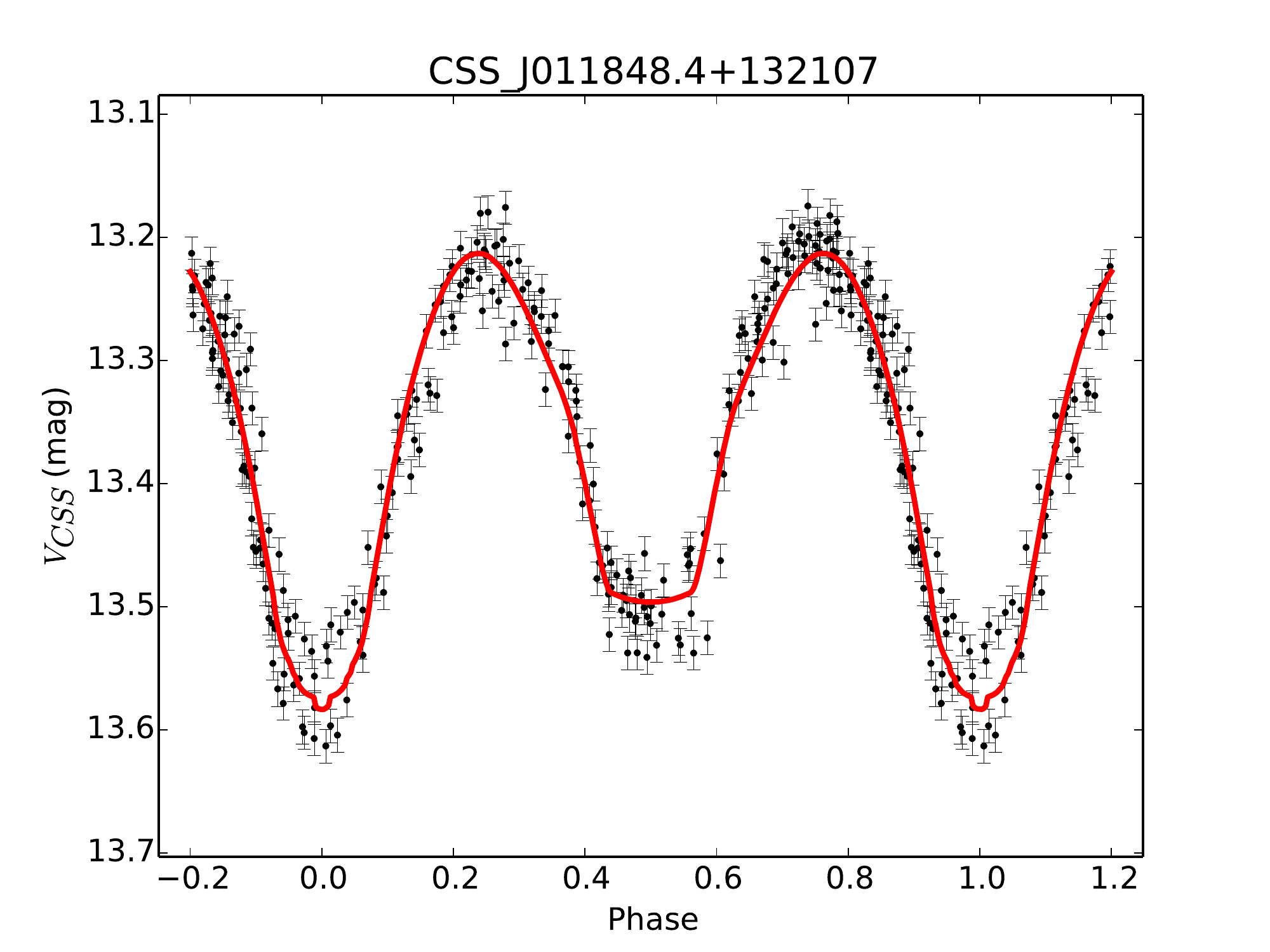} %
\endminipage\hfill
\minipage{0.32\textwidth}
\includegraphics[width=\linewidth]{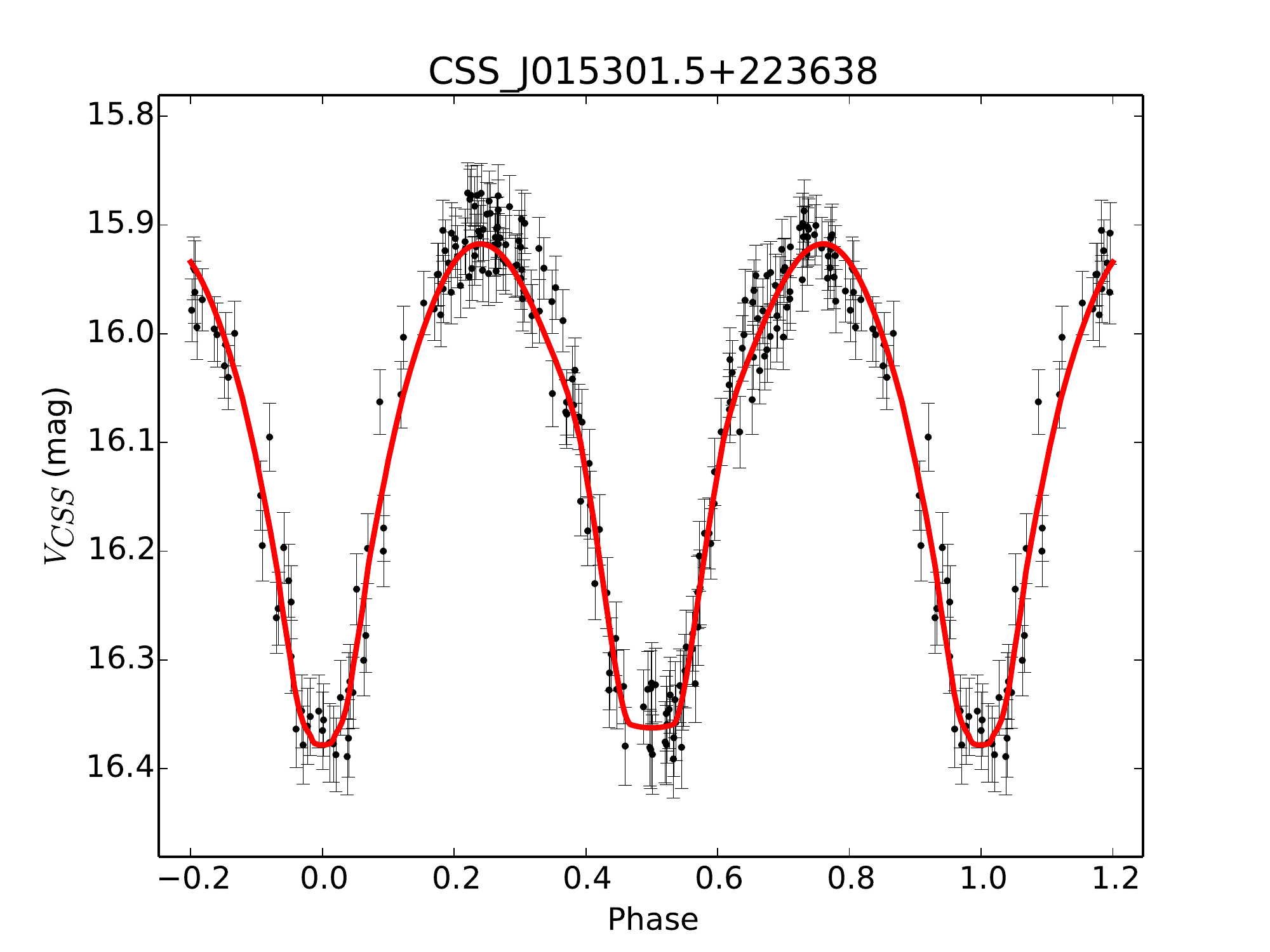} %
\endminipage\hfill
\minipage{0.32\textwidth}
\includegraphics[width=\linewidth]{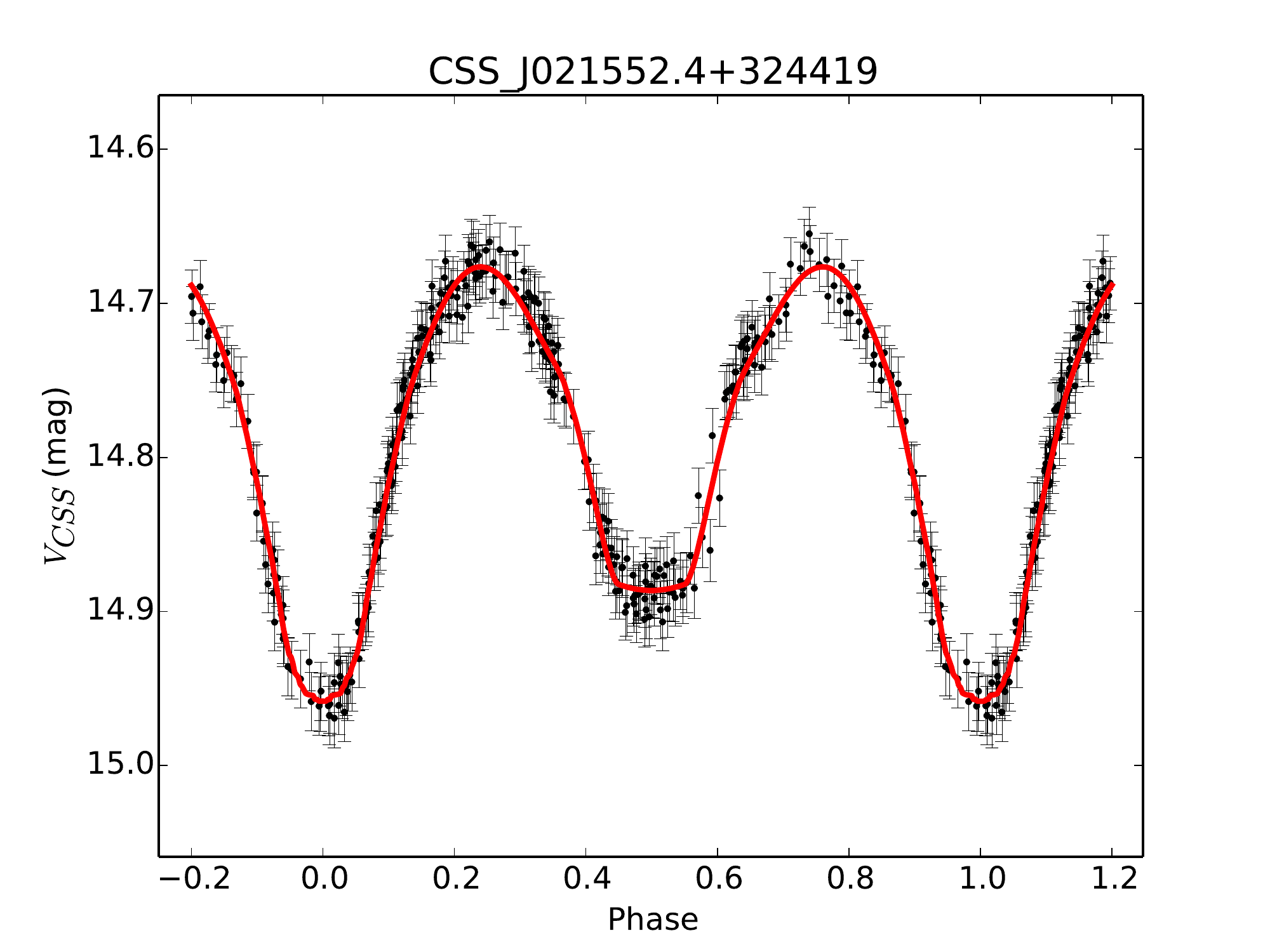} %
\endminipage
 
\minipage{0.32\textwidth}
\includegraphics[width=\linewidth]{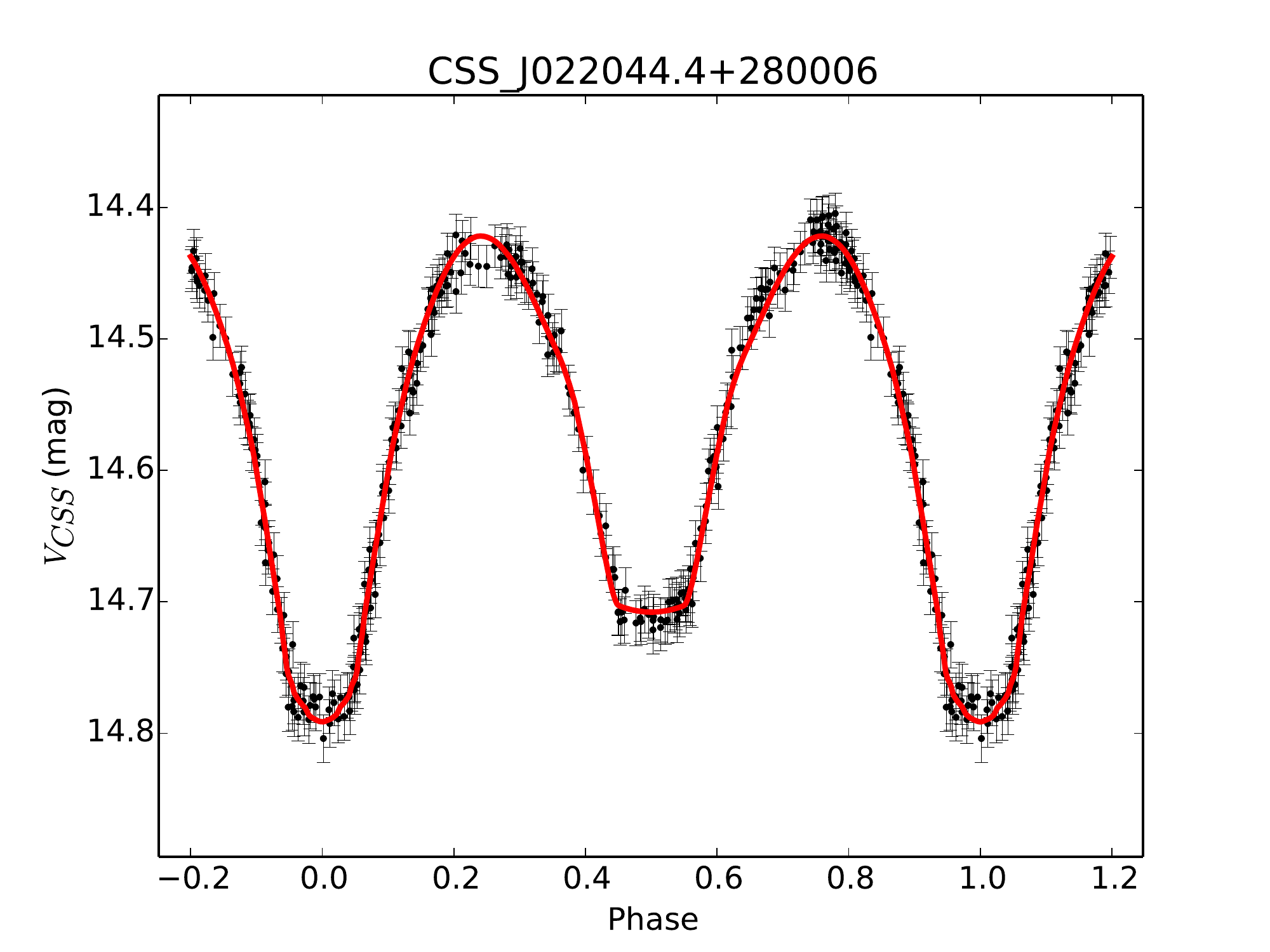} %
\endminipage\hfill
\minipage{0.32\textwidth}
\includegraphics[width=\linewidth]{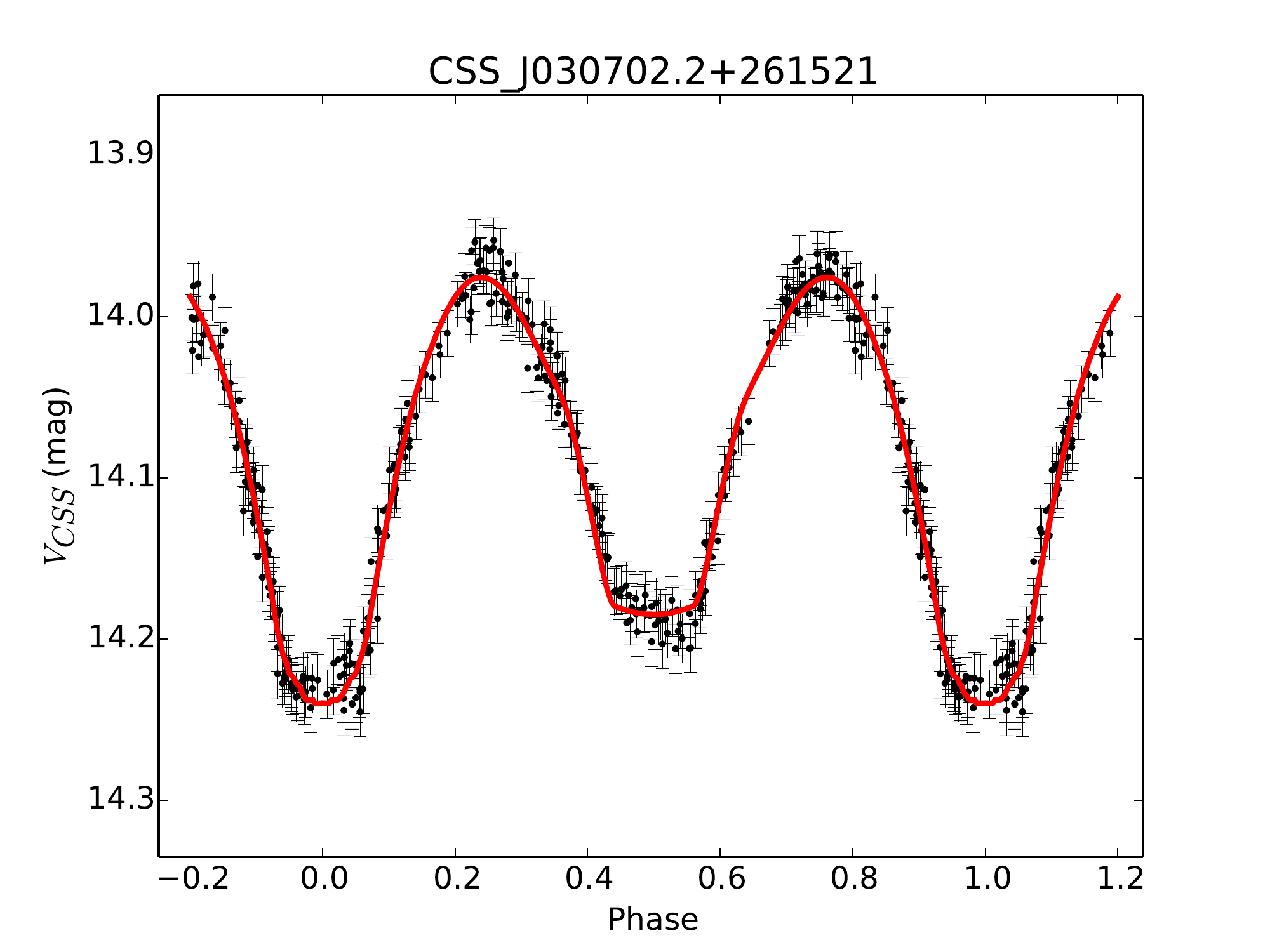} %
\endminipage\hfill
\minipage{0.32\textwidth}
\includegraphics[width=\linewidth]{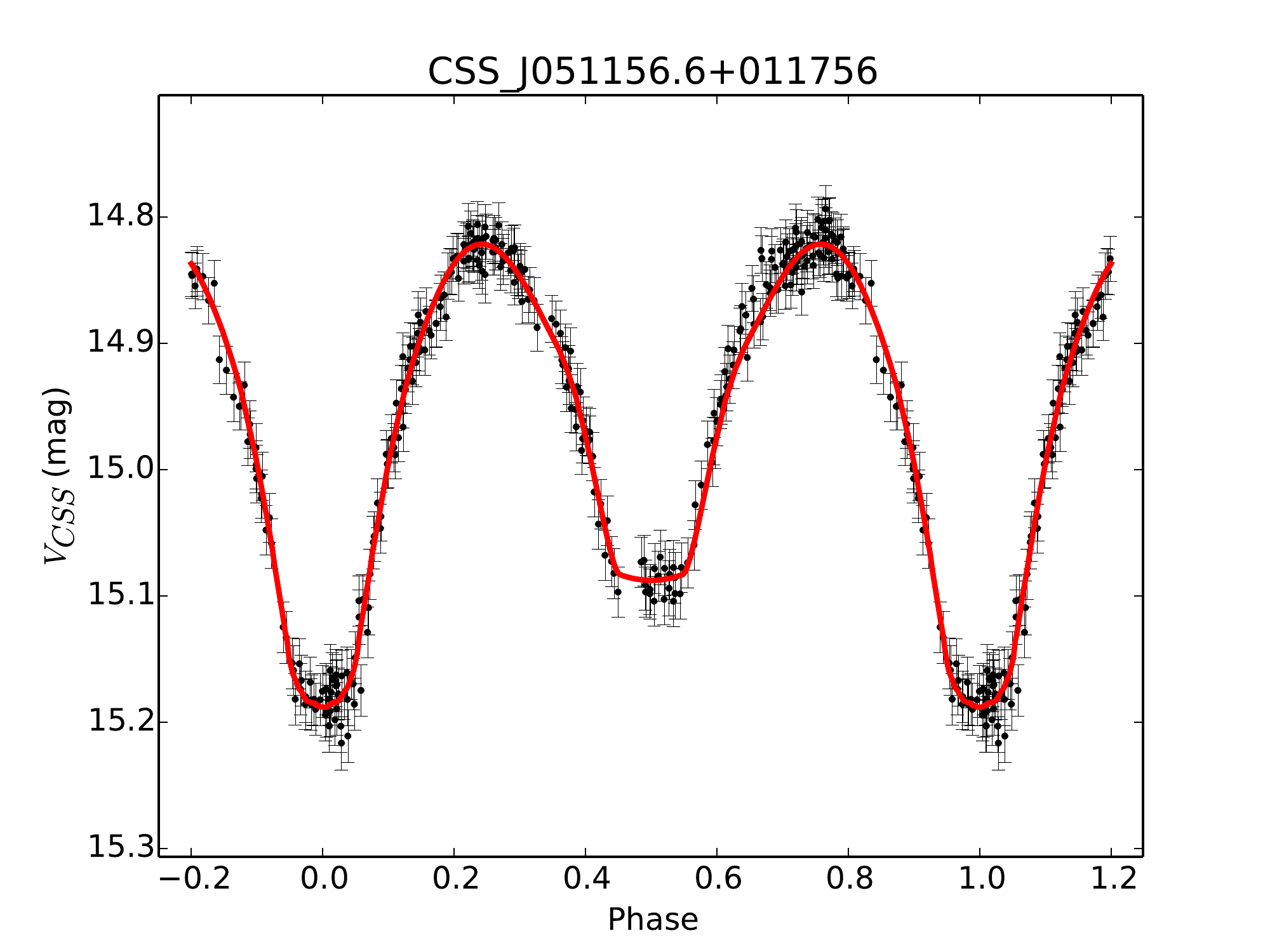} %
\endminipage
 
\minipage{0.32\textwidth}
\includegraphics[width=\linewidth]{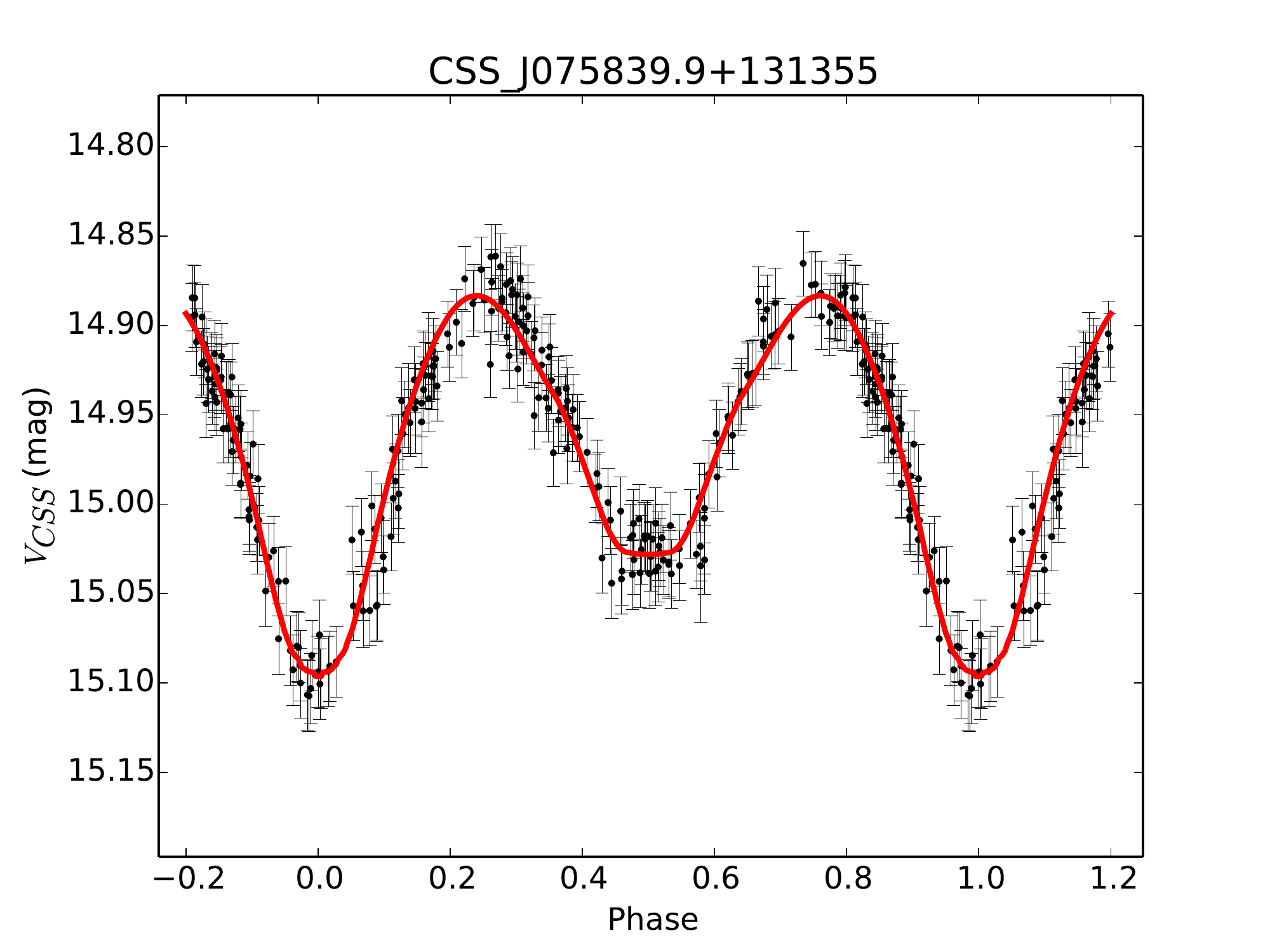} %
\endminipage\hfill
\minipage{0.32\textwidth}
\includegraphics[width=\linewidth]{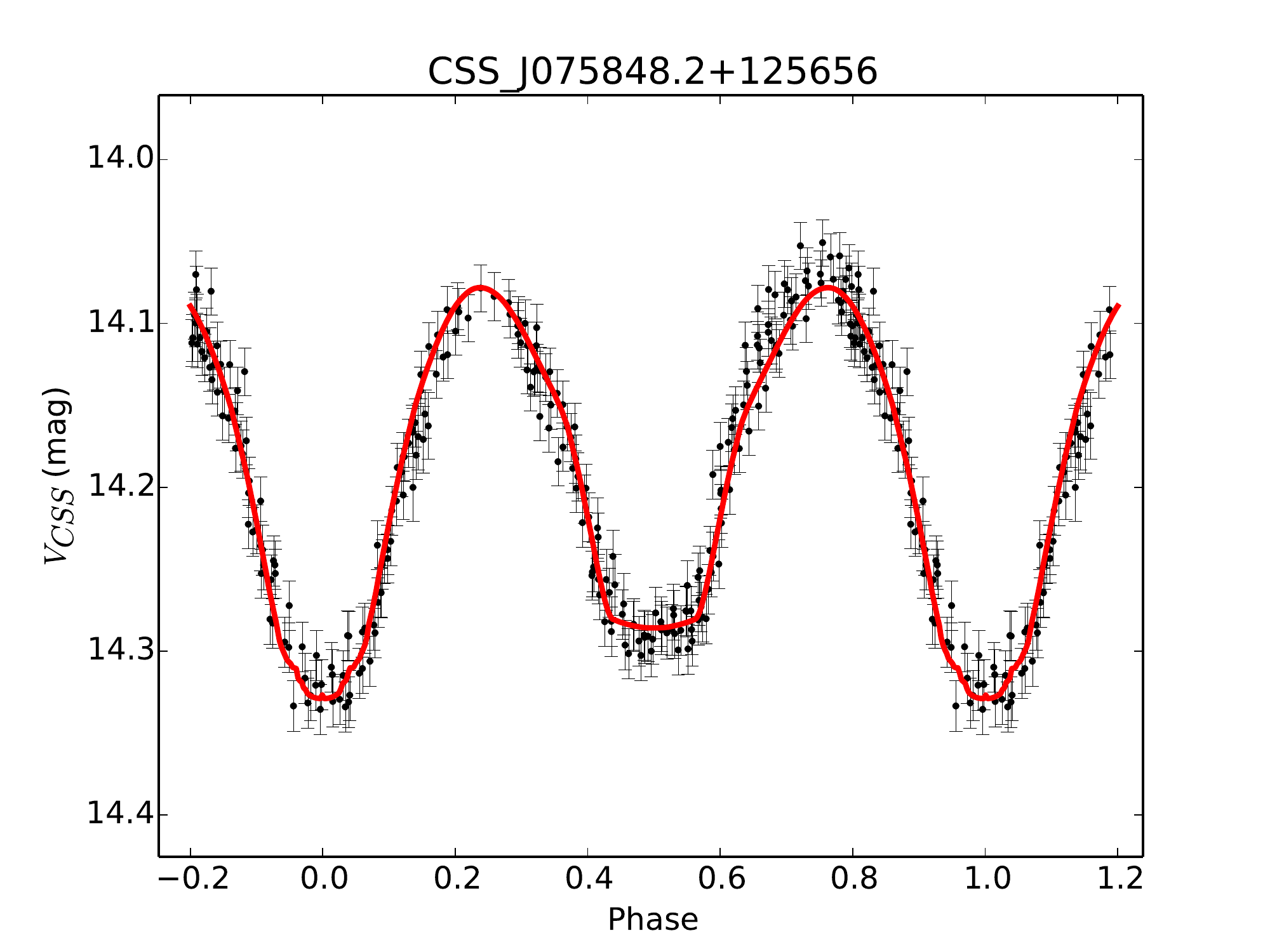} %
\endminipage\hfill
\minipage{0.32\textwidth}
\includegraphics[width=\linewidth]{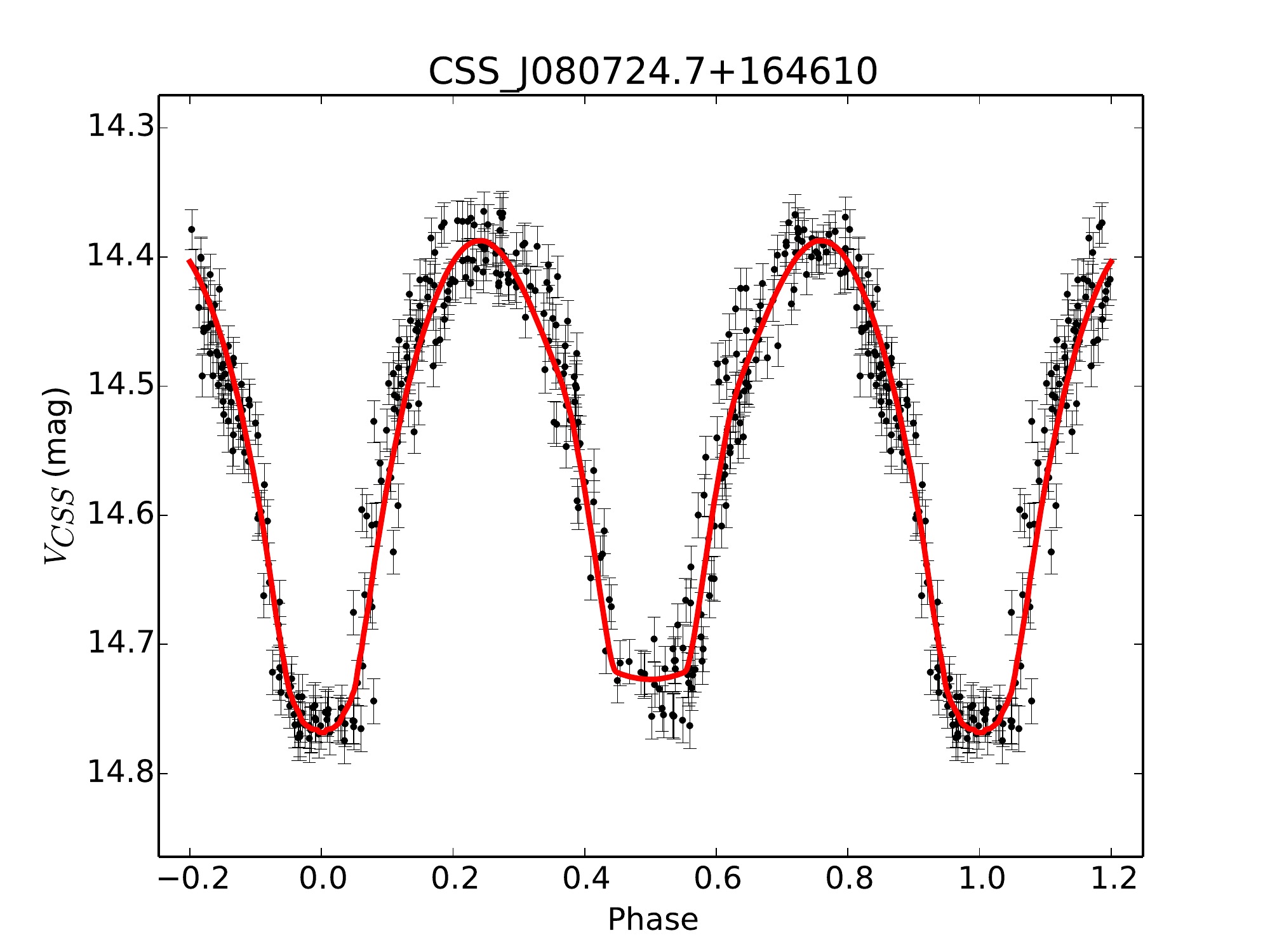} %
\endminipage
 
\minipage{0.32\textwidth}
\includegraphics[width=\linewidth]{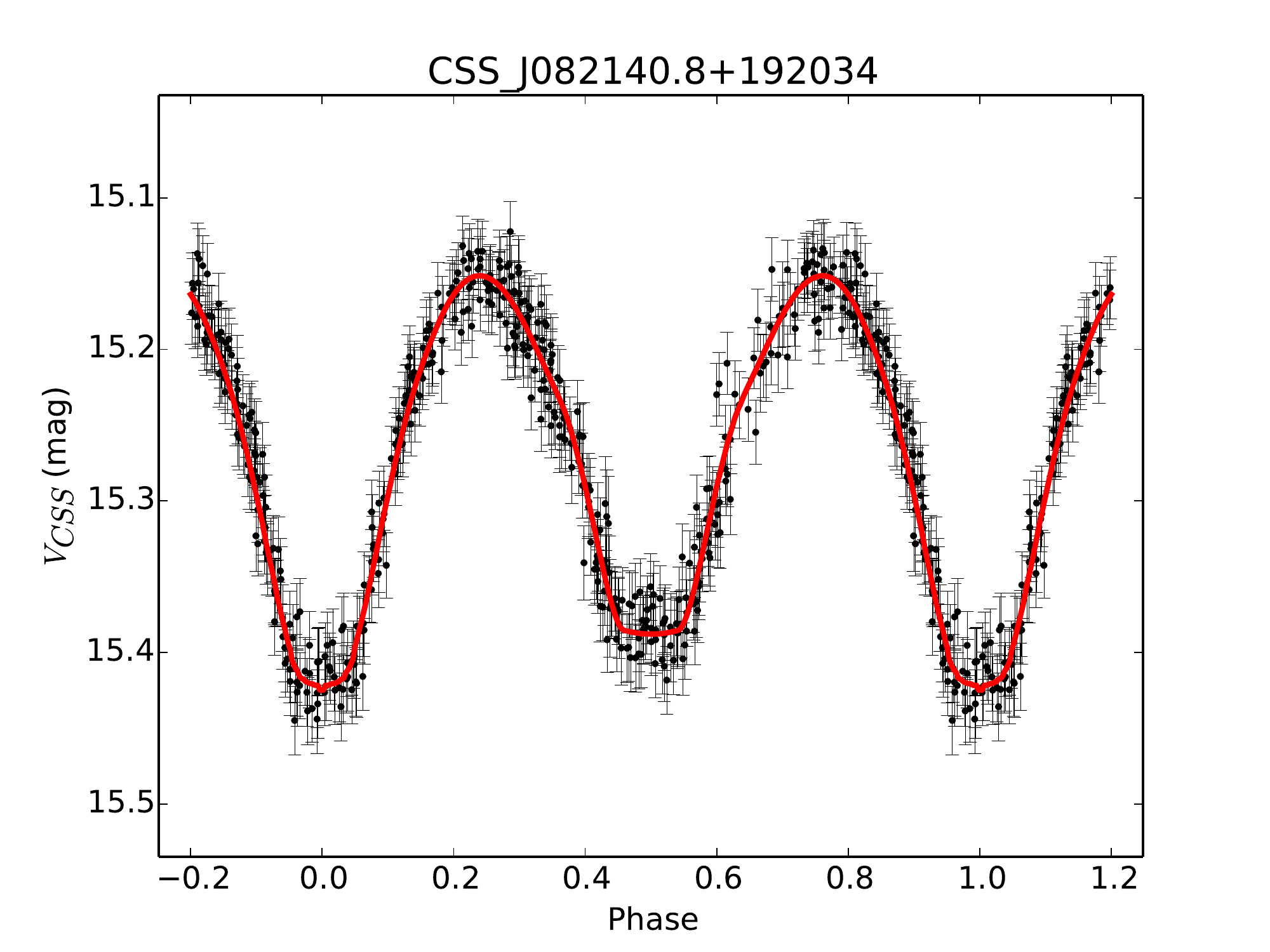} %
\endminipage\hfill
\minipage{0.32\textwidth}
\includegraphics[width=\linewidth]{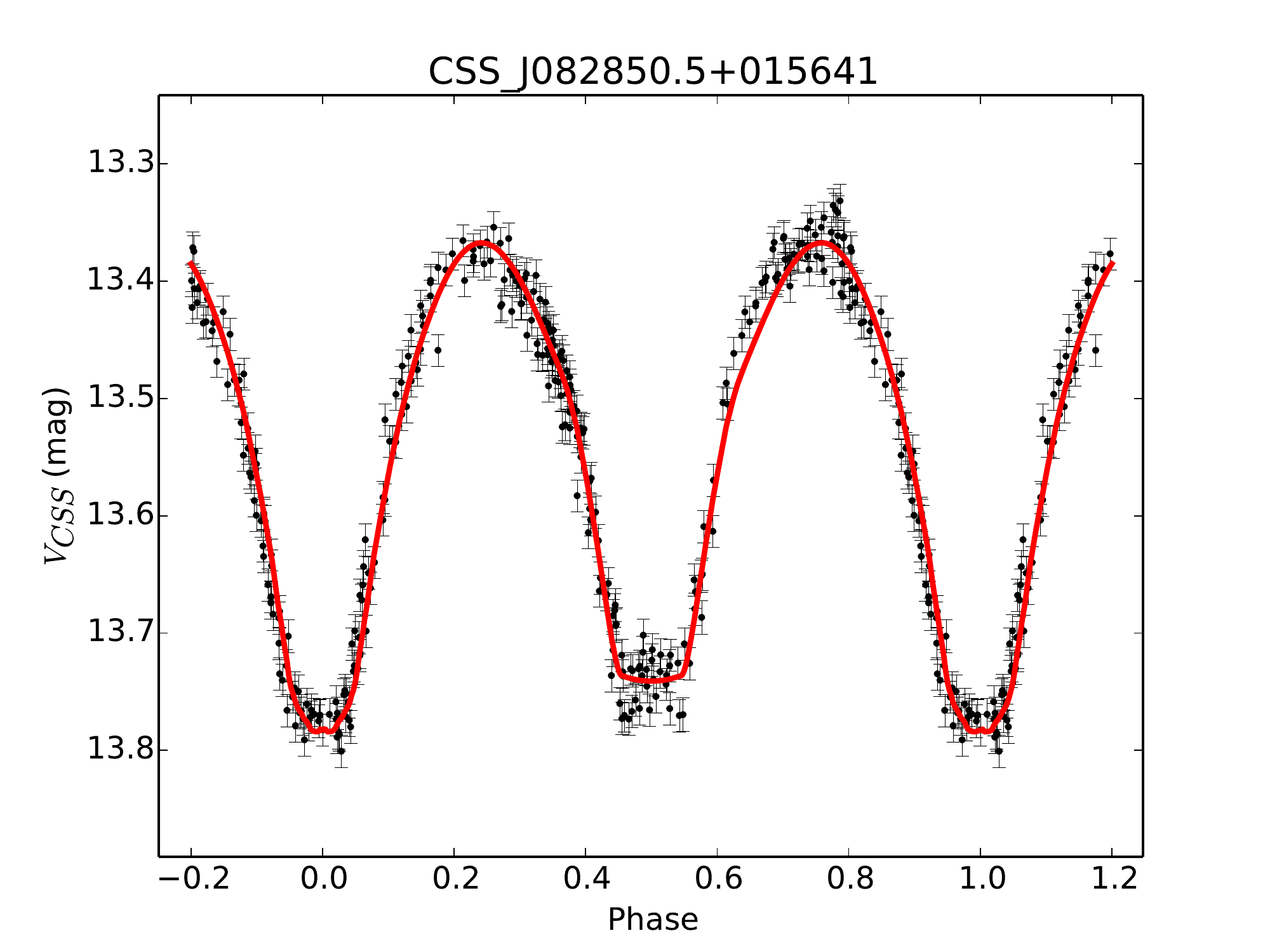} %
\endminipage\hfill
\minipage{0.32\textwidth}
\includegraphics[width=\linewidth]{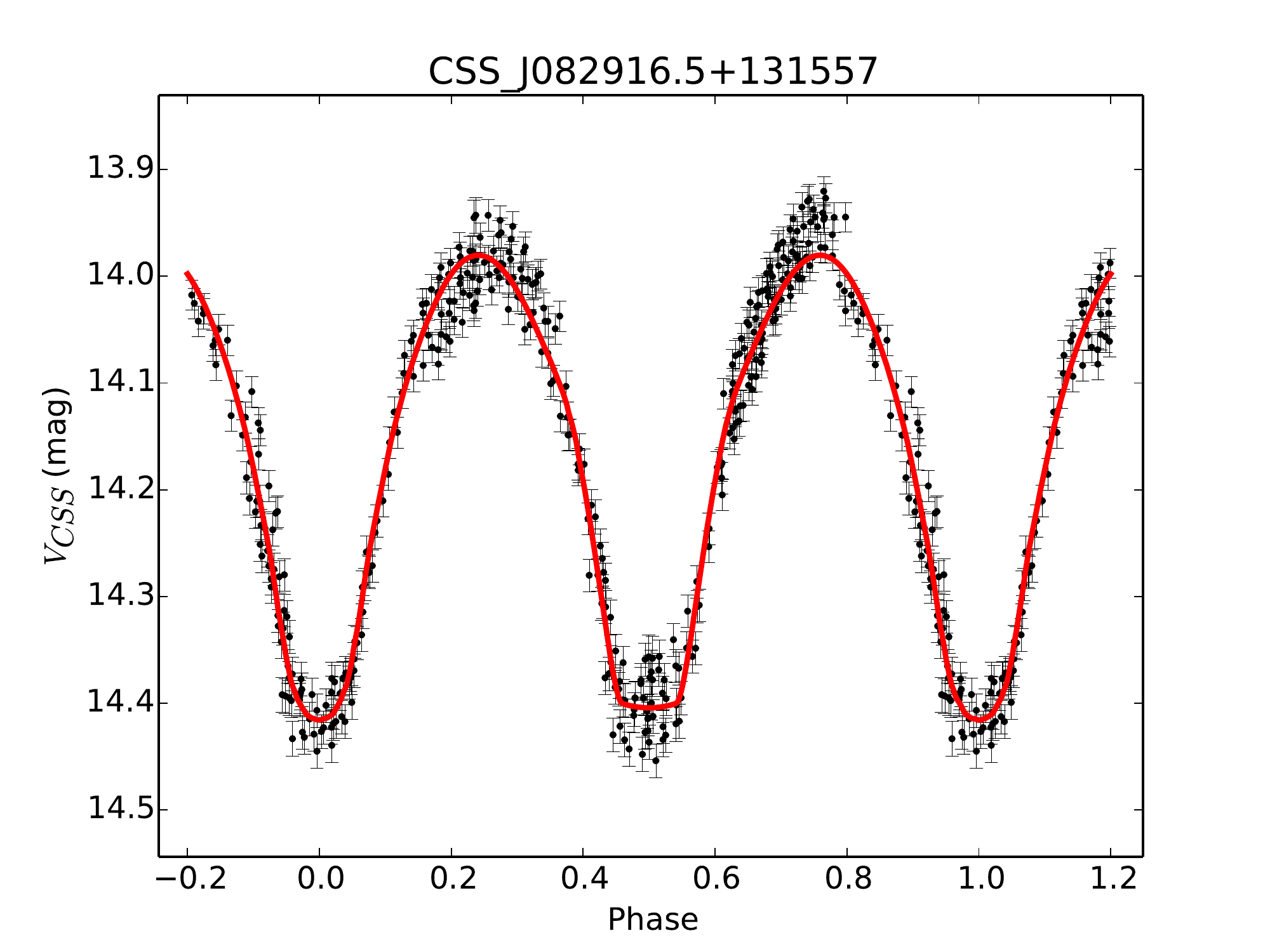} %
\endminipage
 
\minipage{0.32\textwidth}
\includegraphics[width=\linewidth]{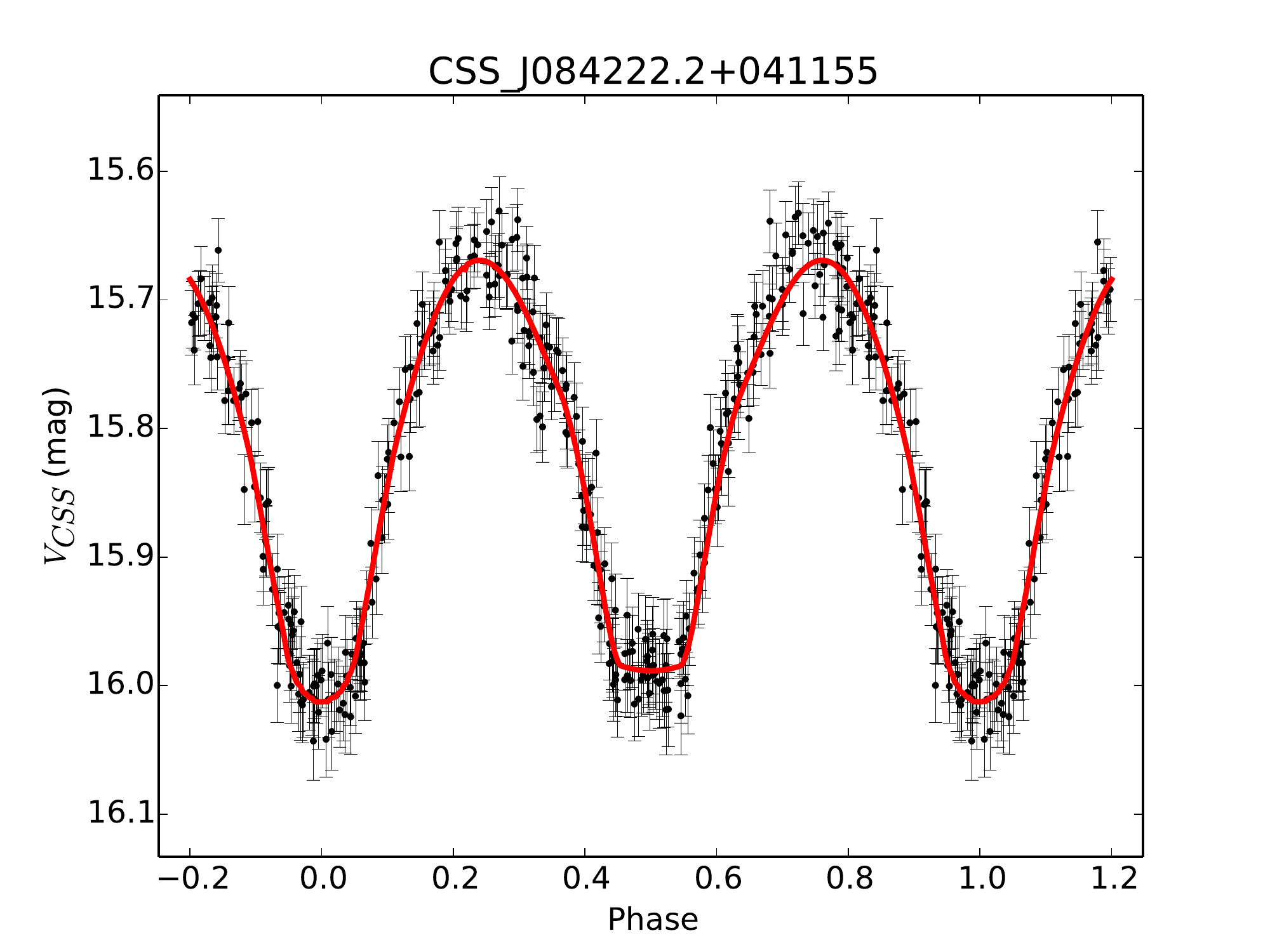} %
\endminipage\hfill
\minipage{0.32\textwidth}
\includegraphics[width=\linewidth]{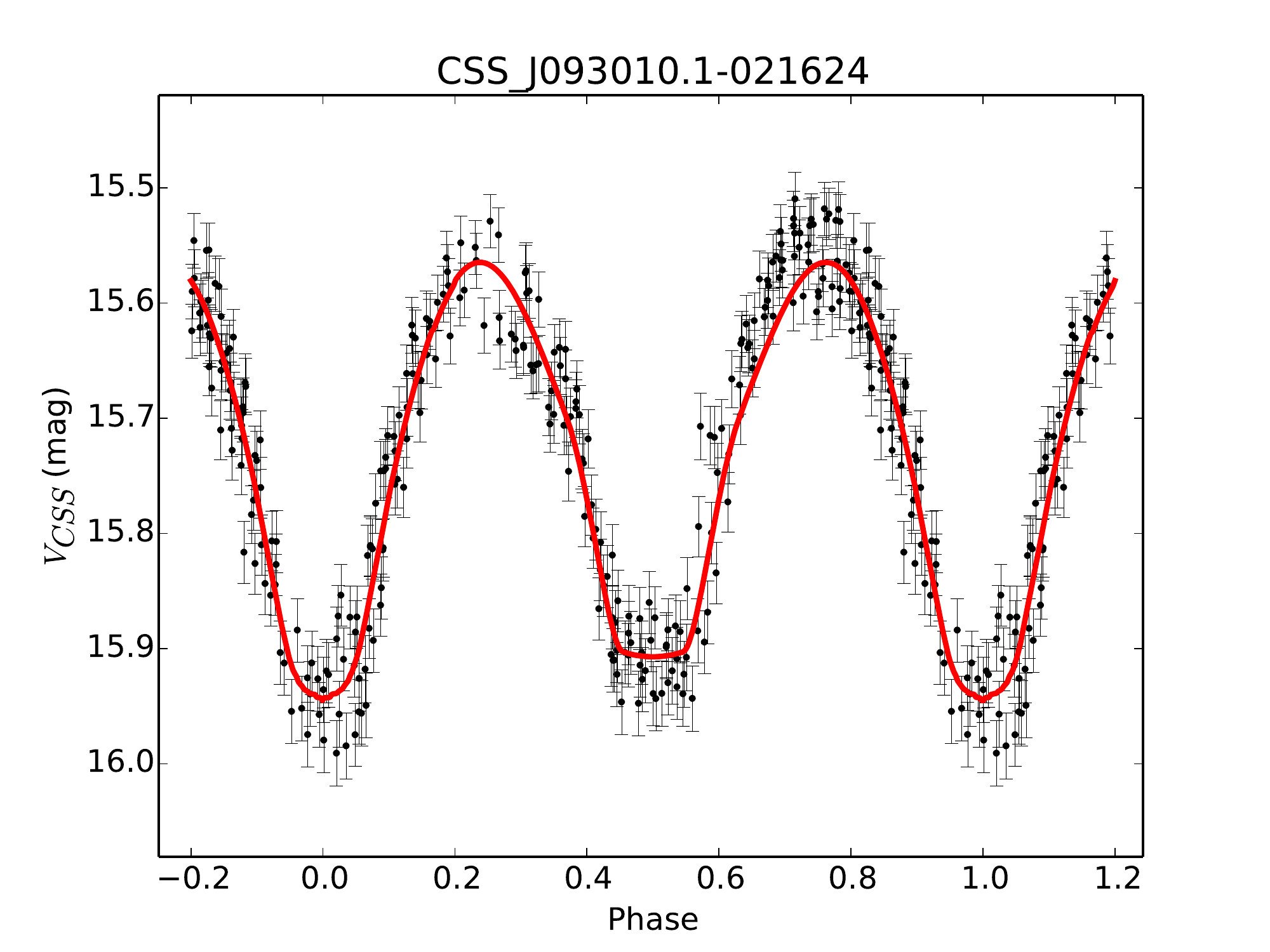} %
\endminipage\hfill
\minipage{0.32\textwidth}
\includegraphics[width=\linewidth]{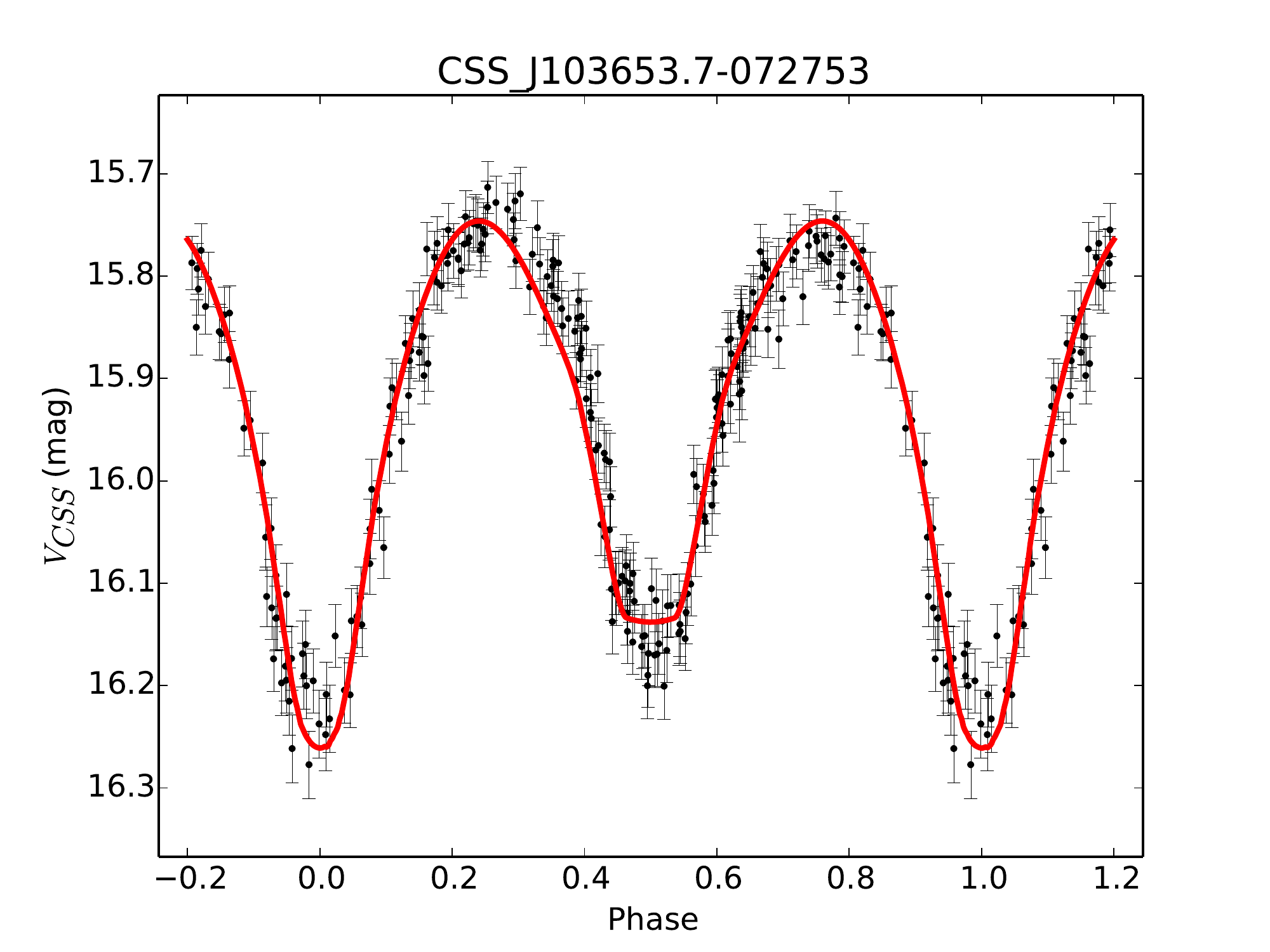} %
\endminipage
 
\end{figure*}
\begin{figure*}
\minipage{0.32\textwidth}
\includegraphics[width=\linewidth]{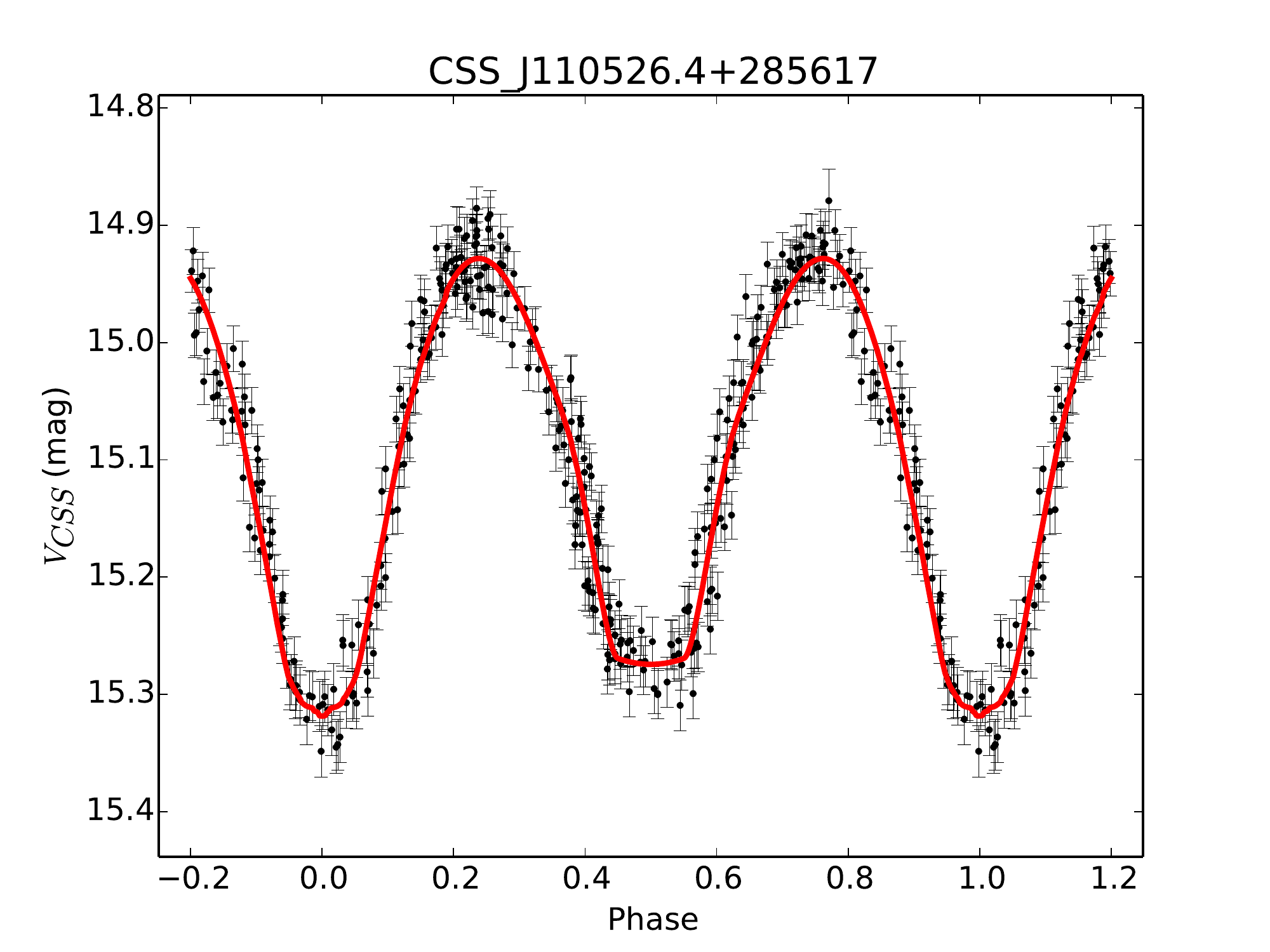} %
\endminipage\hfill
\minipage{0.32\textwidth}
\includegraphics[width=\linewidth]{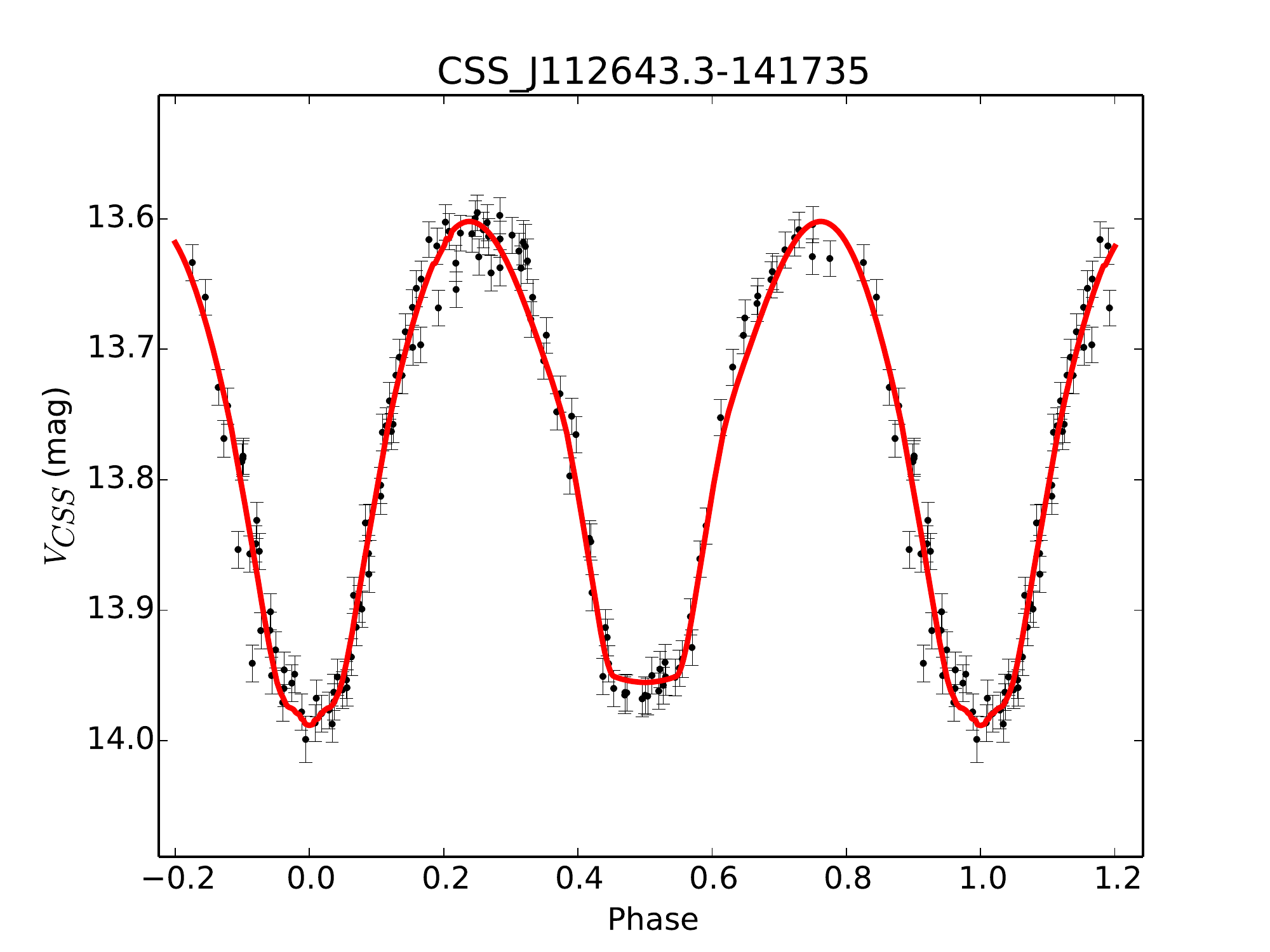} %
\endminipage\hfill
\minipage{0.32\textwidth}
\includegraphics[width=\linewidth]{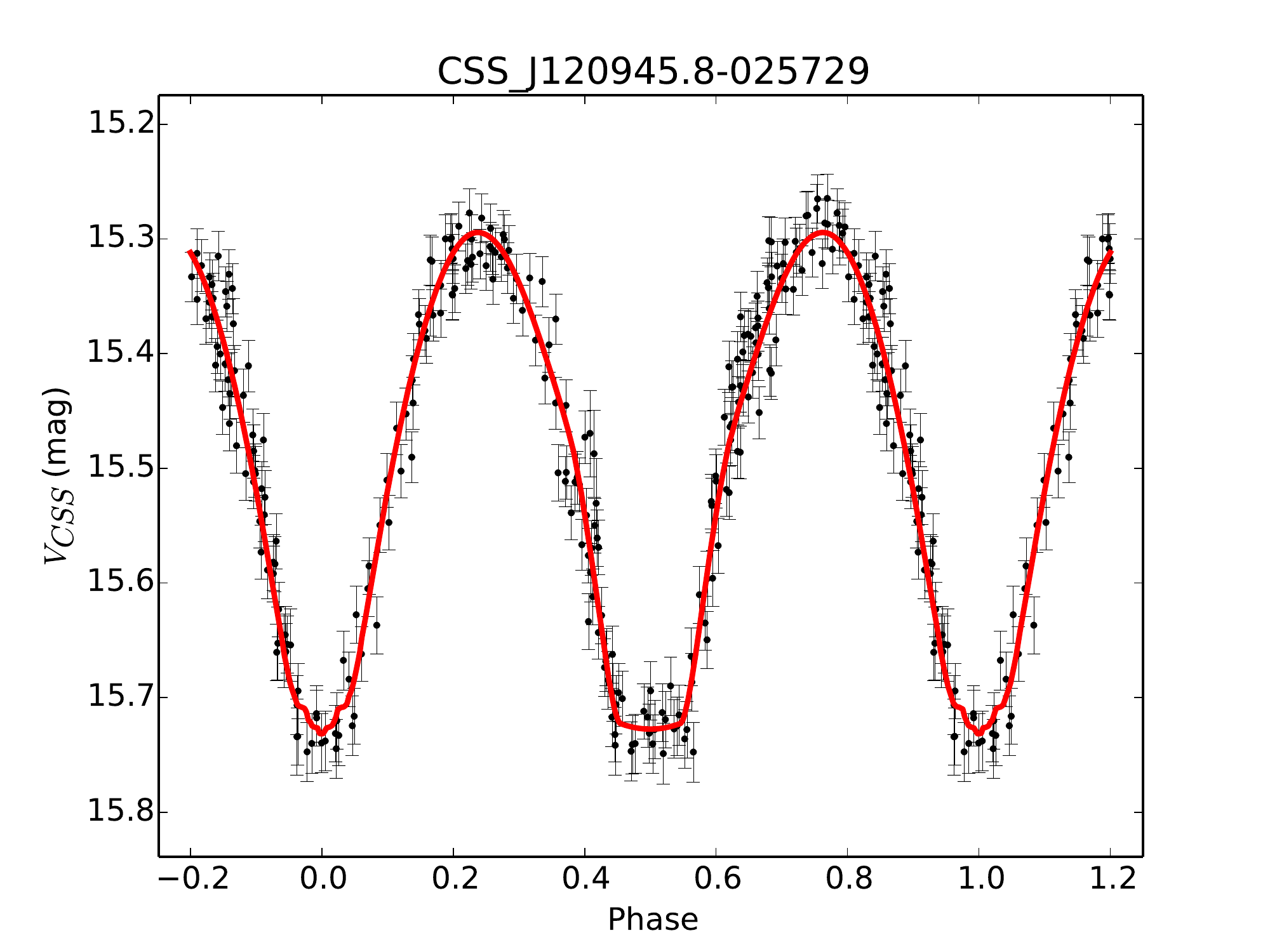} %
\endminipage
 
\minipage{0.32\textwidth}
\includegraphics[width=\linewidth]{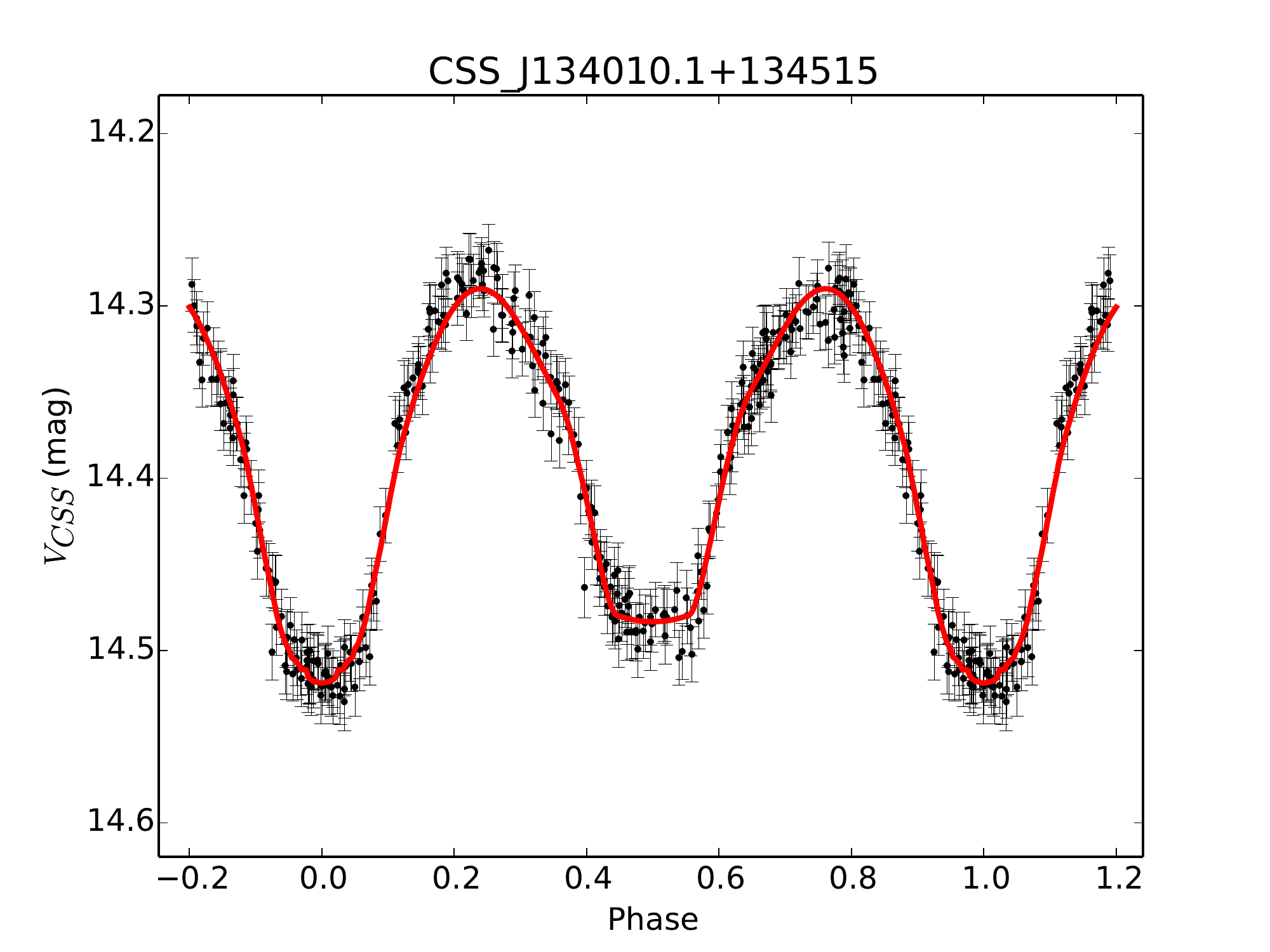} %
\endminipage\hfill
\minipage{0.32\textwidth}
\includegraphics[width=\linewidth]{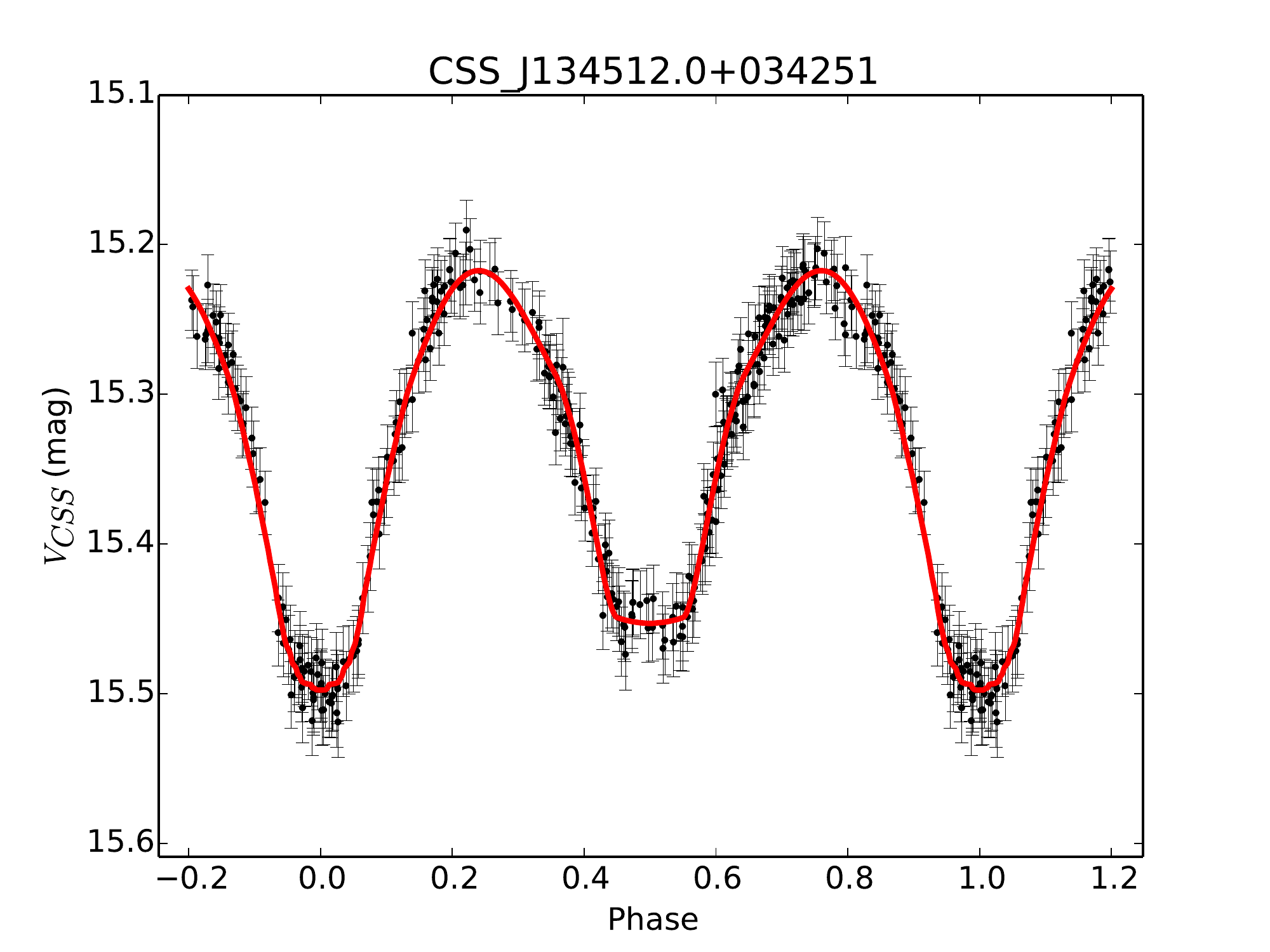} %
\endminipage\hfill
\minipage{0.32\textwidth}
\includegraphics[width=\linewidth]{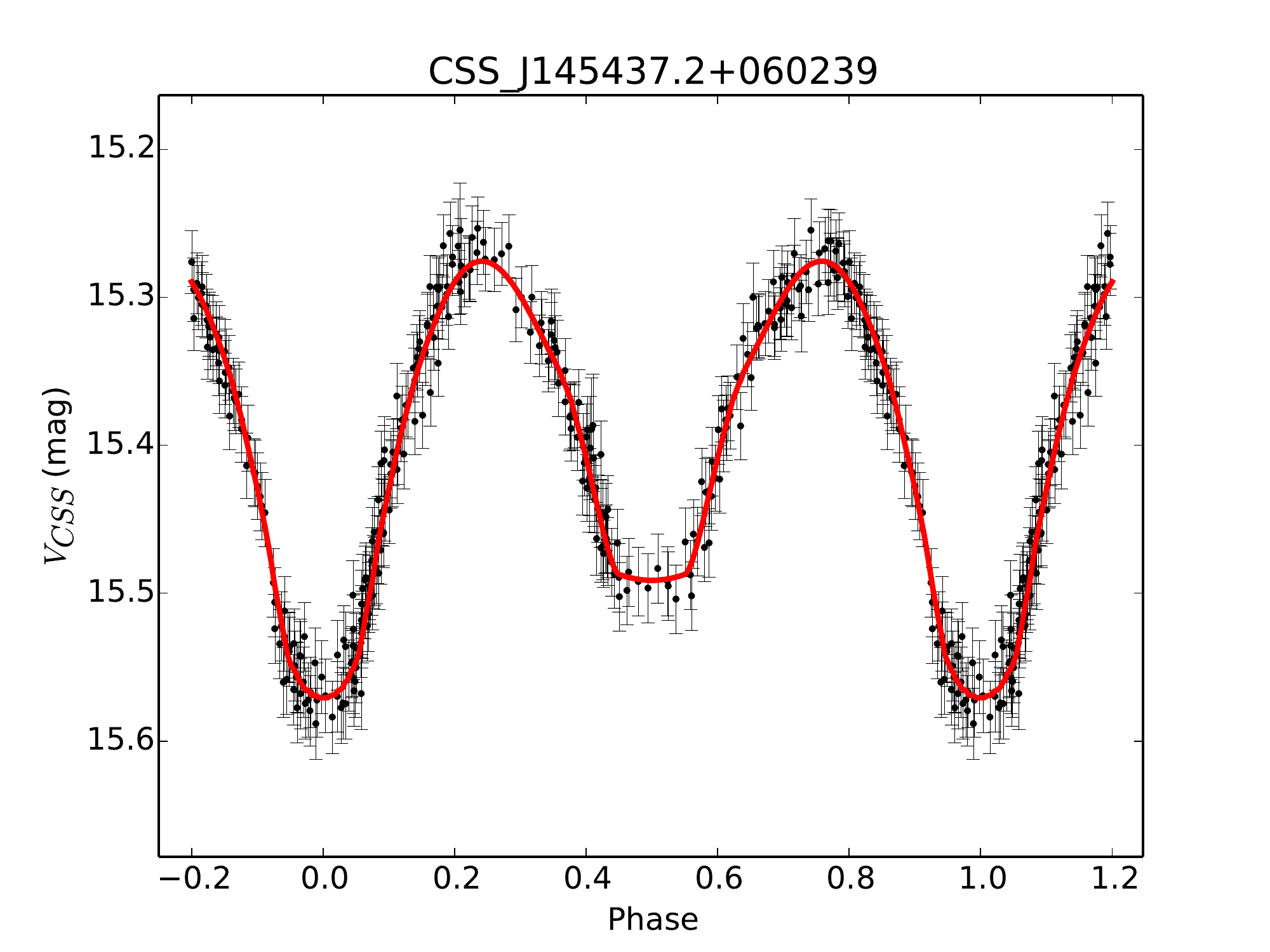} %
\endminipage
 
\minipage{0.32\textwidth}
\includegraphics[width=\linewidth]{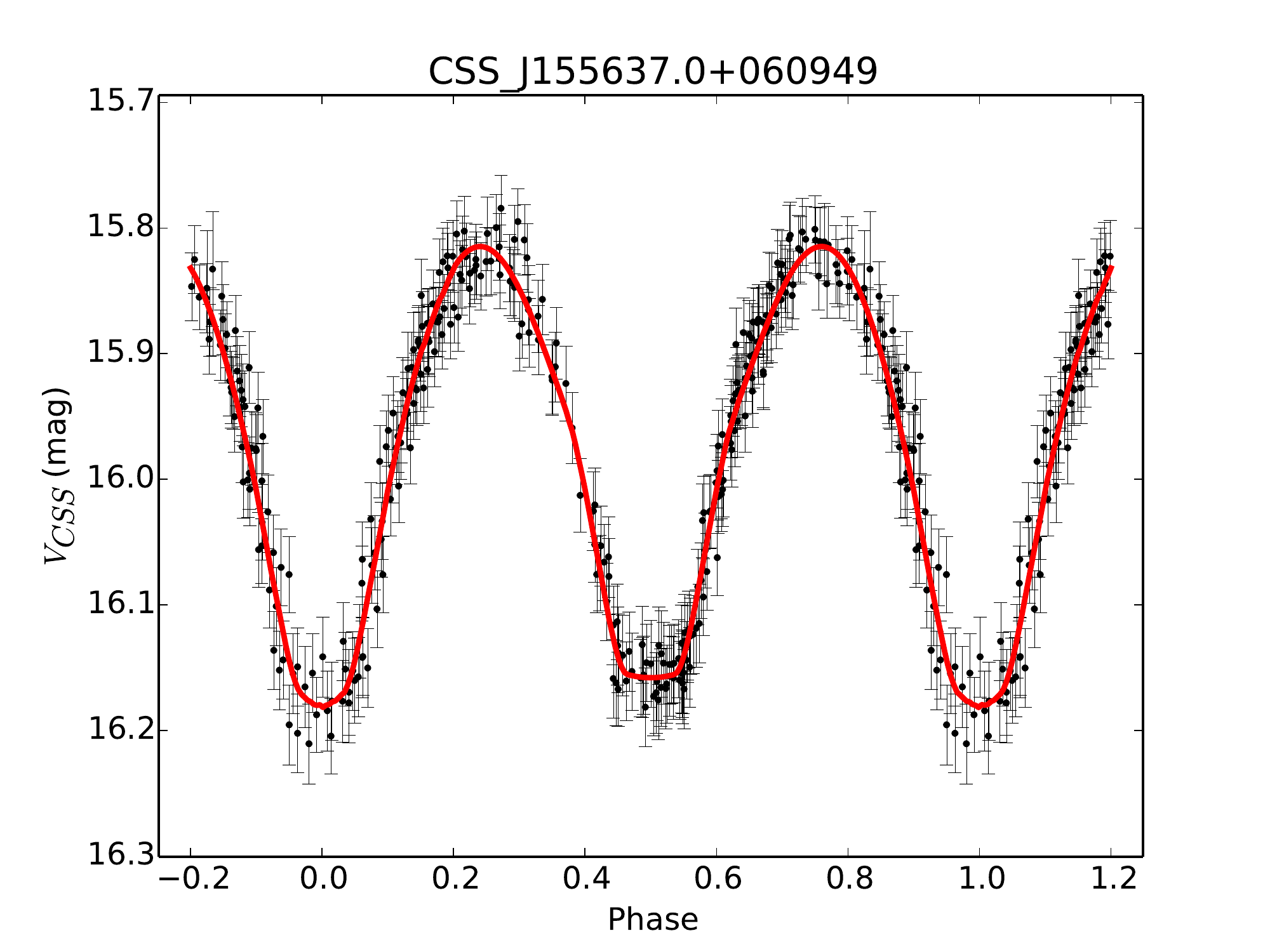} %
\endminipage\hfill
\minipage{0.32\textwidth}
\includegraphics[width=\linewidth]{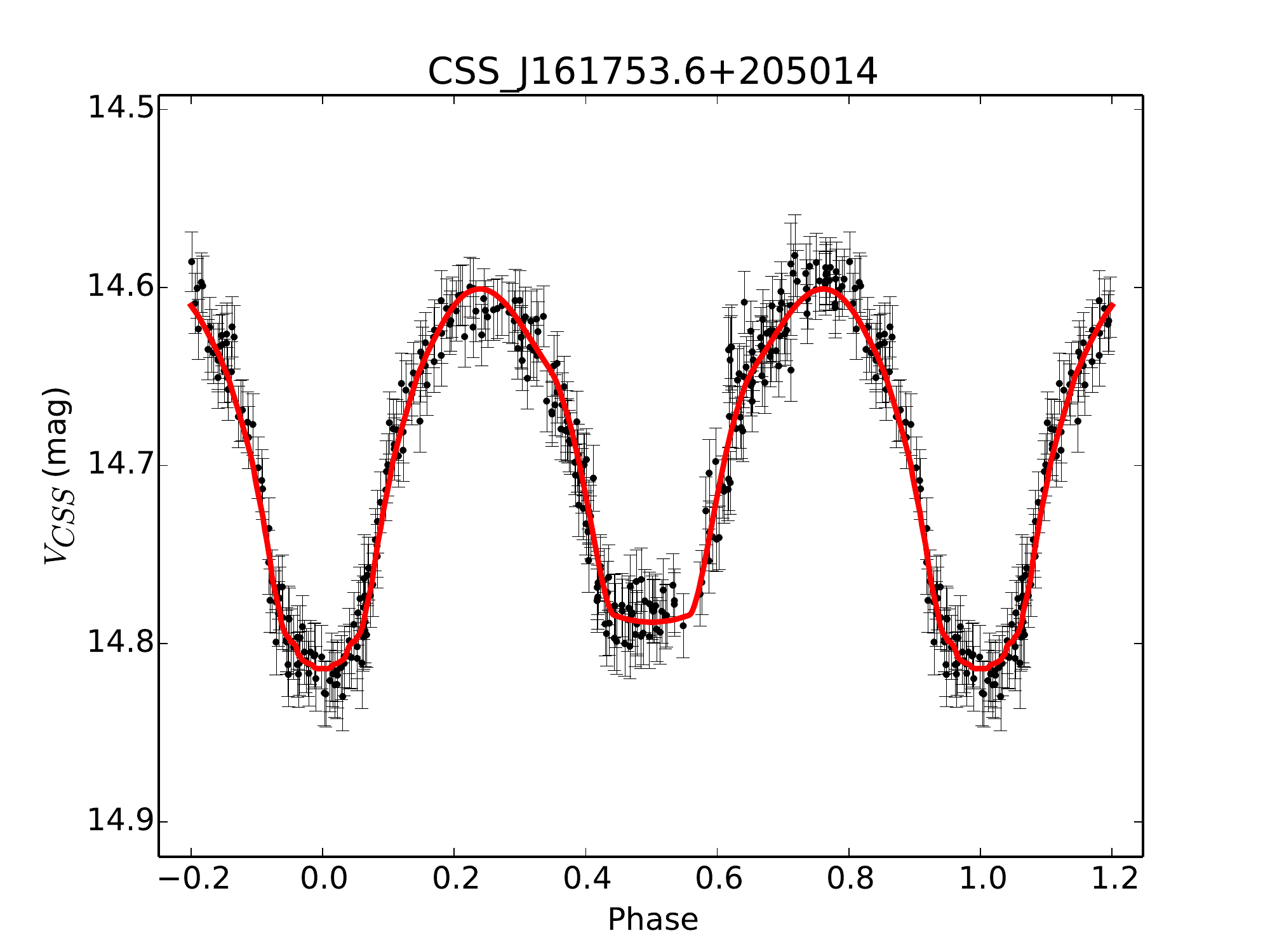} %
\endminipage\hfill
\minipage{0.32\textwidth}
\includegraphics[width=\linewidth]{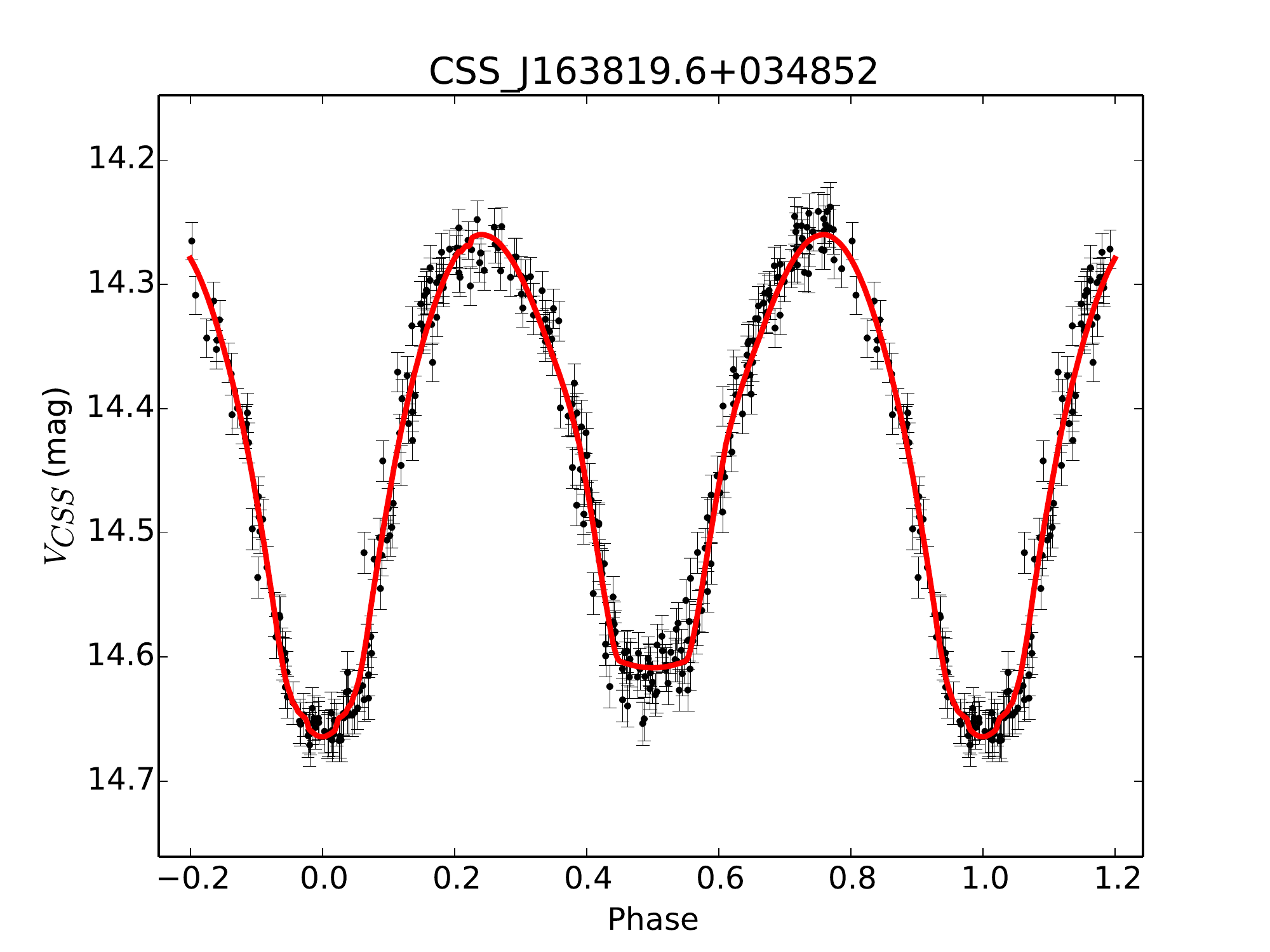} %
\endminipage
 
\minipage{0.32\textwidth}
\includegraphics[width=\linewidth]{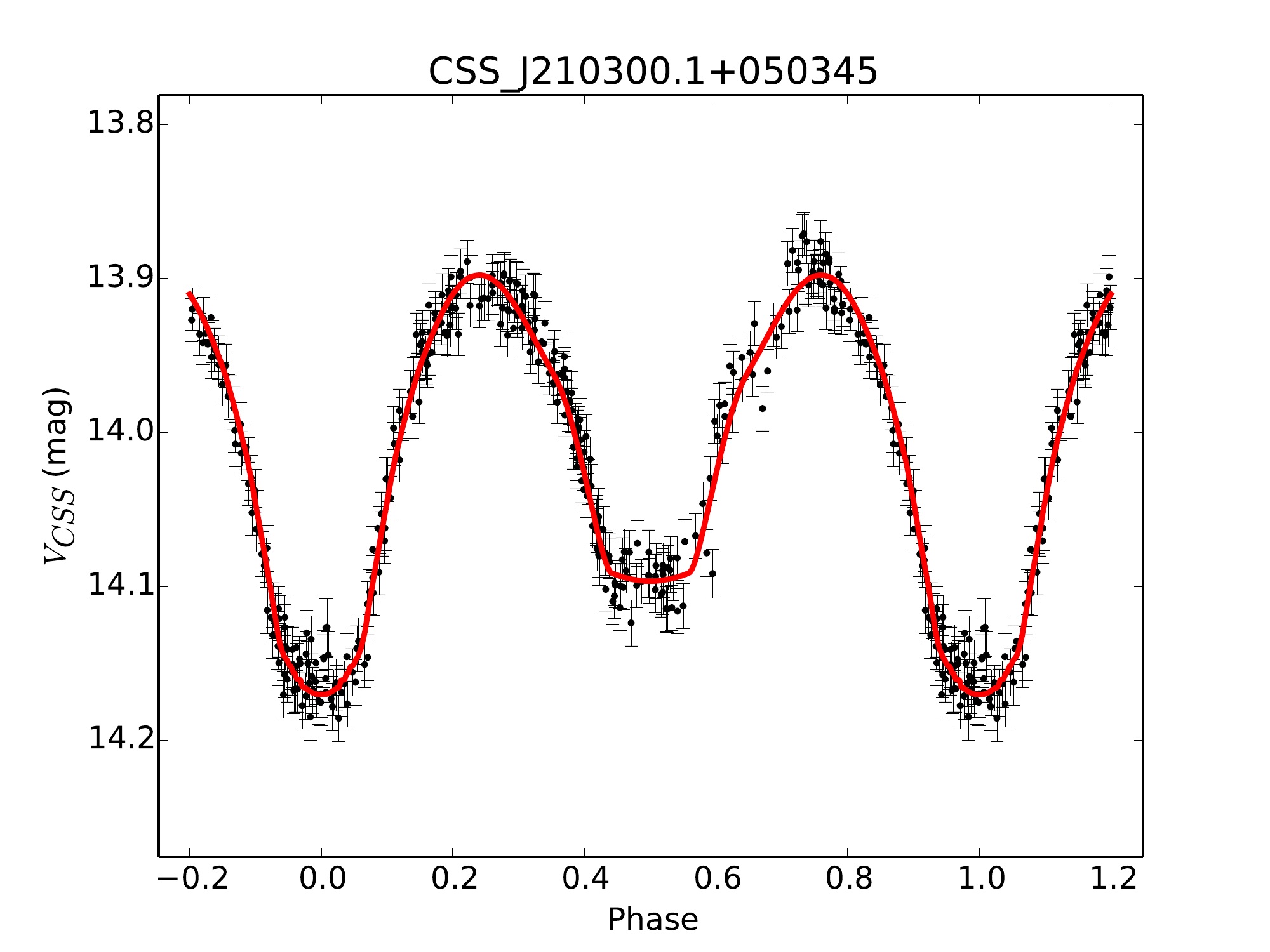} %
\endminipage\hfill
\minipage{0.32\textwidth}
\includegraphics[width=\linewidth]{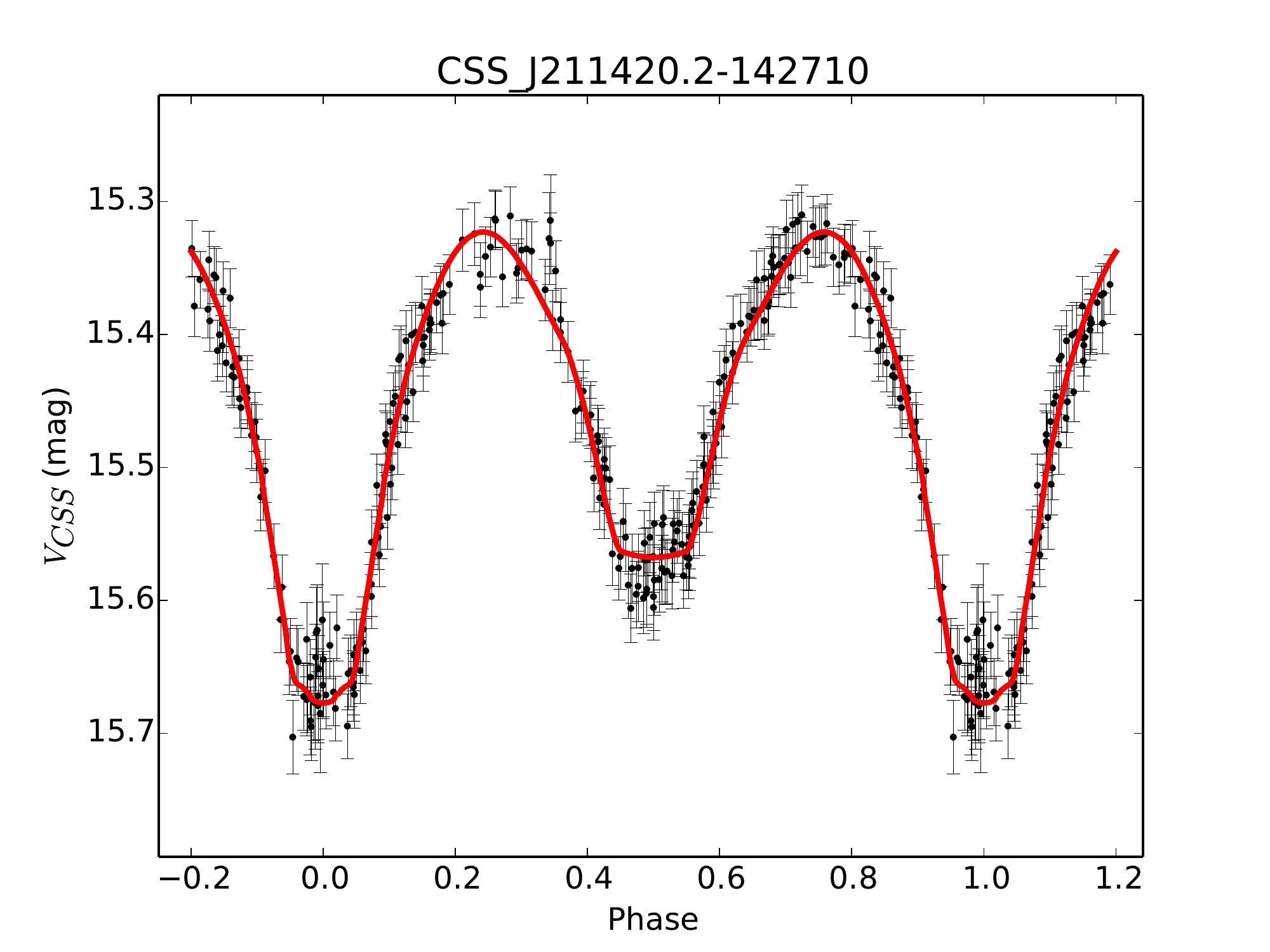} %
\endminipage\hfill
\minipage{0.32\textwidth}
\includegraphics[width=\linewidth]{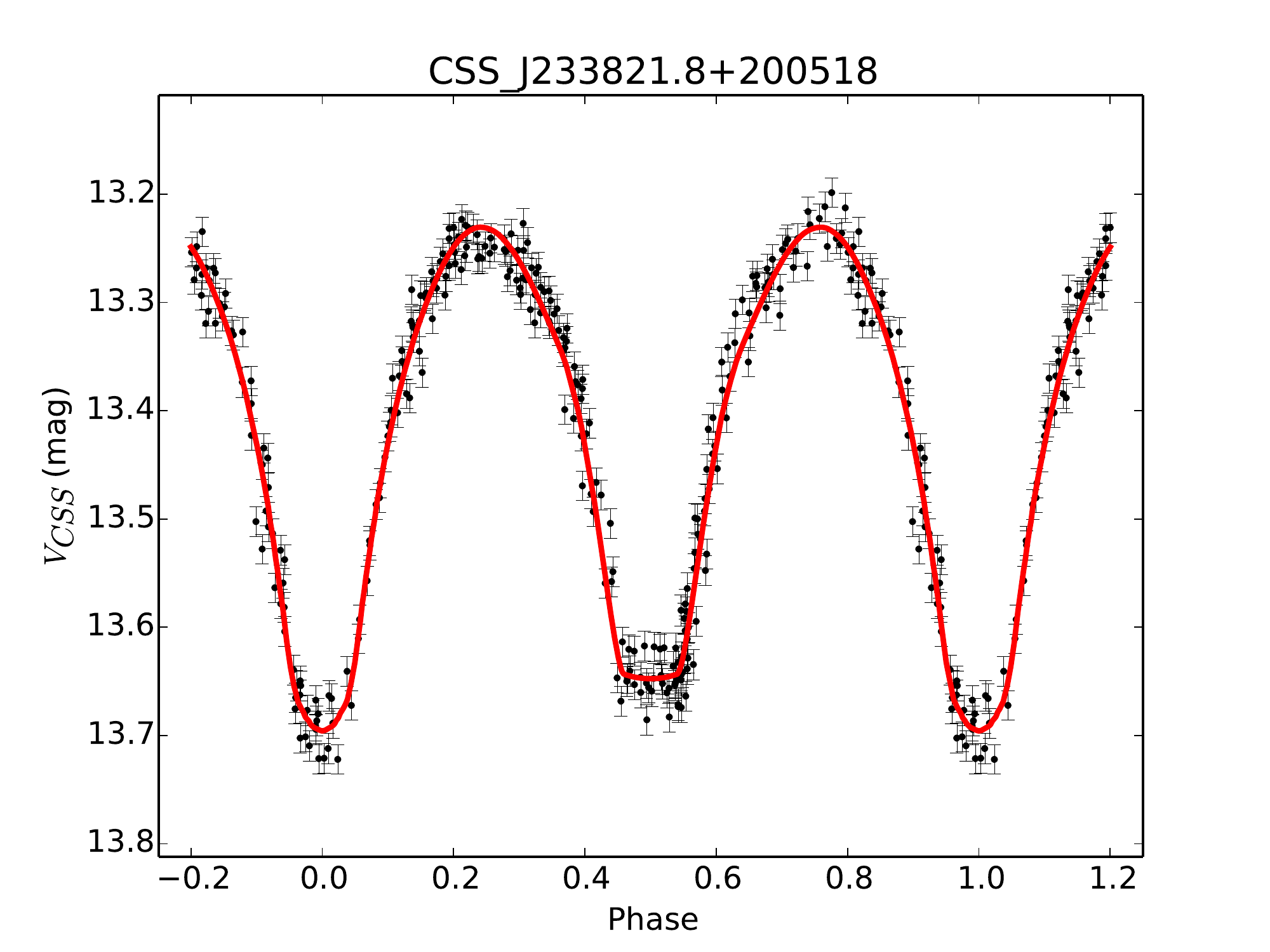} %
\endminipage
 
\minipage{0.32\textwidth}
\includegraphics[width=\linewidth]{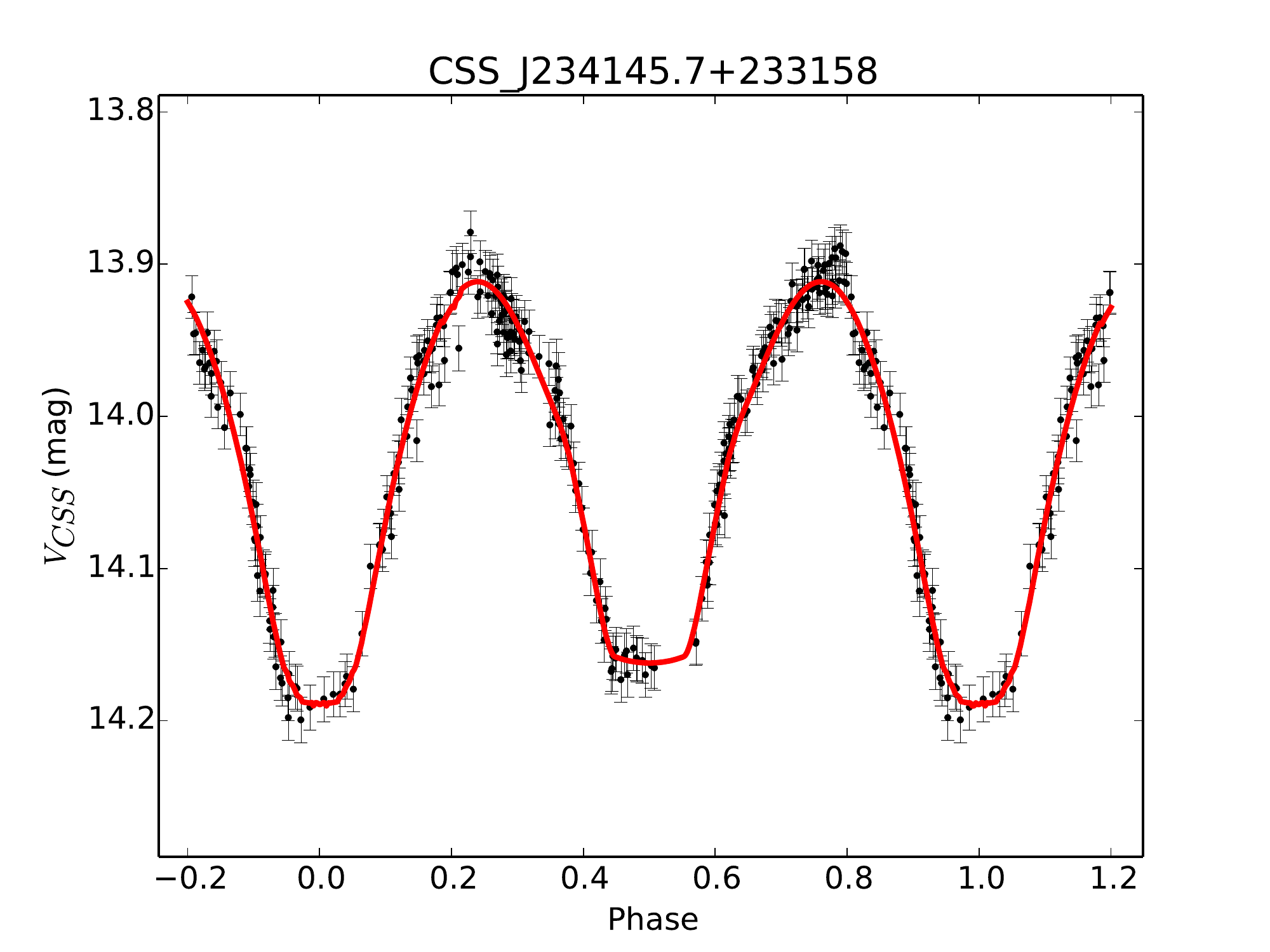} %
\endminipage\hfill
\minipage{0.32\textwidth}
\includegraphics[width=\linewidth]{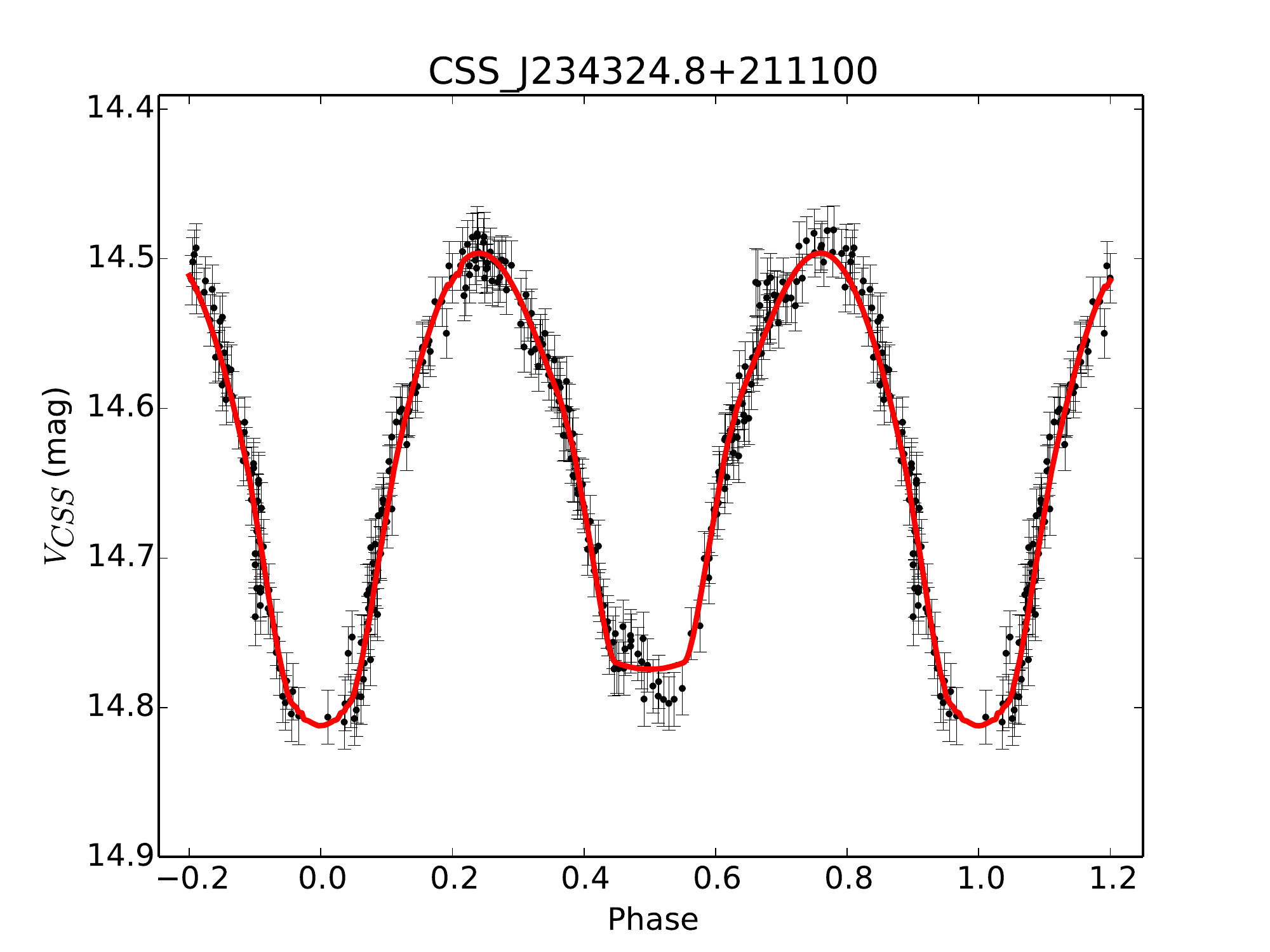} %
\endminipage\hfill
\minipage{0.32\textwidth}
\includegraphics[width=\linewidth]{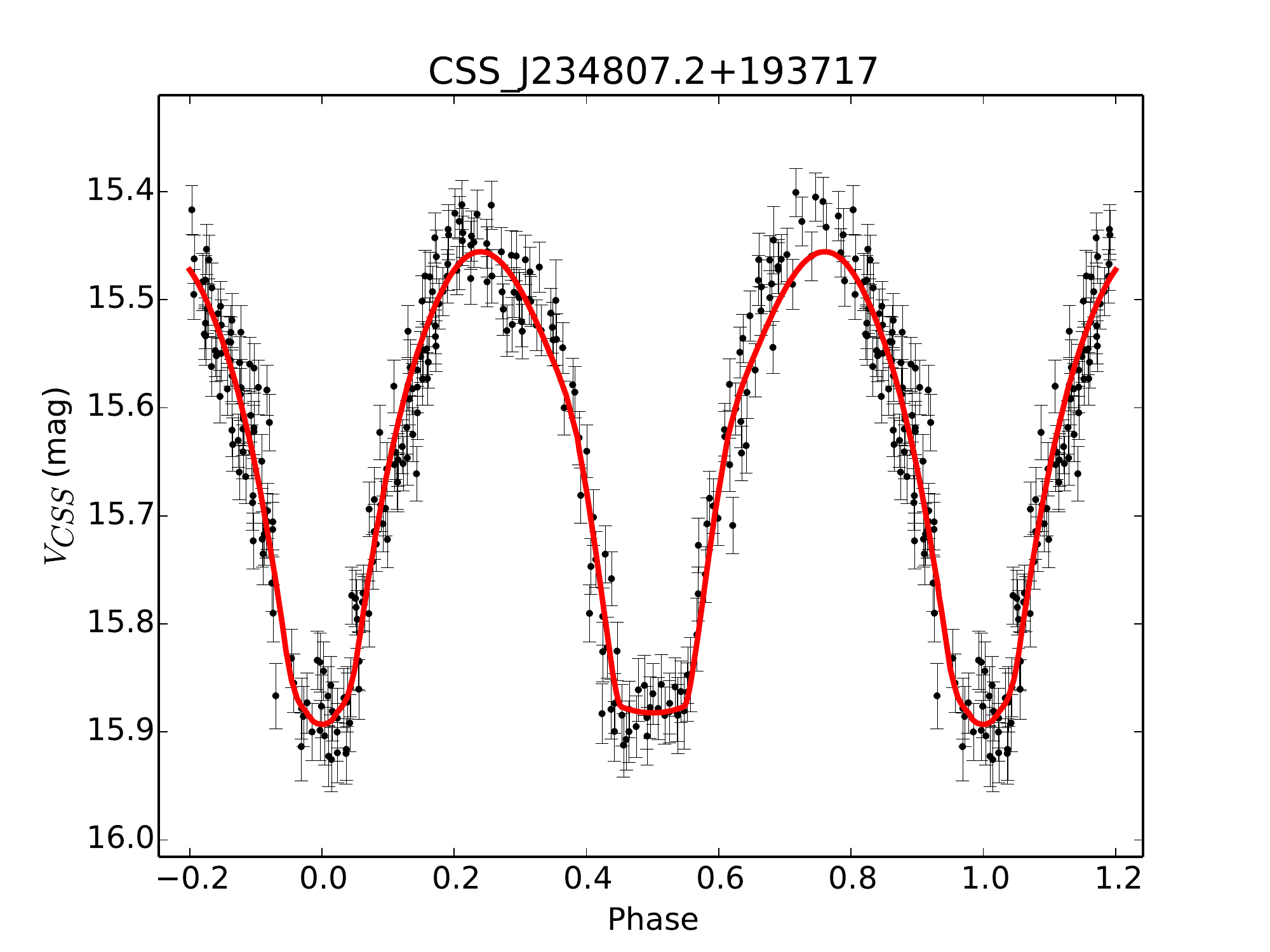} %
\endminipage
 
\end{figure*}


\bsp	
\label{lastpage}
\end{document}